\documentclass[aps,prd,twocolumn,showpacs,superscriptaddress]{revtex4}

\usepackage{graphicx}
\usepackage{color}
\usepackage{natbib}
\usepackage{epsfig}
\usepackage{float}
\usepackage{xcolor}
\usepackage{subfigure}
\usepackage{amsmath,amssymb,amsfonts}

\newcommand{\apjl}{Astrophys. J. Lett.}%
\newcommand{\apjs}{Astrophys. J. Supp.}%
\newcommand{\aap}{Astron. Astrophys.}%
\newcommand{\mnras}{Mon. Not. Roy. Astron. Soc.}%
\newcommand{\lrr}{Living Reviews in Relativity}%
\newcommand{\pasa}{Publications of the Astronomical Society of Australia}

\begin{document}

\title{Constraining the onset density of the hadron-quark phase transition with gravitational-wave observations}

\author{Sebastian Blacker}
\affiliation{GSI Helmholtzzentrum f\"ur Schwerionenforschung, Planckstra{\ss}e 1, 64291 Darmstadt, Germany}
\affiliation{Technische Universit\"at Darmstadt, Fachbereich Physik, Institut f\"ur Kernphysik, Schlossgartenstraße 9, 64289 Darmstadt, Germany}

\author{Niels-Uwe F. Bastian}
\affiliation{Institute of Theoretical Physics, University of Wroc{\l}aw, 50-205 Wroc{\l}aw, Poland}

\author{Andreas Bauswein}
\affiliation{GSI Helmholtzzentrum f\"ur Schwerionenforschung, Planckstra{\ss}e 1, 64291 Darmstadt, Germany}
\affiliation{Helmholtz Research Academy Hesse for FAIR (HFHF), GSI Helmholtz Center for Heavy Ion Research, Campus Darmstadt, Germany}

\author{David B. Blaschke}
\affiliation{Institute of Theoretical Physics, University of Wroc{\l}aw, 50-205 Wroc{\l}aw, Poland}
\affiliation{National Research Nuclear University (MEPhI), 115409 Moscow, Russia}
\affiliation{Bogoliubov Laboratory for Theoretical Physics, Joint Institute for Nuclear Research, 141980 Dubna, Russia}

\author{Tobias Fischer}
\affiliation{Institute of Theoretical Physics, University of Wroc{\l}aw, 50-205 Wroc{\l}aw, Poland}

\author{Micaela Oertel}
\affiliation{LUTH, Observatoire de Paris, PSL Research University, CNRS, Université de Paris, Sorbonne Paris Cité, 5 place Jules Janssen, 92195 Meudon, France}

\author{Theodoros Soultanis}
\affiliation{Heidelberg Institute for Theoretical Studies, Schloss-Wolfsbrunnenweg 35, 69118 Heidelberg, Germany}
\affiliation{Max-Planck-Institut f\"ur Astronomie, K\"onigstuhl 17, 69117 Heidelberg, Germany}

\author{Stefan Typel}
\affiliation{Technische Universit\"at Darmstadt, Fachbereich Physik, Institut f\"ur Kernphysik, Schlossgartenstraße 9, 64289 Darmstadt, Germany}
\affiliation{GSI Helmholtzzentrum f\"ur Schwerionenforschung, Planckstra{\ss}e 1, 64291 Darmstadt, Germany}

\date{\today} 

\begin{abstract}
We study the possible occurrence of the hadron-quark phase transition (PT) during the merging of neutron star binaries by hydrodynamical simulations employing a set of temperature dependent hybrid equations of state (EoSs). Following previous work we describe an unambiguous and measurable signature of deconfined quark matter in the gravitational-wave (GW) signal of neutron star binary mergers including equal-mass and unequal-mass systems of different total binary mass. The softening of the EoS by the PT at higher densities, i.e. after merging, leads to a characteristic increase of the dominant postmerger GW frequency $f_\mathrm{peak}$ relative to the tidal deformability $\Lambda$ inferred during the premerger inspiral phase. Hence, measuring such an increase of the postmerger frequency provides evidence for the presence of a strong PT. If the postmerger frequency and the tidal deformability are compatible with results from purely baryonic EoS models yielding very tight relations between $f_\mathrm{peak}$ and $\Lambda$, a strong PT can be excluded up to a certain density. We find tight correlations of $f_\mathrm{peak}$ and $\Lambda$ with the maximum density during the early postmerger remnant evolution. These GW observables thus inform about the density regime which is probed by the remnant and its GW emission. Exploiting such relations we devise a directly applicable, concrete procedure to constrain the onset density of the QCD PT from future GW measurements. We point out two interesting scenarios: if no indications for a PT are inferred from a GW detection, our procedure yields a lower limit on the onset density of the hadron quark PT. On the contrary, if a merger event reveals evidence for the occurrence of deconfined quark matter, the inferred GW parameters set an upper limit on the PT onset density. Both scenarios would thus have strong implications for high-density matter physics, e.g. determining the range of validity of nuclear physics and constraining the properties for quark deconfinement. These prospects demonstrate the importance of simultaneously measuring pre- and postmerger GW signals to exploit the complementarity of the information encoded in both phases. Hence, our work stresses the value added by dedicated high-frequency GW instruments. 

\end{abstract}

   \pacs{04.30.Tv,26.60.Kp,26.60Dd,97.60.Jd}

\maketitle   
   
\section{Introduction}   
One fundamental property of the theory of strong interactions with quark and gluon degrees of freedom -- quantum chromodynamics (QCD) -- is the running of the strong coupling constant. It results in asymptotic freedom, i.e. at arbitrary high energies quarks and gluons behave as non-interacting and massless particles~\cite{Freedman1976,Gorda2018}. This is the regime where perturbative QCD becomes valid, typically associated with high temperatures $T$ and baryon chemical potentials $\mu_{\rm B}$, respectively.

At vanishing baryon chemical potential, QCD predicts a smooth cross-over from normal nuclear (in general hadronic) matter to the quark-gluon plasma at a pseudocritical temperature of $156.5\pm 1.5$~MeV. This regime has been theoretically and experimentally explored by lattice QCD calculations and particle accelerators~\cite{Bazavov:2019,Andronic2017}.

At finite baryon chemical potential and low temperature the transition from nuclear matter with quarks being confined in hadrons to deconfined quark matter is less understood. In fact, the deconfinement PT is currently not accessible by ab-initio theoretical models or terrestrial experiments. In this non-perturbative regime various phenomenological approaches have been employed to describe quark matter like for instance thermodynamic bag models~\cite{Farhi:1984}, Nambu--Jona--Lasinio type models~\cite{Nambu1961,Klevansky1992,Buballa:2005} or approaches using the Dyson-Schwinger equations, functional renormalisation group developments or lattice developments for finite baryon chemical potential~\cite{Fischer2014,Gao:2020qsj,Philipsen:2019rjq}. But, it is for instance not clear at which baryon density the PT occurs, how exactly deconfinement takes place and what the thermodynamical properties of quark matter are, e.g. the equation of state (EoS) in the non-perturbative regime. Also in the hadronic regime at densities below the PT, the properties of nuclear matter, specifically the EoS, become increasingly uncertain at higher densities.

For these reasons it is also not known whether the hadron-quark PT takes place in neutron stars (NSs), which reach densities of several times nuclear saturation density but have generally low temperatures (see e.g.~\cite{Blaschke:2018mqw,Alford2019} for reviews on quark matter in NSs). The stellar structure of these compact objects is uniquely determined by the EoS, and the presence of a PT can leave a strong and characteristic imprint on the stellar parameters like especially the mass-radius relation. Astronomical observations of compact stellar objects may thus reveal signatures of the hadron-quark PT and elucidate quark deconfinement if it takes place in NSs (see e.g. ~\cite{Baldo2003,Baldo2004,Drago2004,Alford2005,Klaehn2007,Pagliara2008,Blaschke2009,Dexheimer2010,Weissenborn2011,Sedrakian2012,Sasaki2013,Klaehn2013,Weber2014,Kojo2015,Benic2015,Klaehn2015}~\cite{Sandin2007,Pagliara2009,Klaehn2017,Roark2019}~\cite{Dexheimer2018,Chesler2019,Wang2019,Ivanytskyi2019,Han2019a,Cierniak2019,Li2019,Motornenko2019,Kojo2019,Malfatti2019,Xia2019,Motornenko2019a,Montana2019,Xia2019a,Alvarez-Castillo2019,Fraga2019,Ferreira2020,Blaschke2020a,Miao2020,Annala2020,Han2020} for different hybrid EoS models and studies of isolated NSs).
Similarly, the formation of a NS in a core-collapse supernova depends on the EoS and the presence of a PT~\cite{Gentile1993,Nakazato2008,Sagert2009,Dasgupta2010,Nakazato2010,Fischer2011,Nakazato2013,Aloy2019}. The general dynamics and in particular the neutrino signal are affected if quarks occur in these events. 

With the very first detections of gravitational-wave (GW) signals from binary NS mergers \cite{Abbott2017,Abbott2020}, a new possibility to investigate the NS interior has become available. This offers prospects to reveal properties of the EoS at conditions where the hadron-quark PT may take place, e.g.~\cite{Oechslin2004,Bauswein2009,Bauswein2010a,Csaki2018,Paschalidis2018,Most2018a,Bauswein2019,Most2019,Bauswein2019a,Han2019,Christian2018,Sieniawska2018,Burgio2018a,Drago2018,Dexheimer2018,Chatziioannou2019,Chen2019,LlanesEstrada2019,Ecker2019,Weih2019,Orsaria2019,Gieg2019,DePietri2019,Christian2019,Bauswein2020,Friedman2020,Pang2020,Liebling2020}.

Generally, the merger dynamics and corresponding GW signals can be separated into two phases, an inspiral phase prior to the merger and a postmerger phase \cite{Faber2012,Paschalidis2017,Friedman2018,Bauswein2019b,Baiotti2019,DePietri2019a}. The GW signal from the inspiral phase enables a measurement of finite-size effects, which are dominantly described by the tidal deformability $\Lambda$ of the progenitor stars~\cite{Flanagan2008,Hinderer2008,Read2009,Hinderer2010,Read2013,DelPozzo2013,Wade2014,Agathos2015,Chatziioannou2015,Hotokezaka2016,Chatziioannou2018,Miller2019}. Inferring $\Lambda$ with some precision provides insights in the NS EoS as $\Lambda$ depends on the mass and the EoS~\cite{Abbott2017,Chatziioannou2018,De2018,Abbott2020,Landry2020}.

The GW signal from the postmerger phase contains information on the structure and the dynamics of the merger remnant. During the merger the densities and temperatures increase and hence the postmerger phase probes a different regime of the EoS compared to the inspiral phase. A robust feature in NS merger simulations which do not lead to a prompt black hole formation is the excitation of fluid oscillations in the remnant and associated GW emission. The dominant oscillation mode generates a pronounced peak in the GW spectrum at a frequency $f_{\rm peak}$, which is typically in the range between 2 and 4~kHz \cite{Bauswein2012,Bauswein2012a,Hotokezaka2013a,Takami2014,Bernuzzi2015,Bauswein2015}. Being not in the optimal frequency range of current GW detectors, this most prominent feature of the postmerger GW emission was not detected in GW170817 or GW190425. GW data analysis studies with simulated injections show that $f_{\rm peak}$ can be measured with good accuracy with current instruments operating at design sensitivity or with projected upgrades to the current detectors \cite{Clark2014,Clark2016,Chatziioannou2017,Bose2018,Yang2018,Torres-Rivas2018,Breschi2019,Easter2019,Martynov2019,Haster2020}.

In \cite{Bauswein2019} we demonstrated that the simultaneous detection of $\Lambda$ and $f_\mathrm{peak}$ provides an unambiguous signature of a strong first-order PT occurring in the remnant during the merger~\footnote{For the sake of clarity, in this introduction we do not explicitly distinguish between the tidal deformability $\Lambda$ of an individual star and the combined tidal deformability $\tilde{\Lambda}$ of a binary.}. The presence of such a PT leads to a softening of the EoS at high baryon density and hence to a more compact remnant with higher postmerger oscillation frequencies. This results in a significant deviation from an empirical relation between $\Lambda$ and $f_{\rm peak}$ which holds for purely hadronic EoSs~\cite{Bauswein2019}. Note that both quantities will be measurable with sufficient precision in the near future to identify the presence or absence of a strong PT. Furthermore, in \cite{Bauswein2019} we discovered a relation between the maximum density that is obtained during the early postmerger evolution and $f_{\rm peak}$, and we pointed out that this relation can be used to constrain the onset densities of the deconfinement PT.

In this work we follow up on our findings and present a detailed procedure to constrain the onset density of a strong PT. We develop a scheme, which is ready to use and which can be immediately applied when observational data become available. The constraint is based on the aforementioned observation that a sufficiently strong PT leads to a characteristic increase of the dominant postmerger GW frequency $f_\mathrm{peak}$ and that $f_{\rm peak}$ scales with the maximum density $\rho_{\rm max}^{\rm max}$ which is encountered during the early postmerger evolution. Thus, the GW signal contains information about the highest density reached in the remnant. See also~\cite{Bauswein2014a} for a discussion on how the GW signal can provide information on the highest densities reached in isolated, static NSs at the maximum mass.

After a detection and measurement of $\Lambda$ and $f_{\rm peak}$ two outcomes are possible~\cite{Bauswein2019}:
\begin{enumerate}
\item[(1)] The measured values of $\Lambda$ and $f_{\rm peak}$ do not provide evidence for the occurrence of a PT, i.e. $\Lambda$ and $f_{\rm peak}$ are compatible with an empirical relation between both quantities, which holds for purely hadronic EoSs. In this case no PT occurred during merging and the measured $f_{\rm peak}$ in combination with the $\rho_{\rm max}^{\rm max}(f_{\rm peak})$ relation yields a \textit{lower limit} for the onset density. 
\item[(2)] If the measured dominant postmerger GW frequency is increased, compared to the empirical  $f_{\rm peak}(\Lambda)$ relation, this provides strong evidence for the occurrence of quark matter during merging. In this case the measured $f_\mathrm{peak}$ or $\Lambda$, respectively, yield an \textit{upper limit} on the onset density of the PT. 
\end{enumerate}
We also note that for a quantitatively reliable procedure, it is essential to develop an effective description to incorporate additional effects such as the temperature and composition dependence of the phase boundaries.

There have been previous studies focused on identifying a PT solely from the behavior of the tidal deformability (e.g. \cite{Chen2019,Chatziioannou2019,Pang2020}), i.e. investigating the prospects to identify a kink in the tidal deformability as a function of $M$ (see e.g. Fig.~3 in \cite{Bauswein2019a}). These methods require many observations of NS mergers and rather accurate measurements of $\Lambda$ to resolve a potential kink in $\Lambda(M)$. Finally, it is not clear, how often the inspiraling stars fall into the mass range around the first occurrence of quark matter although the total mass in GW190425 was relatively high. The advantage of our procedure is that it allows for a constraint on the transition density from one single simultaneous detection of $\Lambda$ and $f_{\rm peak}$. We exploit the fact that merging leads to a density increase and thus the postmerger phase naturally probes higher densities of the EoS than the premerger stage. We also refer to \cite{Bauswein2019a} showing that the occurrence of a PT alters the mass ejecta of a NS merger and thus the electromagnetic counterpart in the IR and optical range (see \cite{Metzger2019} for a review on these so called kilonovae). The impact on the mass ejection is however neither systematic nor overly strong compared to binary NS merger simulations with hadronic EoSs. It may thus be challenging to unambiguously identify the signature of a PT in the electromagnetic emission of a NS merger (see also \cite{Bauswein2019a} for a discussion of possibly more subtle effects). We also refer to~\cite{Bauswein2020}, where we stress that a determination of the threshold binary mass $M_\mathrm{thres}$ for prompt black-hole formation can be indicative of a PT if combined with information on the combined tidal deformability. In summary, these different methods to identify a PT can yield additional information which can in principle be incorporated in the procedure to constrain the onset density.

This paper consists of two parts. We first discuss in detail the influence of a strong PT on the GW signal and which signature provides an unambiguous indication of the presence or absence of such a transition. Here we generalize our findings of \cite{Bauswein2019} to systems with arbitrary mass. In the second part we distinguish two cases depending on whether or not evidence for a PT in the merger is identified. For each case we describe the resulting constraint, i.e. an upper or lower bound on the onset density of the QCD PT. We stress that our procedure is only applicable to identify or exclude a sufficiently strong PT. A weak transition will hardly alter the compactness and hence the postmerger oscillation frequencies of the remnant will not be shifted to higher values. Although we do not provide explicit tests for other scenarios, we emphasize that the PT does not necessarily need to be first order to lead to an observable impact. Any transition resulting in a significant softening of the EoS, and subsequent stiffening towards higher density, will likely lead to the effects described here. Note that our procedure requires the remnant to be at least temporarily stable to obtain a sufficiently strong postmerger GW signal. A too strong softening of the high-density EoS may however lead to an immediate collapse of the remnant to a black hole. In the remaining article we will use the term ``strong'' PT in the sense of a transition with large latent heat leading to a softening of the EoS and a significant impact on the stellar structure.

Throughout this work, the term hybrid star refers to stable stars with a pure quark matter core.

This paper is structured as follows: In Sect.~\ref{Simstuff} we describe the used EoS models, the simulation setup as well as simulation results. The signature of a strong first-order PT is discussed in Sect.~\ref{fpeaklambda}. In Sect.~\ref{postdensities} we explain how $f_{\mathrm{peak}}$ is linked to the maximum densities reached in the remnant soon after the merging. Sect.~\ref{constraining} contains the procedure to constrain the onset density of the PT from a simultaneous measurement of $f_{\mathrm{peak}}$ and $\Lambda$. We summarize and conclude in Sect.~\ref{summary}.

Unless noted otherwise stellar masses refer to the gravitational mass in isolation. For binary systems we consider the gravitational binary mass at infinite separation.

\section{EoS models, setup and simulations} \label{Simstuff}

\textit{Equations of state} --- In this work we use the same set of hybrid EoSs as in \cite{Bauswein2019}. They are based on the microscopic hadron-quark EoS DD2-SF of \cite{Fischer2018,Bastian2018}, featuring a strong first-order PT to deconfined quark matter. A phase transition is obtained fulfilling the Gibbs conditions including both charges (electric and baryonic) at every point of the phase boundary and a mixed phase construction while preserving global charge neutrality~\cite{Bastian2020}.
Alternative approaches to construct the PT have been discussed in Refs.~\cite{Yasutake2014,Yasutake2019}. The pure quark matter phase is described by the microscopic two-flavor string-flip model (SF) obtained within the density-functional formalism (further details can be found in \cite{Kaltenborn2017} and references therein). Higher order terms of repulsive vector interactions among quarks are also considered to guarantee sufficient stiffness of the quark phase in order to allow for stable hybrid stars with maximum masses above 2~M$_\odot$, in agreement with the observations of the presently most massive NSs~\cite{Antoniadis2013,Cromartie2019}. The hadronic phase is described by the DD2F EoS~\cite{Typel2010,Alvarez-Castillo2016}. It is is based on the relativistic mean-field approach with density dependent-couplings yielding quantitative agreement with constraints provided by nuclear physics~\cite{Lattimer2013,Fischer2017}. At densities below nuclear saturation density and at low temperatures, light and heavy nuclear clusters are present, for which we apply the modified nuclear statistical equilibrium approach of \cite{Hempel2009}, which is based on several 1000 nuclear species taking into account nuclear shell effects.

By varying the SF parameters, hybrid EoSs with different properties are constructed. The hadronic regime, however, is based on the same DD2F EoS in all cases. Different onset densities of the PT result from different parameterization of the SF model. The seven sets of parameters we use here can be found in the supplement material of \cite{Bauswein2019} together with selected properties of the resulting model EoSs. Following the notation of this reference we label the hybrid EoS models DD2F-SF-n with n $\in\{1,2,3,4,5,6,7\}$. To refer to the whole set of all seven hybrid models we use the acronym DD2F-SF. The different hybrid models differ in the onset density $\rho_\mathrm{onset}$ of the PT, the latent heat and the stiffening of the pure quark matter phase. The latter relates to the maximum mass of the hybrid EoS. Consequently, these different DD2F-SF EoSs lead to different mass-radius relations for cold, non-rotating hybrid stars, in particular, with different maximum masses.

All DD2F-SF models employ  a microscopic temperature dependence at the level of Fermi-Dirac distribution functions, as well as isospin dependence. The latter aspect is important due to the fact that matter in simulations of NS mergers can feature arbitrary isospin asymmetry. 
Note further that the phase boundaries of the DD2F-SF EoSs show a mild temperature dependence in the relevant temperature range, e.g. for the DD2F-SF-1 EoS we obtain:
\begin{equation*}
\begin{array}{c|c|c}
T~[\rm MeV] & \rho_{\rm onset}~[\rho_{\rm sat}] & \rho_{\rm final}~[\rho_{\rm sat}] \\
\hline
0 & 3.26 & 3.87 \\
30 & 2.52 & 3.68
\end{array}
\end{equation*}
with nuclear saturation density, $\rho_{\rm sat}\simeq 2.7\times10^{14}$~g~cm$^{-3}$. $\rho_\mathrm{onset}$ and $\rho_\mathrm{final}$ specify the density jump across the PT with $\rho_\mathrm{final}$ referring to the rest-mass density, where pure quark matter is present.

In this study we also perform simulations with the purely hadronic DD2F model as a reference model without a PT.

Additionally, we use a set of 15 other EoSs which serves as a representative sample of purely hadronic EoSs. These EoSs are APR \cite{Akmal1998}, BHBLP \cite{Banik2014}, BSK20 \cite{Goriely2010}, BSK21 \cite{Goriely2010}, DD2 \cite{Typel2010,Hempel2010}, DD2Y \cite{Marques2017}, eosUU \cite{Wiringa1988}, GS2 \cite{Shen2011}, LS220 \cite{Lattimer1991}, LS375 \cite{Lattimer1991}, SFHO \cite{Steiner2013}, SFHOY \cite{Fortin2018}, SFHX \cite{Steiner2013}, Sly4 \cite{Douchin2001} and TMA \cite{Hempel2012,Toki1995} (see \cite{Bauswein2012a,Bauswein2013a,Bauswein2014a}, for more details on the different EoSs and the meaning of the acronyms). Except for GS2, LS375 and TMA, all EoSs are compatible with the tidal deformability limits inferred from GW170817 at the 90\% credible interval. All EoS models are consistent with radius constraints derived from a multi-messenger interpretation of GW170817 \cite{Bauswein2017,Bauswein2019c} and with the NS maximum mass limit set by \cite{Antoniadis2013}. Some models are in tension with the one-sigma limit of \cite{Cromartie2019}.

The three models BHBLP, SFHOY and DD2Y include a PT to hyperonic matter. In these EoSs the hyperonic interactions are modeled to be compatible with a maximum mass of 2~M$_{\odot}$ for a cold, non-rotating NS~\cite{Antoniadis2013,Cromartie2019} as well as with data from hypernuclei~\cite{Fortin2017a,vandalen_14,SchaffnerBielich:2000wj,Schaffner:1993qj}.

\textit{Setup} --- We consider the following symmetric binary systems 1.2--1.2~M$_{\odot}$, 1.35--1.35~M$_{\odot}$, 1.4--1.4~M$_{\odot}$ and 1.5--1.5~M$_{\odot}$. In order to explicitly study the effect of the binary mass ratio, $q=M_{1}/M_{2}\le 1$, we perform simulations with  1.3--1.4~M$_{\odot}$ binaries, i.e. $q\approx 0.929$.

We simulate NS mergers with a general relativistic, smoothed particle hydrodynamics (SPH) code using the conformal flatness condition \cite{Isenberg1980,Wilson1996} to solve the field equations (see \cite{Oechslin2002,Oechslin2007,Bauswein2010a,Bauswein2012a} for more information). The stars are initially set up as cold, irrotational stars in neutrino less beta-equilibrium. The simulation starts from circular quasi-equilibrium orbits a few revolutions before merging. The system is relaxed for a short time to ensure the distribution of the SPH particles is in equilibrium before the actual simulation of the merger begins.

If provided by the EoS, temperature effects are taken into account consistently during the simulation. This is in particular the case for all hybrid DD2F-SF models and the hadronic reference model DD2F.

For those EoSs where the temperature dependencies are not available, we include thermal effects by an approximate treatment (see \cite{Bauswein2010} for an in-depth discussion). This treatment requires to choose a coefficient $\Gamma_{\mathrm{th}}$ regulating the strength of the thermal pressure contribution. We use $\Gamma_{\mathrm{th}}=1.75$ in all simulations where we employ this treatment. This value has been picked to reproduce results with fully temperature dependent EoSs relatively well (see \cite{Bauswein2010}).

\begin{figure*}[ht]  
\centering
\subfigure[]{\includegraphics[width=0.48\linewidth]{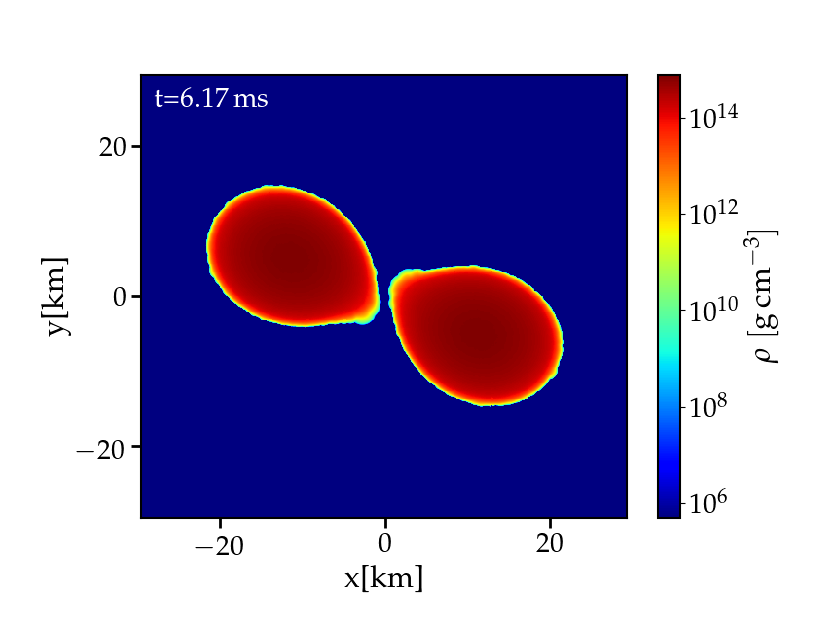}\label{denistyevolution_a}}
\hfill
\subfigure[]{\includegraphics[width=0.48\linewidth]{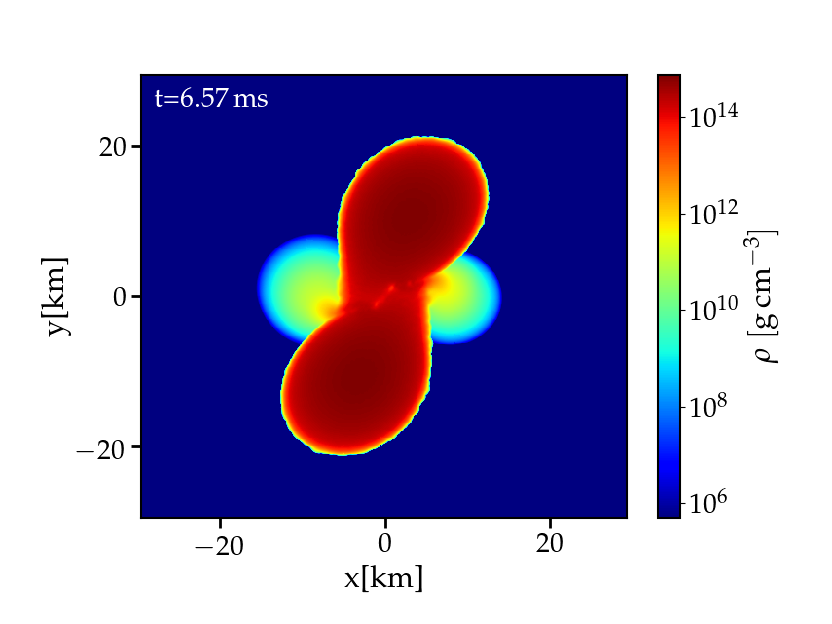}\label{denistyevolution_b}}
\\
\subfigure[]{\includegraphics[width=0.48\linewidth]{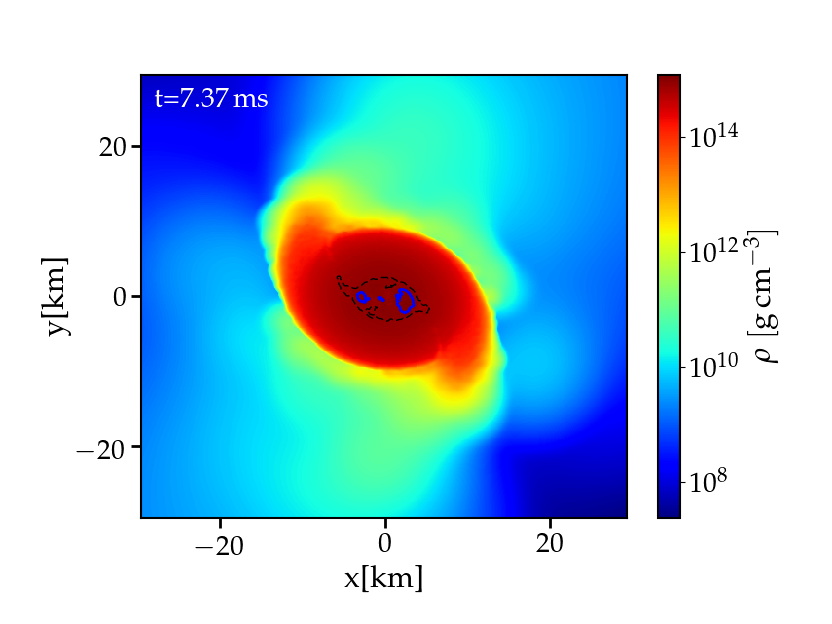}\label{denistyevolution_c}}
\hfill
\subfigure[]{\includegraphics[width=0.48\linewidth]{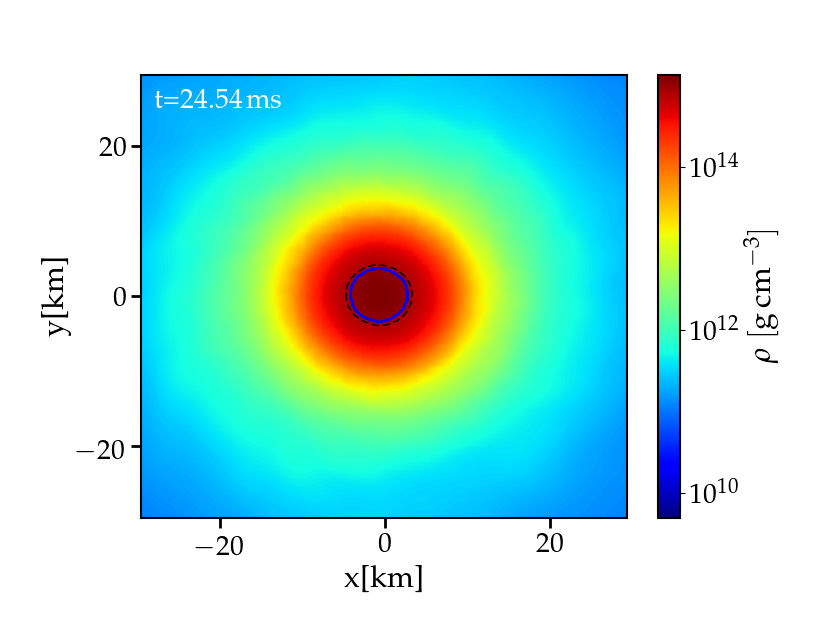}\label{denistyevolution_d}}
\caption{Rest-mass density (color-coded) in the equatorial plane of a merger simulation of two 1.35~M$_{\odot}$ NSs described by the DD2F-SF-6 EoS. The dashed, black line marks the region corresponding to the onset of the hadron-quark PT. The solid, blue line encloses regions with pure quark matter.}
\label{densityevolution}
\end{figure*}

\begin{figure*}[ht]  
\centering
\subfigure[]{\includegraphics[width=0.48\linewidth]{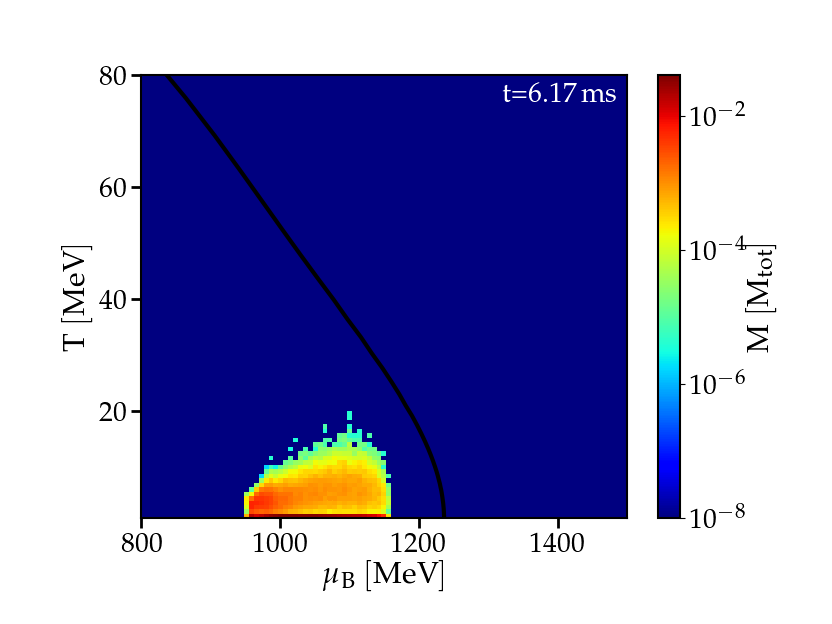}\label{tmuevolution_a}}
\hfill
\subfigure[]{\includegraphics[width=0.48\linewidth]{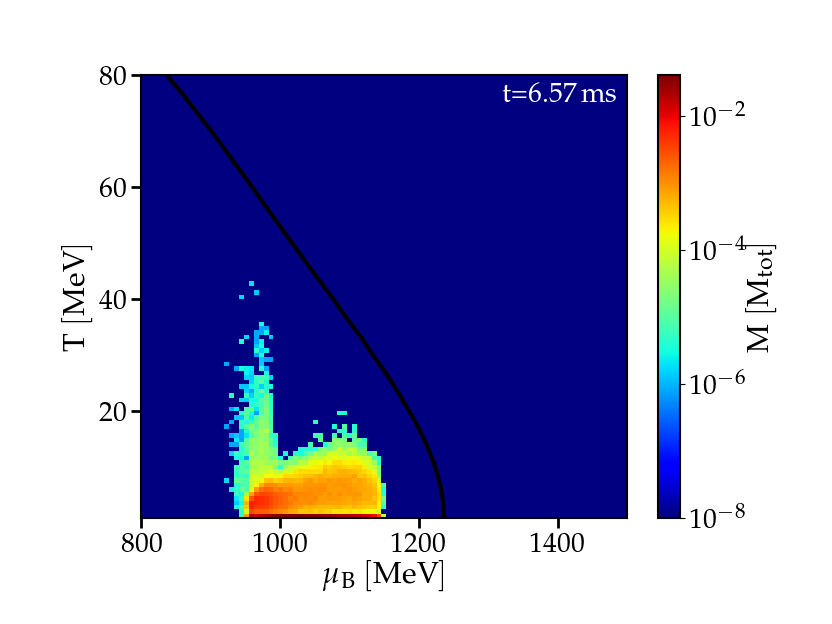}\label{tmuevolution_b}}
\\
\subfigure[]{\includegraphics[width=0.48\linewidth]{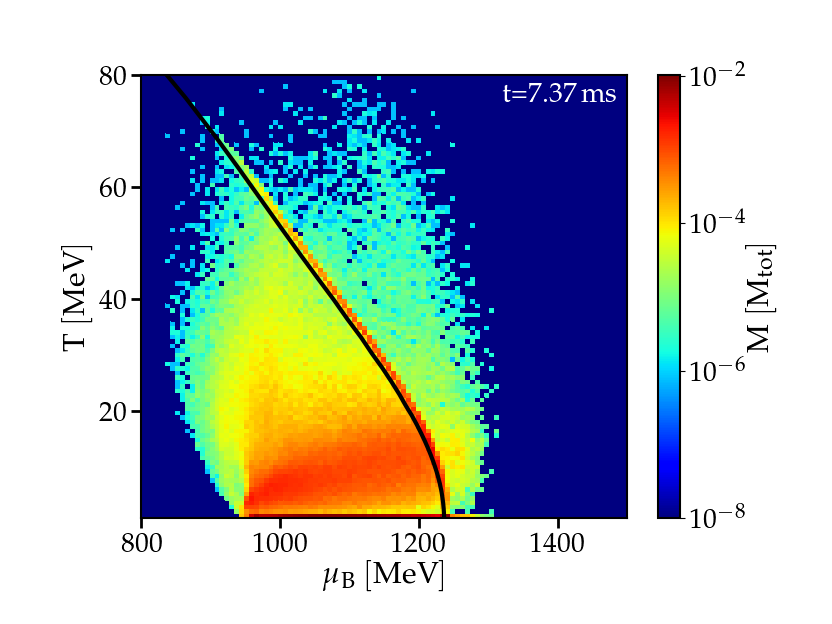}\label{tmuevolution_c}}
\hfill
\subfigure[]{\includegraphics[width=0.48\linewidth]{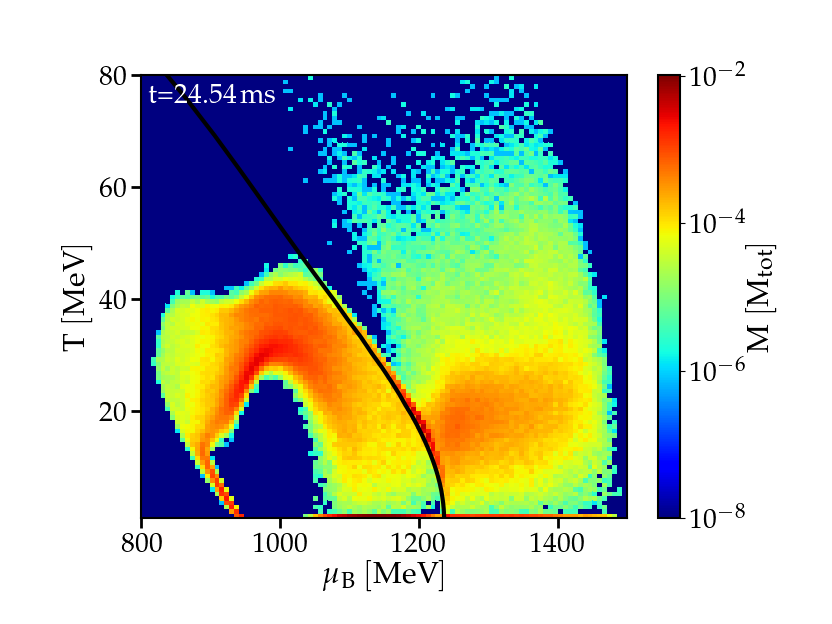}\label{tmuevolution_d}}
\caption{Rest mass distribution of matter in the baryon chemical potential--temperature plane for the merger simulation from Fig.~\ref{densityevolution} normalized to the total mass of the system. Note the logarithmic mass scale. The black line is the temperature dependent phase boundary between the hadronic phase at low chemical potential and the deconfined quark phase at high chemical potential. The total amount of matter at temperatures below 5\,MeV in the graphs (a)-(d) are about 92\%, 92\%, 42\% and 19\%, respectively.}
\label{tmuevolution}
\end{figure*}

\textit{Simulations} --- Figure~\ref{densityevolution} illustrates the evolution of the rest-mass density in the equatorial plane for the system with initial masses of 1.35--1.35~M$_{\odot}$ using the DD2F-SF-6 EoS. These plots were done by mapping the SPH-data onto a grid using the S-normed SPH binning method described in \cite{Roettgers2018}. The two contour lines indicate the presence of deconfined quark matter. The dashed, black line indicates the location corresponding to the onset of the hadron-quark PT, while the blue, solid line marks the area with pure quark matter. Note that the chosen mass configuration is comparable with the total mass of GW170817~\cite{Abbott2017,TheLIGOScientificCollaboration2018a,Abbott2018} and represents a likely binary configuration according to population studies and pulsar observations, e.g.~\cite{Lattimer2012,Dominik2012}. In Fig.~\ref{tmuevolution} we show the distribution of matter at the same times as in Fig.~\ref{densityevolution} in the $\mu_\mathrm{B}$--T plane, where $\mu_\mathrm{B}$ is the baryon chemical potential. The solid, black lines in these figures represent the temperature dependent phase boundary between the hadronic and the deconfined quark matter phase. We also provide the matter distributions in the $\rho$--T plane in Appendix~\ref{apprhot}.

Figure~\ref{denistyevolution_a} shows the system shortly before merging. The individual stars are still separated, however, one can clearly recognize tidal deformations. Because the densities are still below the transition density, no deconfined quark matter is present, as can also be seen from Fig.~\ref{tmuevolution_a}. Note that numerical heating is present in the simulation, which explains the small amount of matter with finite temperatures at this time. However, about 92\% of the matter is still at low temperature (below 5\,MeV). Figure~\ref{denistyevolution_b} depicts the merging of the stars into a single, rapidly rotating object. The densities and temperatures in this merger remnant increase significantly.

The temperatures reach several tens of MeV and the largest densities surpass the temperature dependent onset density of the hybrid DD2F-SF-6 EoS (see also Refs.~\cite{Fischer2018,Bastian2020} for more information on the phase boundaries of the EoSs). Although the total mass of the remnant exceeds the maximum mass of a non-rotating NS, rapid, differential rotation and the thermal pressure stabilize the object against the gravitational collapse. Initially, the remnant strongly oscillates producing postmerger GW emission. Figure~\ref{denistyevolution_c} shows the remnant a few milliseconds after merging. One can clearly see the distortion of the whole remnant. Also, the hadron-quark PT has now taken place in the central region. Due to the non-congruent character of the hadron-quark transition, a small but nonzero pressure gradient is observed from hadron to quark matter as function of baryon density $n_\mathrm{b}$ and for constant hadronic charge fraction  $Y_\mathrm{c}$~\cite{Hempel2009a,Gulminelli2012,Hempel2013,Gulminelli2013}. This pressure gradient vanishes only for symmetric matter.

The increase in temperature and the appearance of deconfined quark matter can also bee seen in Fig.~\ref{tmuevolution_b} and Fig.~\ref{tmuevolution_c}. However, a significant amount of matter ({\raise.17ex\hbox{$\scriptstyle\mathtt{\sim}$}}92\% in Fig.~\ref{tmuevolution_b} and {\raise.17ex\hbox{$\scriptstyle\mathtt{\sim}$}}42\% in Fig.~\ref{tmuevolution_c}) is still at low temperature.
Note that the apparent accumulation of matter at the phase boundary in Fig.~\ref{tmuevolution_c} is caused by the fact that the matter in the mixed phase has a constant chemical potential. Hence, for constant temperature all mass in the mixed phase appears at a single value of $\mu_\mathrm{B}$. The matter distribution in the mixed phase can be seen more clearly in Fig~\ref{trhoevolution}.

\begin{figure*}
\centering
\subfigure[]{\includegraphics[width=0.48\linewidth]{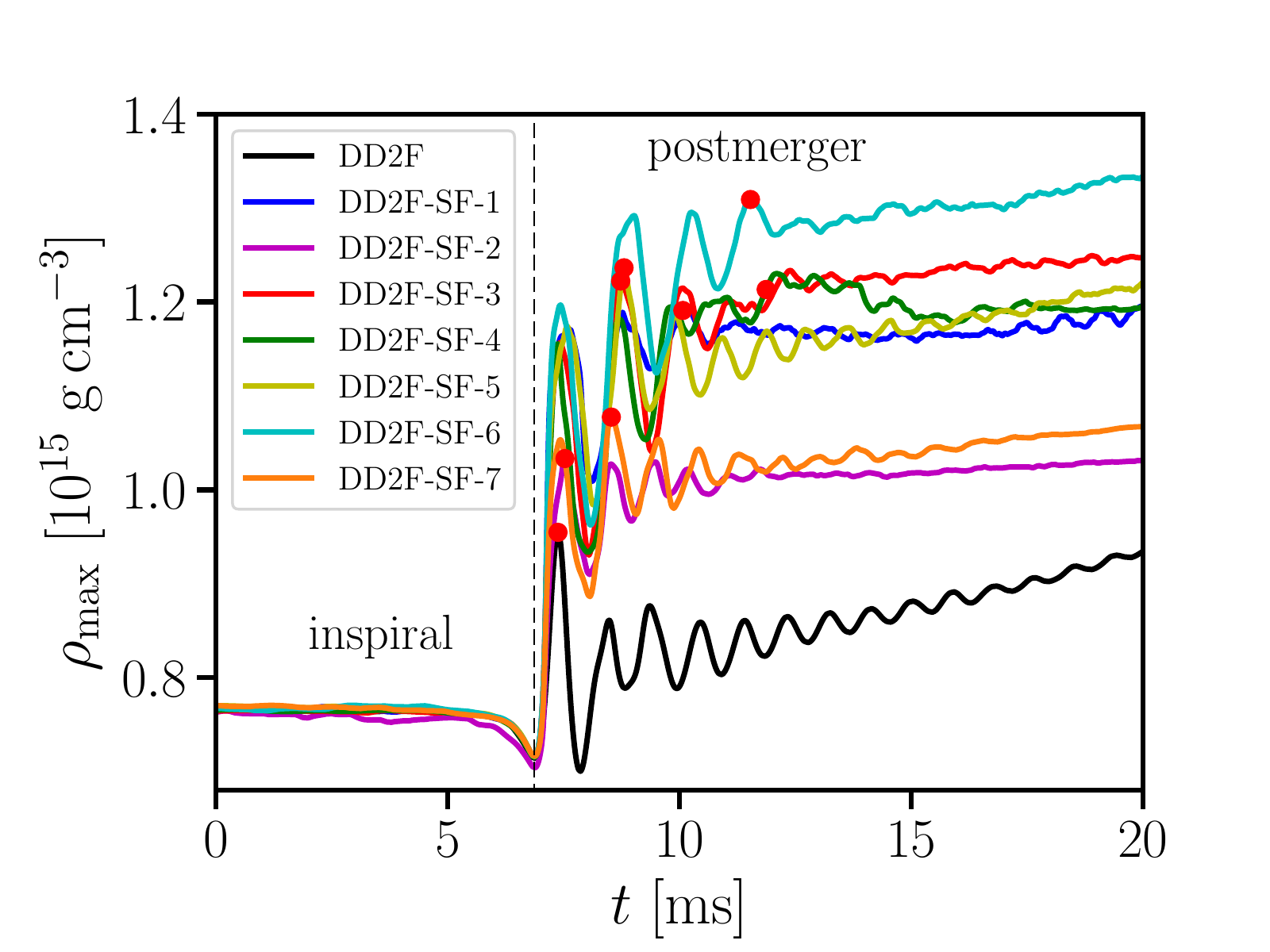}\label{allrhomax}}
\hfill
\subfigure[]{\includegraphics[width=0.48\linewidth]{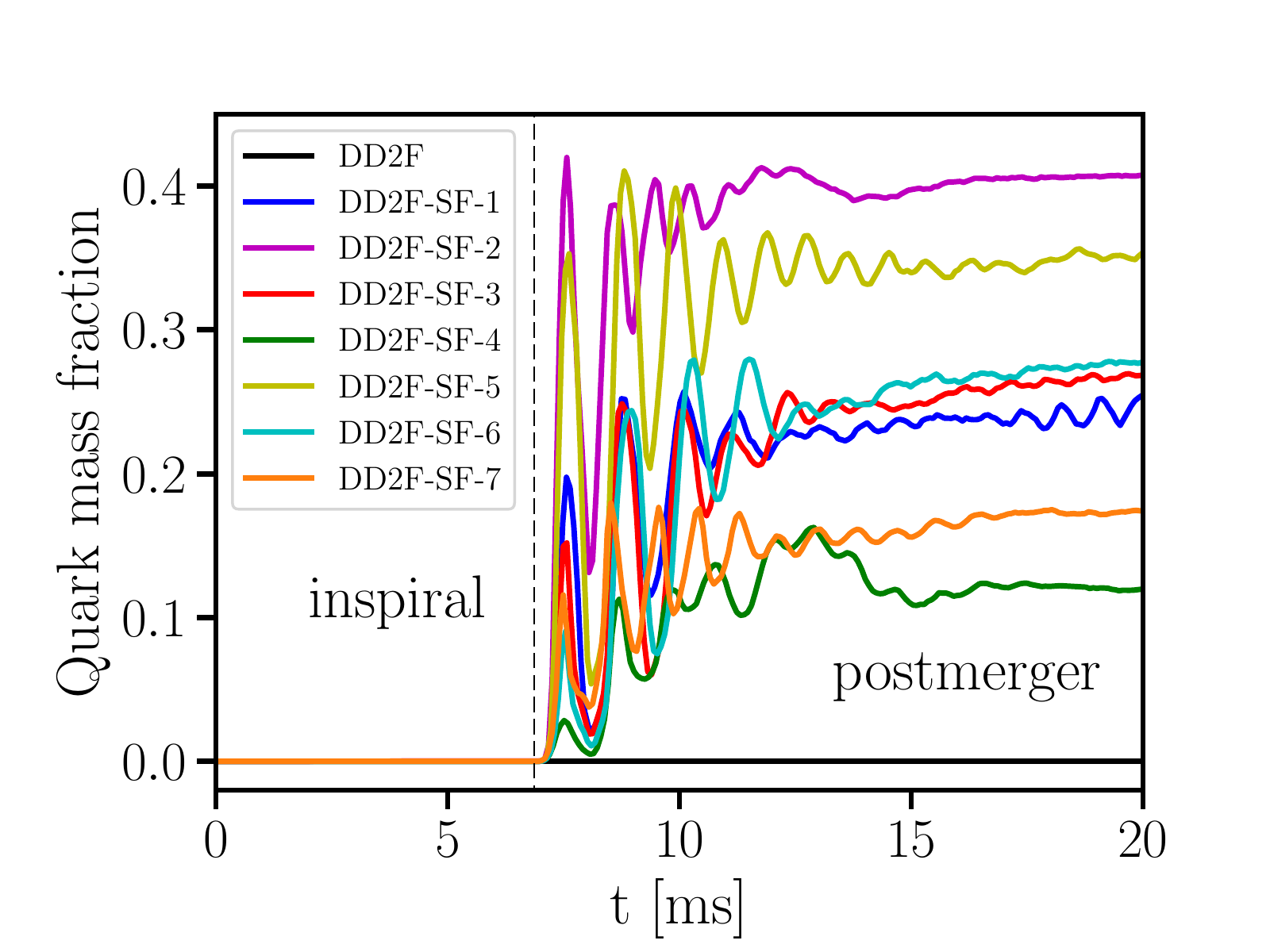}\label{allquarks}}

\caption{Left panel: Maximum rest-mass density as a function of time in binary NS merger simulations of two 1.35~M$_{\odot}$ NSs with different hybrid DD2F-SF EoSs (colored curves) in comparison to the hadronic DD2F reference model (black curve). The merging time is shown by the vertical dashed line. Red dots show the value of $\rho_{\rm max}^{\rm max}$ (the highest maximum density within the first 5~ms after merging) for every simulation. Right panel: Total fraction of mass being present as deconfined quark matter as a function of time for the same binary systems as in the left panel.}
\label{allrhomaxallq}
\end{figure*}

After a few tens of milliseconds the oscillations have become less pronounced and the remnant has settled into a more axial-symmetric configuration. This is shown in Fig.~\ref{denistyevolution_d}. Here, a clear, almost axial-symmetric pure quark matter core surrounded by a thin shell of a mixed phase is visible. The temperature in the shell with the mixed phase is rather high and follows a sequence of constant entropy (compare Fig.~\ref{trhoevolution_d} and Fig.~1 from \cite{Fischer2018}); cf. also Fig.~4 from \cite{Weih2019}. Fig.~\ref{tmuevolution_d} shows the presence of hot and cold deconfined quark matter in the remnant. We find that at this time {\raise.17ex\hbox{$\scriptstyle\mathtt{\sim}$}}29\% of the matter is contained in the quark core (see also Fig.~\ref{allquarks}). 

Note that the simulations with the other DD2F-SF EoSs behave similarly for this binary mass configuration.

In Fig.~\ref{allrhomax} we show the evolution of the maximum rest-mass density $\rho_{\mathrm{max}}(t)$ as a function of time in the simulations with all DD2F-SF models together with the hadronic reference model DD2F for 1.35--1.35~M$_{\odot}$ binaries. Fig.~\ref{allquarks} shows the time evolution of the total deconfined quark matter mass fraction being present in these NS binary systems (both from the mixed and the pure quark matter phase). During the inspiral the densities are almost identically for nearly all models since the maximum densities in the initial stars are below the transition densities, i.e. no deconfined quark matter is present. Very small differences originate from statistical fluctuations, which are a result of the computation of the initial data that involves a random component in the initial distribution of the SPH particles. The calculation with the DD2F-SF-2 EoS shows slightly larger deviations. 
For this EoS we used a different variant of the hadronic DD2F with an excluded volume modeling \cite{Typel2016} resulting in a slightly stiffer hadronic phase, which affects the stellar structure of stars with masses of about 1.35~$M_\odot$ (see Fig.~1 in the supplemental material of Ref.~\cite{Bauswein2019}). This implies a slightly reduced central density of 1.35~$M_\odot$ NSs during the inspiral for this particular EoS model.

The stars merge after about 7~ms. The densities increase and exceed the transitions densities of the hybrid DD2F-SF models leading to the formation of quark cores. In these calculations with hybrid models the PT effectively softens the EoS and thus leads to considerably higher maximum densities in the remnants compared to the DD2F model. From Fig.~\ref{allquarks} one can see that the masses of the forming quark cores sensitively depend on the underlying hybrid model. We observe quark cores ranging from roughly 10\% to 40\% of the total mass.

Note that the total fraction of quark matter is strongly affected by the onset density. The EoSs with the lowest onset densities are DD2F-SF-2 and DD2F-SF-5, which also lead to the largest postmerger quark cores. The maximum densities in the remnant, are not necessarily correlated with the quark core size. The DD2F-SF-6 model leads to the largest postmerger densities, however, the quark core in this model is considerably smaller than in  simulations with the DD2F-SF-2 or DD2F-SF-5 model. This results from the different stiffness of the the quark phase in these models.

After the merger all curves in Fig.~\ref{allrhomaxallq} show oscillating behavior. These oscillations are linked to the quasi-radial oscillations of the remnant (see e.g. \cite{Bauswein2015,Bauswein2019b}). Below we discuss the maximum rest-mass density  $\rho_{\mathrm{max}}^{\mathrm{max}}$ during the early postmerger evolution to determine the density regime of the EoS which is actually probed by the remnant. Specifically, we define $\rho_{\mathrm{max}}^{\mathrm{max}}$ as the maximum of $\rho_{\mathrm{max}}(t)$ during the first 5 milliseconds after merging. In Fig.~\ref{allrhomax} $\rho_{\mathrm{max}}^{\mathrm{max}}$ is marked by red points for every EoS. Note that at late times $\rho_{\mathrm{max}}(t)$ can exceed $\rho_{\mathrm{max}}^{\mathrm{max}}$. However, the emission of GWs is strongest right after merging which is why the GW signal is dominantly determined by the density regime up to $\rho_{\mathrm{max}}^{\mathrm{max}}$.

We note that $\rho_{\mathrm{max}}^{\mathrm{max}}$ is somewhat affected by the numerical resolution. We find in simulations with different SPH particle numbers that $\rho_{\mathrm{max}}^{\mathrm{max}}$ can vary by some per cent (for the 1.35--1.35~M$_\odot$ merger with the DD2F EoS we determine $\rho_{\rm max}^{\rm max}=9.53\times10^{14}~\rm{g~cm}^{-3}$ using about 300'000 particles, $\rho_{\rm max}^{\rm  max}=1.013\times10^{15}~\rm{g~cm}^{-3}$ using about 500'000 particles, and $\rho_{\rm max}^{\rm max}=1.013\times10^{15}~\rm{g~cm}^{-3}$ using about 600'000 particles). For the same setup of simulations the dominant postmerger frequency changes by less than one per cent.

In Appendix~\ref{asymrhofpeak} we briefly describe results from some additional calculations with a grid-based simulation tool, which we employ to validate the robustness of the relations presented in this paper.

\section{Signature of first-order PTs: $\Lambda-f_\mathrm{peak}$ relations}\label{fpeaklambda}

\begin{figure*}[ht]
\centering
\subfigure[]{\includegraphics[width=0.48\linewidth]{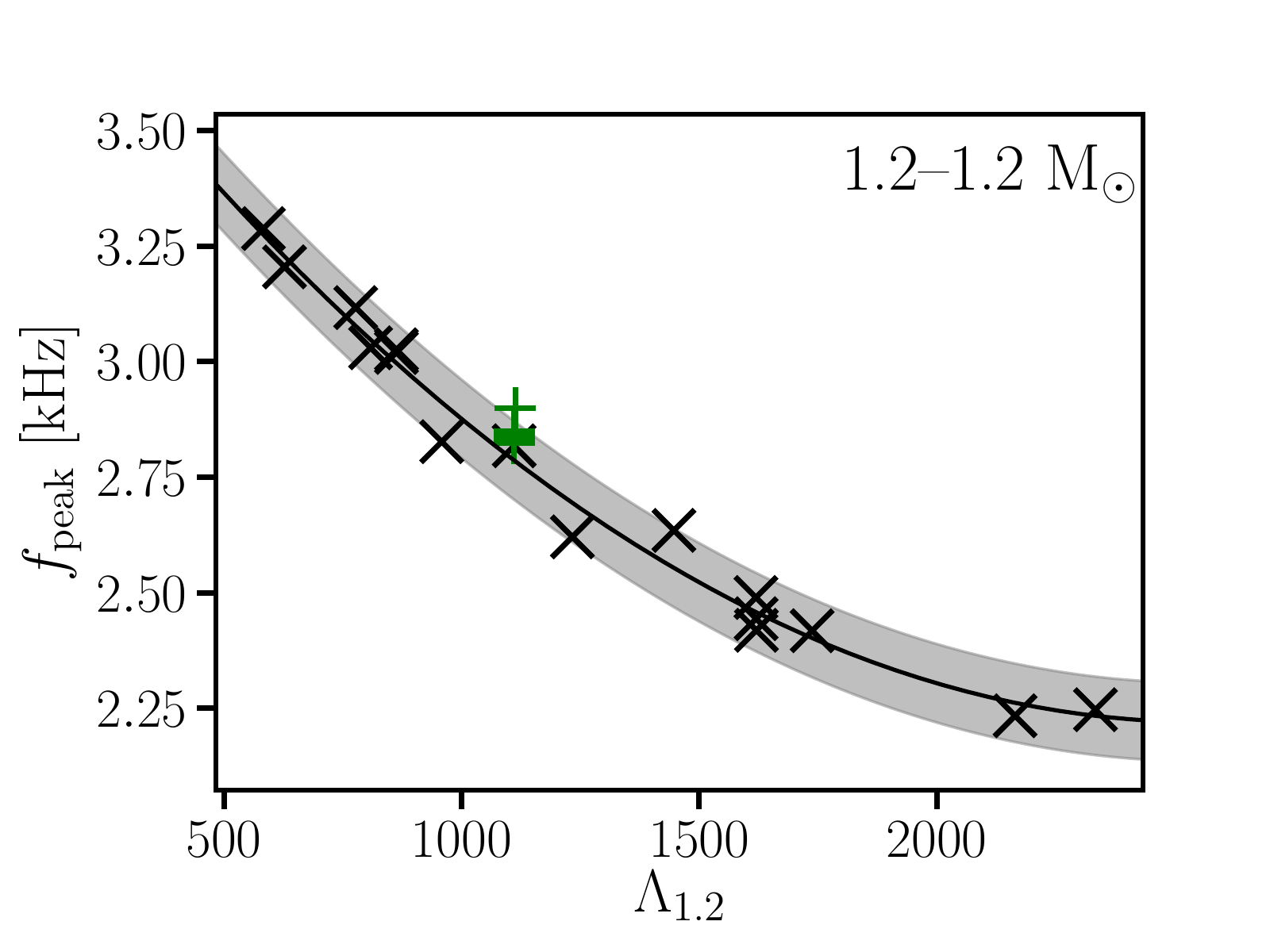}\label{fpeaklampdarelations_a}}
\hfill
\subfigure[]{\includegraphics[width=0.48\linewidth]{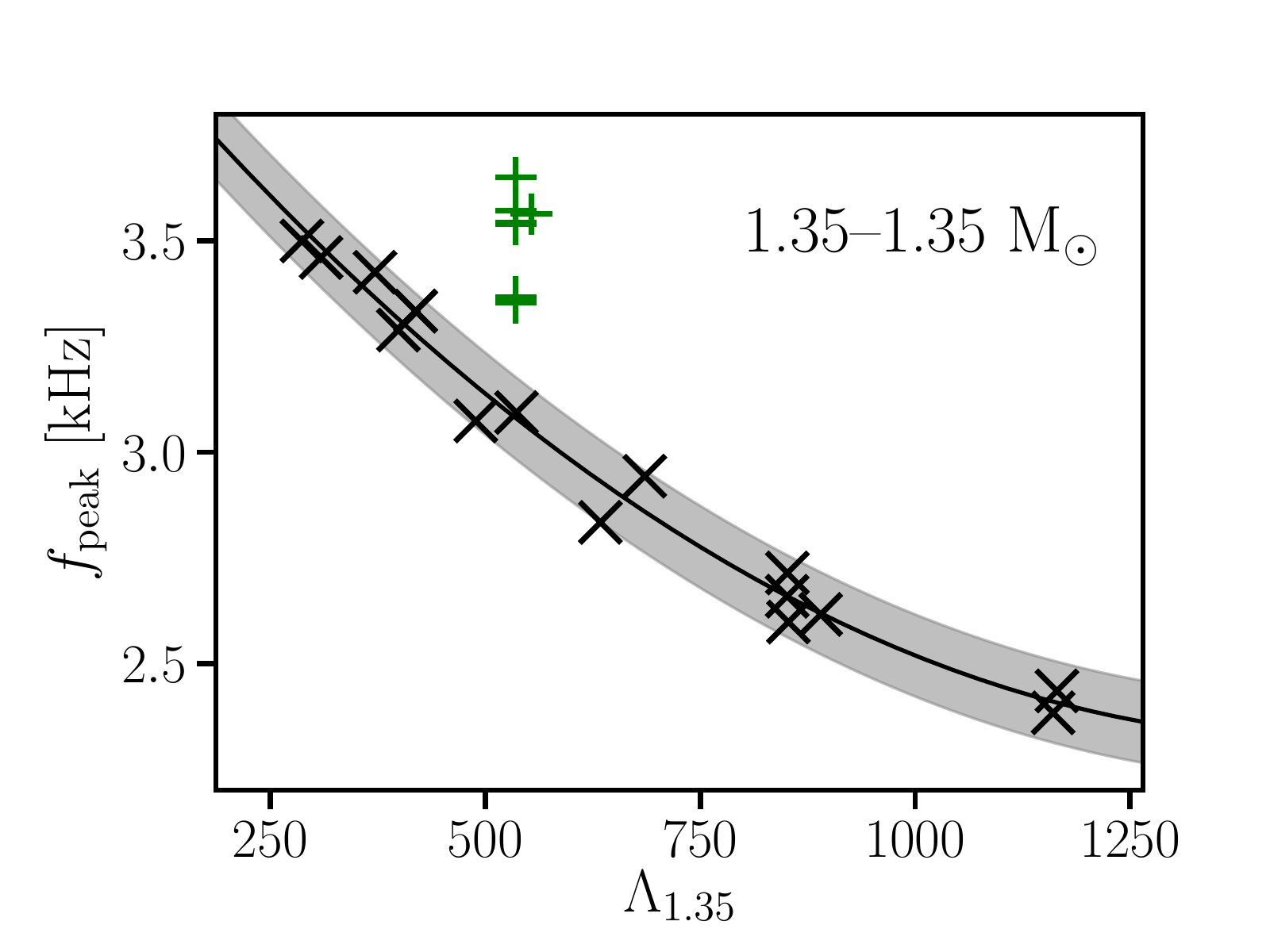}\label{fpeaklampdarelations_b}}
\\
\subfigure[]{\includegraphics[width=0.48\linewidth]{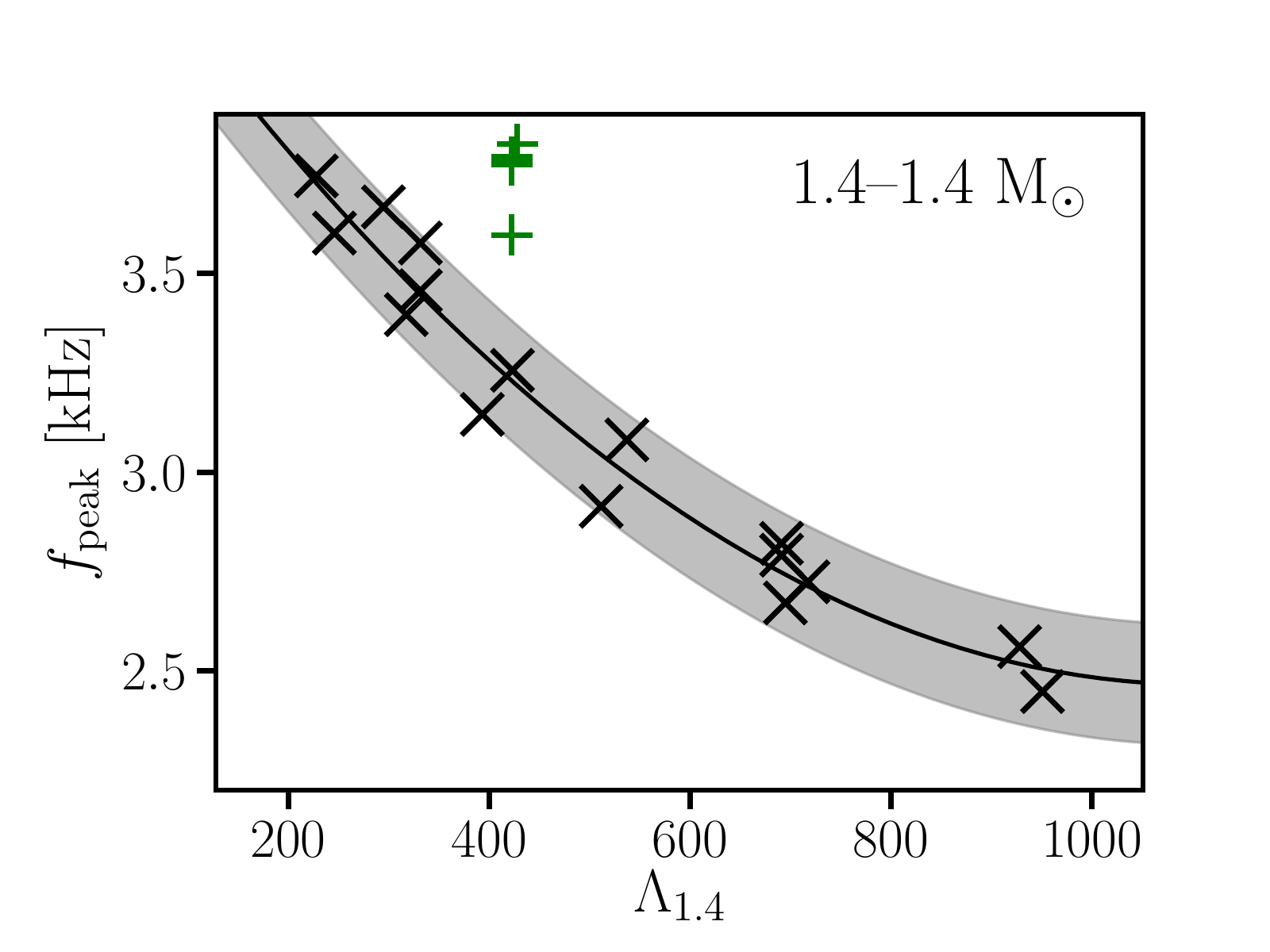}\label{fpeaklampdarelations_c}}
\hfill
\subfigure[]{\includegraphics[width=0.48\linewidth]{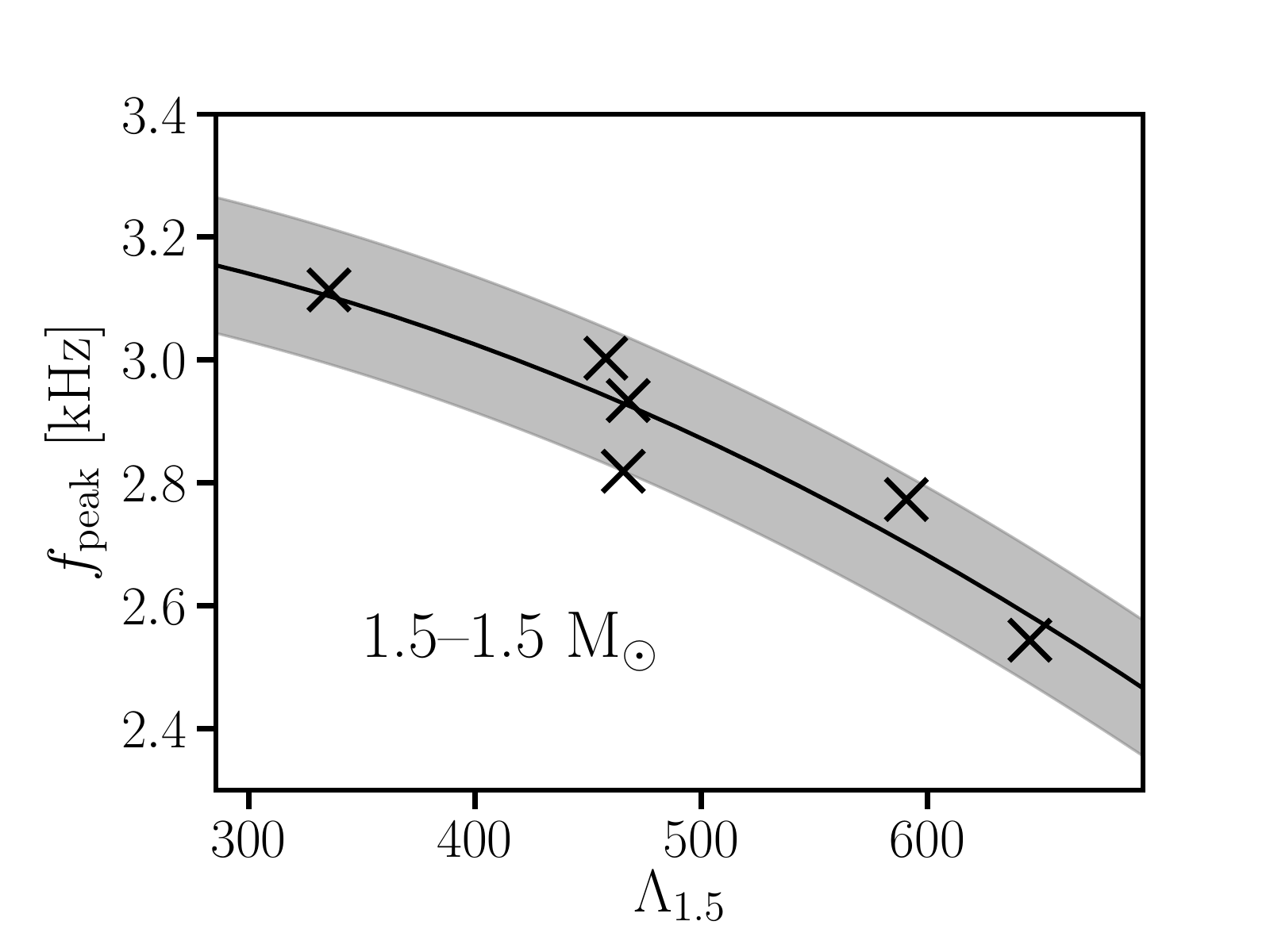}\label{fpeaklampdarelations_d}}
\caption{Dominant postmerger GW frequency $f_{\mathrm{peak}}$ as a function of the tidal deformability $\Lambda$ for 1.2--1.2~M$_{\odot}$ (graph~(a)), 1.35--1.35~M$_{\odot}$ (graph~(b)), 1.4--1.4~M$_{\odot}$ (graph~(c)) and 1.5--1.5~M$_{\odot}$ (graph~(d)) mergers with different microphysical EoSs. Black crosses display results with purely hadronic EoSs, while green plus signs depict results with the hybrid DD2F-SF models. The solid curves are least squares fits to data points from purely hadronic EoSs. The gray shaded area illustrates the largest deviation of the data of purely hadronic models from the least squares fit. For 1.35--1.35~M$_{\odot}$ and 1.4--1.4~M$_{\odot}$ binaries the results from the hybrid models appear as clear outliers at higher GW frequencies.}
\label{fpeaklampdarelations}
\end{figure*}

In \cite{Bauswein2019} we demonstrated that a strong first-order PT leads to clear deviations from a tight relation between the dominant postmerger oscillation frequency $f_{\mathrm{peak}}$ and the tidal deformability $\Lambda=\frac{2}{3}k_{2}\left(\frac{R}{M}\right)^{5}$. Here $R$ and $M$ are the radius and the gravitational mass of a NS, respectively, and $k_{2}$ is the tidal Love number \cite{Hinderer2008,Hinderer2010}. $f_{\mathrm{peak}}$ and $\Lambda$ both can be inferred from the GW signal of a NS merger and are expected to be measurable with sufficient precision in the future either with the LIGO-Virgo-Kagra network operating at design sensitivity, with upgraded GW instruments like \cite{Miller2015} or with third generation GW detectors like the Einstein Telescope or Cosmic Explorer \cite{Punturo2010,Hild2011,Reitze2019}

More specifically, the parameter describing finite-size effects in waveform models is the combined tidal deformability $\tilde{\Lambda}$ defined as
\begin{align}
  \tilde{\Lambda}=\frac{16}{13}\frac{(M_1+12 M_2)M_1^4\Lambda_{1}+(M_2+12 M_1)M_2^{4}\Lambda_{2}}{(M_1+M_2)^{5}}. 
  \label{lambdatilde}
\end{align} 
Here, $\Lambda_{1,2}$ refer to the tidal deformabilities of the individual stars with masses $M_{1,2}$. For equal-mass binaries $\tilde{\Lambda}$ coincides with $\Lambda$ of the individual stars. Using this fact, we often do not explicitly distinguish $\tilde{\Lambda}$ and $\Lambda$ for equal-mass binaries in the remainder of this work. In the relations discussed below $\Lambda$ can be replaced by $\tilde{\Lambda}$ for $M_1=M_2$. Thus, these relations in fact include the quantity which is actually inferred from measurements of binary mergers.

For a more detailed analysis of empirical relations between postmerger oscillation frequencies and tidal deformabilities as well as other physical properties such as NS radii, total binary masses and mass ratios we refer the reader to \cite{Vretinaris2019}.

We remark that for the sake of simplicity we typically discuss our findings referring to their total binary mass instead of the chirp mass of a binary. The latter is the quantity which is actually obtained with high precision from a measurement. For a fixed binary mass ratio, the total mass and the chirp mass are fully equivalent. We emphasize that for detections with sufficiently large signal-to-noise ratio, where the methods discussed here are applicable, the mass ratio will be measured with good precision. Hence, the total mass and the individual masses of the binary components can be derived with high accuracy. We thus discuss our results for the physically more intuitive total binary mass. These considerations justify to focus on systems with equal masses or only moderate binary mass asymmetry and to consider sets of simulations with fixed total binary mass.

\subsection{Mass-dependent relations}\label{fpeaklambdamassdep}

In Fig.~\ref{fpeaklampdarelations} we show $f_{\rm peak}$ as a function of $\Lambda$ for four different binary configurations, 1.2--1.2~M$_{\odot}$ in Fig.~\ref{fpeaklampdarelations_a}, 1.35--1.35~M$_{\odot}$ in Fig.~\ref{fpeaklampdarelations_b}, 1.4--1.4~M$_{\odot}$ in Fig.~\ref{fpeaklampdarelations_c} and 1.5--1.5~M$_{\odot}$ in Fig.~\ref{fpeaklampdarelations_d}. $\Lambda$ refers to the tidal deformability of a single, inspiraling NS, i.e. $\Lambda=\Lambda(M_\mathrm{tot}/2)$, which for equal-mass binary equals the combined tidal deformability of the system. Hence, we plot the postmerger frequency as a function of the combined tidal deformability of the binary system. Black crosses represent results from merger simulations using different, purely hadronic microphysical EoSs, while green plus signs exhibit data obtained from the hybrid DD2F-SF models. Solid black lines display least squares fits of the data using a second order polynomial
\begin{align}
f^\mathrm{had}_{\mathrm{peak}}=(a_{M}\Lambda^{2}+b_{M}\Lambda+c_{M})~\mathrm{kHz}\label{QuadraticLambdaFit}
\end{align}
(excluding the hybrid DD2F-SF models). Gray shaded areas illustrate the maximum deviation of the data points from the fit considering only hadronic EoS models. The fit parameters $a_{M},~b_{M},~c_{M}$ together with the mean and the maximum deviation of the purely hadronic models from the fit can be found in Tab.~\ref{fpeaklambdafits1}. (Note that below we use $a_{M},~b_{M},~c_{M}$ as parameters in different fit formulae, but every time we explicitly state their values for the respective relation.)

\begin{table}
\caption{Dimensionless fit parameters $a_{M},~b_{M},~c_{M}$ for the empirical relation Eq.~\eqref{QuadraticLambdaFit}, which is shown in Figs.~\ref{fpeaklampdarelations_a}--\ref{fpeaklampdarelations_d} together with the mean and the maximum deviation of the data from the fit. These fits and the resulting residuals include only data from purely hadronic EoSs.}
\begin{tabular}{c r r c c c}
\hline\hline
$M_{\mathrm{tot}}$ & $a_{M}$ & $b_{M}$ & $c_{M}$ & mean dev.  & max dev.\\
$[$M$_{\odot}]$ & $[10^{-7}]$ & $[10^{-3}]$ & & [Hz]& [Hz]\\
\hline
2.4 & $2.704$ & $-1.383$ & 3.989 & 35 & 85 \\
2.7 & $8.463$ & $-2.509$ & 4.182 & 44 & 97 \\
2.8 & $16.35$ & $-3.616$ & 4.465 & 71 & 152 \\
3.0 & $-18.79$ & $0.164$ & 3.261 & 50 & 111 \\
\hline
\hline
\end{tabular}
\label{fpeaklambdafits1}
\end{table}

One can see that for each binary configuration the $f^\mathrm{had}_{\mathrm{peak}}(\Lambda)$ relation for the hadronic models is well described by the fits with maximum residuals of the order of 100~Hz.

As shown in \cite{Bauswein2019} for a binary configuration of 1.35--1.35~M$_{\odot}$ (Fig.~\ref{fpeaklampdarelations_b}) the data points from the hybrid DD2F-SF models appear as clear outliers at larger frequencies. This is understandable since $f_{\mathrm{peak}}$ is expected to scale with the compactness of the remnant \cite{Bauswein2012a},
and the PT leads to significantly more compact remnants. As stated in \cite{Bauswein2019} this behaviour is an unambiguous signature of a strong PT since all other models including those with a transition to hyperonic matter closely follow the fit to purely hadronic models. 

For binaries of two 1.4~M$_{\odot}$ NSs the situation is similar. Again the postmerger frequencies obtained with the hybrid DD2F-SF models are significantly larger than those of the respective hadronic model at the same value of $\Lambda$.

For 1.2--1.2~M$_{\odot}$ binaries (Fig.~\ref{fpeaklampdarelations_a}) the situation is different. At this relatively low binary mass the densities in the remnant are smaller and the fraction of matter that undergoes the PT is not large enough to have a noticeable impact on $f_{\mathrm{peak}}$ (or is even zero). Therefore, for this total binary mass the postmerger frequencies from the simulations with the hybrid DD2F-SF models are consistent with the respective $f^\mathrm{had}_{\mathrm{peak}}(\Lambda)$ relation of purely hadronic EoSs. In this case, the DD2F-SF EoSs cannot be clearly distinguished from hadronic EoSs because essentially only the hadronic part of the DD2F-SF models is probed. We address this point in more detail below and discuss to which extent a consistency of a measurement with Eq.~\eqref{QuadraticLambdaFit} within the maximum residual of all hadronic models implies the absence of a PT.

Fig.~\ref{fpeaklampdarelations_d} does not contain any data points from simulations with the hybrid DD2F-SF models because for this total binary mass prompt collapse to black hole occurs for all DD2F-SF EoSs. However, for different hybrid models that would not immediately collapse to a black hole we also expect strong deviations from the respective $f^\mathrm{had}_{\mathrm{peak}}(\Lambda)$ relation.

We also emphasize that the hybrid models in this study are based on only one hadronic model for the density regime below the PT. This is the reason for all hybrid models occurring at the same $\Lambda$. We expect that other choices for the hadronic regime of hybrid models will lead to a very similar increase of the postmerger GW frequency relative to the respective $\Lambda$. Note that the properties of the chosen hadronic model DD2F for the density regime below the PT fall roughly in the middle of current constraints on the EoS. 
\subsection{Asymmetric binaries}\label{fpeaklambdaasym}
So far we have only considered symmetric binaries. We expect the previous discussion to also hold for not too asymmetric systems. To explicitly study the effect of the mass ratio $q=M_{1}/M_{2}$ on the $f^\mathrm{had}_{\mathrm{peak}}(\Lambda)$ relation we compare results for different mass ratios at a constant chirp mass $M{_\mathrm{chirp}}$. Note that $M{_\mathrm{chirp}}$ defined as
\begin{align}
    M{_\mathrm{chirp}}=\frac{(M_{1} M_{2})^{3/5}}{(M_{1}+M_{2})^{1/5}}\label{chirpmass}
\end{align}
can be directly inferred from the inspiral GW signal with high precision, while the determination of $q$ has larger uncertainties.

For this comparison we perform additional simulations with 1.3--1.4~M$_{\odot}$ binaries for every EoS in our sample. We then interpolate our results from symmetric binaries to the chirp mass of a 1.3--1.4~M$_{\odot}$ binary ($M{_\mathrm{chirp}}\approx$ 1.174~M$_{\odot}$) \footnote{It suffices to linearly interpolate the data of equal-mass binaries to the system with $M{_\mathrm{chirp}}$=1.174~M$_{\odot}$ instead of performing an additional simulation. The corresponding equal-mass system is a binary with two stars of 1.349~M$_{\odot}$ and is thus very close to a 1.35--1.35~M$_{\odot}$ simulation.}.

\begin{figure}
\centering
\includegraphics[width=1.0\linewidth]{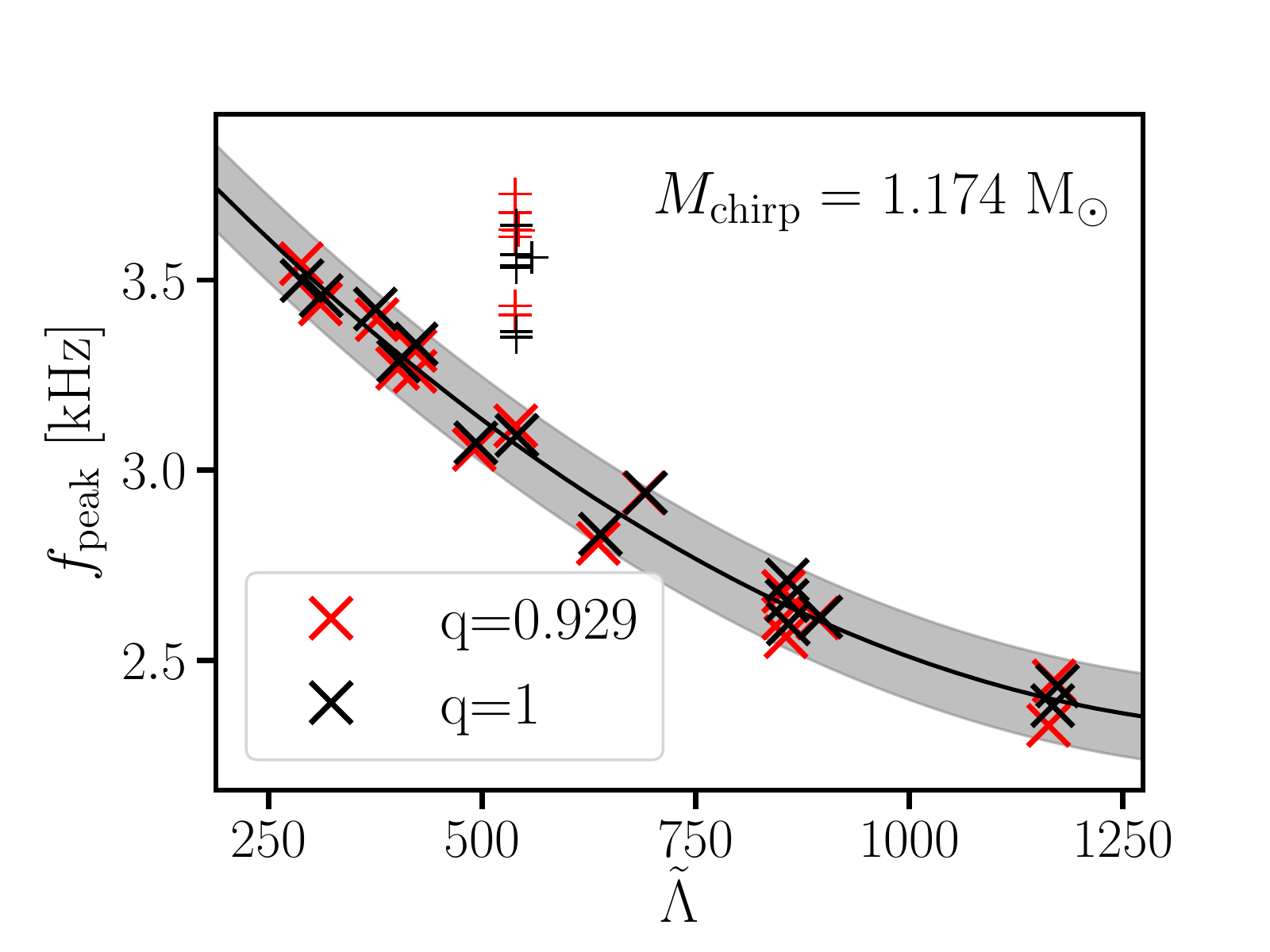}
\caption{Dominant postmerger GW frequency $f_{\mathrm{peak}}$ as a function of the combined tidal deformability $\tilde{\Lambda}$ of the respective binary system. Red symbols refer do data from 1.3--1.4~M$_{\odot}$ binaries, while black symbols refer do data from equal-mass binaries with the same chirp mass of a 1.3--1.4~M$_{\odot}$ binary. Crosses represent data from purely hadronic EoSs while plus signs display data obtained with hybrid DD2F-SF models. The solid black line shows a least squares fit with a second order polynomial to the data (excluding the hybrid models). The gray shaded areas illustrate the maximum deviation of the data of hadronic models from the fit.}
\label{fpeaklambdachirp}
\end{figure}

The resulting values for $f_{\mathrm{peak}}$ and $\tilde{\Lambda}$ are shown in Fig.~\ref{fpeaklambdachirp}. 
The red symbols mark data from 1.3--1.4~M$_{\odot}$ binary simulations, while black symbols refer to interpolated data from symmetric binaries. Crosses represent data obtained with purely hadronic EoSs, while plus signs refer to data obtained with hybrid DD2F-SF EoSs. As before, the black solid line shows a least squares fit to all hadronic data with a second order polynomial (see Eq.~\eqref{QuadraticLambdaFit}) and the gray shaded area illustrates the maximum deviation of the data from the fit (excluding hybrid models). The fit parameters are given by 
$a_{M}=8.710\times 10^{-7},~b_{M}=\mathrm{-}2.553\times 10^{-3}$ and $c_{M}=4.192$, while the mean and the maximum deviation of hadronic data from the fit are 40~Hz and 113~Hz, respectively. 

One can see that a variation of the mass ratio does not have a large impact on the $f^\mathrm{had}_{\mathrm{peak}}(\tilde{\Lambda})$ relation at a constant chirp mass. For both values of $q$ the data points from hybrid EoSs appear as clear outliers. The deviation from the fit is even somewhat larger for asymmetric binaries than for symmetric binaries. Note that postmerger GW measurements will become available with high signal-to-noise ratios implying that $q$ can be inferred with a precision better than the variation of $q$ in Fig.~\ref{fpeaklambdachirp}.

A further discussion on the impact of the binary mass ratio can be found in Appendix~\ref{asymrhofpeak}.
\subsection{Mass-independent relations}\label{fpeaklambdamassindep}
Future binary merger observations will most likely have total masses different from the 4 cases discussed above. We therefore derive universal relations between $\Lambda$ and $f^\mathrm{had}_{\mathrm{peak}}$ independent of a specific mass. Note, however, that it will be easily possible and, in fact, advantageous to simulate a new set of binary mergers for actually measured binary masses after a detection and to obtain corresponding fits for this specific setup. We here describe procedures which can be directly applied to upcoming measurements of not too asymmetric binary mergers (explicitly we show that simulations with $q=0.929$ lead to nearly identical results).

\begin{figure}
\centering
\includegraphics[width=1.0\linewidth]{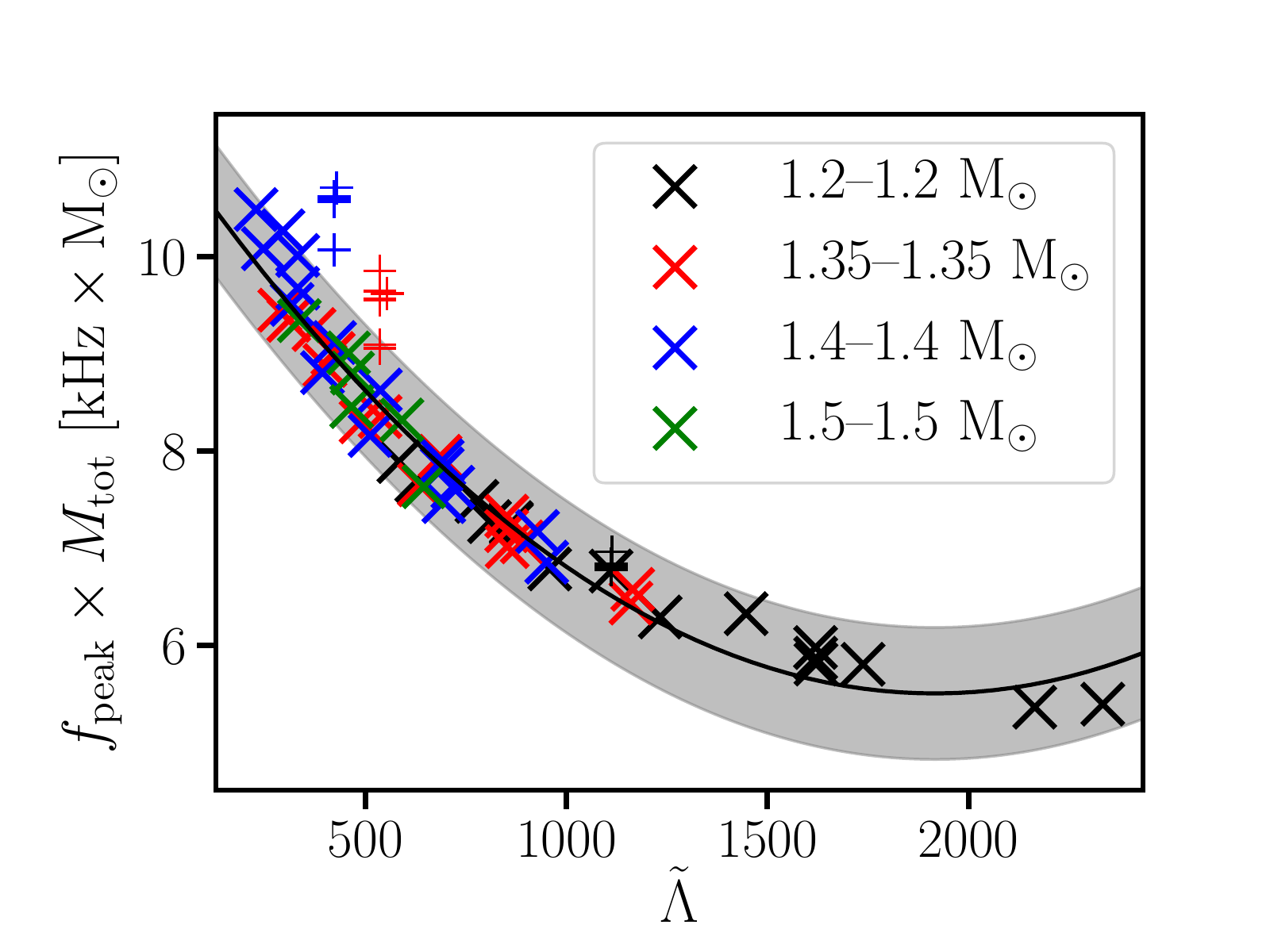}
\caption{Dominant postmerger GW frequency $f_{\mathrm{peak}}$ scaled by the total binary mass $M_{\mathrm{tot}}$ as a function of the combined tidal deformability $\tilde{\Lambda}$. Different colors refer to data from different total binary masses. Crosses refer to data from purely hadronic models, while plus signs represent data with hybrid DD2F-SF models. Solid black line is a least squares fit with a second order polynomial to the data (excluding the DD2F-SF models). The gray shaded area illustrates the maximum deviation of the data of hadronic models from the fit.}
\label{fpeaklampdauniversal}
\end{figure}

As in \cite{Bernuzzi2015} we multiply $f^\mathrm{had}_{\mathrm{peak}}$ with the total binary mass $M_{\mathrm{tot}}$, which yields a relatively tight relation between $f^\mathrm{had}_{\mathrm{peak}}\times M_{\mathrm{tot}}$ and $\Lambda$. This relation is shown in Fig.~\ref{fpeaklampdauniversal}. Different colored crosses refer to data from hadronic EoSs with different binary masses. Colored plus signs represent results for the hybrid DD2F-SF models. The solid black line shows a least squares fit to the data with a second order polynomial of the form
\begin{align}
f^\mathrm{had}_{\mathrm{peak}}\times M_{\mathrm{tot}}=(a\Lambda^{2}+b\Lambda+c)~\mathrm{kHz}~ \mathrm{M}_{\odot}  \label{QuadraticLambdaFitUni}
\end{align}
excluding the data from the DD2F-SF models. The gray shaded area illustrates the maximum deviation of data from hadronic models from the fit. The fit parameters in Eq.~\eqref{QuadraticLambdaFitUni} are given by $a=1.554\times 10^{-6}$, $ b=\mathrm{-}5.954\times 10^{-3} $ and $c=11.21$. The mean deviation of the data from the fit is 206~Hz$~ $M$_\odot$ and the maximum residual is 680~Hz$~ \mathrm{M}_\odot$. The increase of the fit at very large $\Lambda$ is an artefact of the chosen fit function and the equation should not be employed for even larger $\Lambda$.

As in Fig.~\ref{fpeaklampdarelations} the DD2F-SF data at a total mass of 1.2--1.2~M$_{\odot}$ is in good agreement with the relation for hadronic models since the densities are too low to form a sufficiently large quark core to significantly influence $f_{\mathrm{peak}}$.

For binary masses of 1.4--1.4~M$_{\odot}$ the data points from simulations with the DD2F-SF EoSs appear as clear outliers.
For 1.35--1.35~M$_{\odot}$ mergers some of the hybrid models (with smaller density jumps across the PT) are marginally consistent with the band defined by the purely hadronic models. This is simply a consequence of the larger scatter, which results from combining results for different binary masses in a single relation. This was not the case for the relations for fixed binary mass. Therefore universal relations including different binary masses over a large mass range like Eq.~\eqref{QuadraticLambdaFitUni} (Fig.~\ref{fpeaklampdauniversal}) are not the optimal choice for the identification of a PT.

In addition, we thus introduce $f^\mathrm{had}_{\mathrm{peak}}\times M_{\mathrm{tot}}$($\Lambda$) relations restricted to tighter binary mass ranges. To obtain these relations, we consider three subsets of data and fit the data in each set using Eq.~\eqref{QuadraticLambdaFitUni}. The subsets consist of data from simulations with 1.2--1.2~M$_{\odot}$ and 1.35--1.35~M$_{\odot}$, 1.35--1.35~M$_{\odot}$ and 1.4--1.4~M$_{\odot}$ as well as from 1.4--1.4~M$_{\odot}$ and 1.5--1.5~M$_{\odot}$ binaries, respectively. The resulting parameters of Eq.~\eqref{QuadraticLambdaFitUni} as well as the mean and the maximum deviation of the data for each range of binary masses are given in Tab.~\ref{fpeaklambdafits}. Plots of these relations can be found in Appendix~\ref{moreplotsfpeaklambda}. The advantage of these relations, which hold for a smaller binary mass range, is that they result in a smaller scatter. This is helpful if one employs such mass-independent relations to infer the presence or absence of a PT.

Below we use the universal relations for smaller binary mass ranges assuming that the relations are valid for any total binary mass within the range. In particular, we assume that the maximum residual is representative for the respective range. For instance, for a measured binary mass of $M_{\mathrm{tot}}$=2.5~M$_{\odot}$ we would consider the relation resulting from the subset of 1.2--1.2~M$_{\odot}$ and 1.35--1.35~M$_{\odot}$ data.

The restriction to smaller binary mass ranges reduces the deviations of the data of hadronic models from the universal relations while still allowing to analyze signals from NS binaries for any total binary mass.
\begin{table}
\begin{tabular}{c c c c c c}
\hline\hline
$M_{\mathrm{tot}}$ & $a$ & $b$ & $c$ & mean dev.  & max dev.\\
$[$M$_{\odot}]$ & $[10^{-6}]$ & $[10^{-3}]$ & & [Hz$~ \mathrm{M}_\odot$]& [Hz$~ \mathrm{M}_\odot$]\\
\hline\
2.4--2.7 & $1.201$ & $-4.974$ & 10.650 & 151 & 357 \\
2.7--2.8 & $3.405$ & $-8.620$ & 12.014 & 181 & 499 \\
2.8--3.0 & $4.608$ & $-10.12$ & 12.472 & 206 & 400 \\
\hline
\hline
\end{tabular}
\caption{Fit parameters $a,~b,~c$ as defined by Eq.~\eqref{QuadraticLambdaFitUni} for the empirical relations shown in Figs.~\ref{fpeaklampdacombirelations_a}--\ref{fpeaklampdacombirelations_c} together with the mean and the maximum deviation of the data from the fit. These fits and the resulting residuals include only data from purely hadronic EoSs.}
\label{fpeaklambdafits}
\end{table}

A measured postmerger frequency $f_{\mathrm{peak}}$ strongly conflicting with the universal relations discussed in this section (Tab.~\ref{fpeaklambdafits}) would provide strong evidence for the occurrence of a strong PT during merging. We emphasize, that the binary masses will be measured with high precision. As we have shown, one will obtain tighter relations for fixed binary masses permitting more stringent comparisons between a measured $f_{\mathrm{peak}}$ and the $f^\mathrm{had}_{\mathrm{peak}}$ expected for hadronic EoSs based on the measured $\Lambda$. For this, one has to perform new simulations with the measured binary masses $M_{1}$ and $M_{2}$ for a set of hadronic EoSs and determine the $f^\mathrm{had}_{\mathrm{peak}}(\Lambda)$ relation as we have done in Fig.~\ref{fpeaklampdarelations}. In particular, this may be necessary for very asymmetric binaries, whereas the relations derived above hold for roughly symmetric binaries. The effects of slightly asymmetric binaries on the mass-independent relations are further discussed in Appendix~\ref{asymbinar}.

\section{Postmerger densities: $\rho_\mathrm{max}^\mathrm{max}-f_\mathrm{peak}$ relations}\label{postdensities}
\subsection{Mass-dependent relations} \label{fpeakrhomassdep}
In \cite{Bauswein2019} we also found that for hadronic EoSs the maximum rest-mass density $\rho_{\mathrm{max}}^{\mathrm{max}}$ during the first 5 milliseconds after merging (see Fig.~\ref{allrhomax}) correlates with $f_{\mathrm{peak}}$. 
The densities in the remnant might exceed $\rho_{\mathrm{max}}^{\mathrm{max}}$ at later times, but the gravitational radiation from the remnant at later times is weaker because its oscillations are damped. Hence, the postmerger GW emission is shaped during the early evolution of the remnant, and the characteristics of the signal only inform about the density regime up to $\rho_{\mathrm{max}}^{\mathrm{max}}$.
For this reason we consider $\rho_{\mathrm{max}}^{\mathrm{max}}$ and not the overall highest value of the density.

\begin{figure*}[ht]
\centering
\subfigure[]{\includegraphics[width=0.48\linewidth]{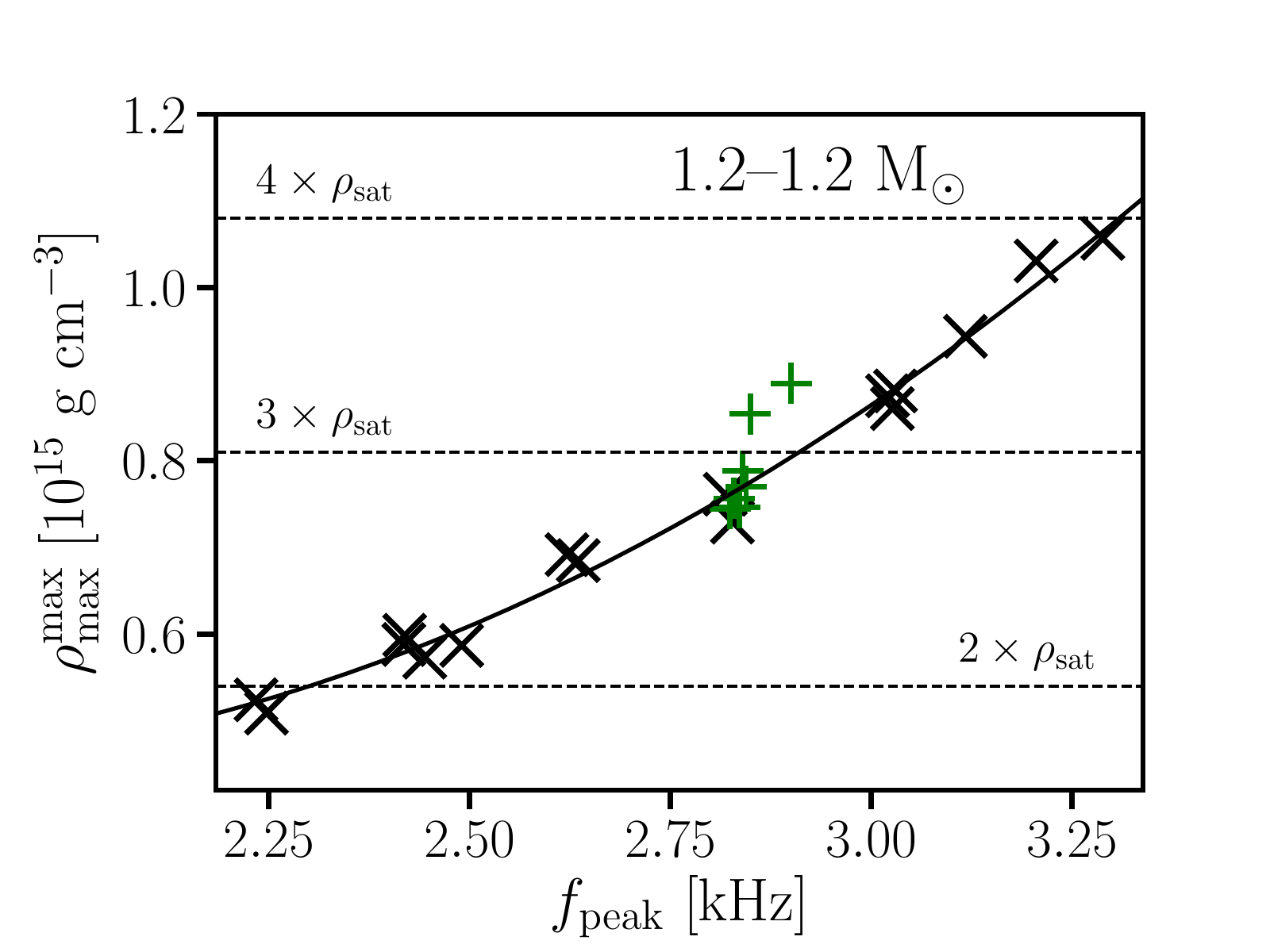}\label{fpeakrhomamaxrelations_a}} \hfill
\subfigure[]{\includegraphics[width=0.48\linewidth]{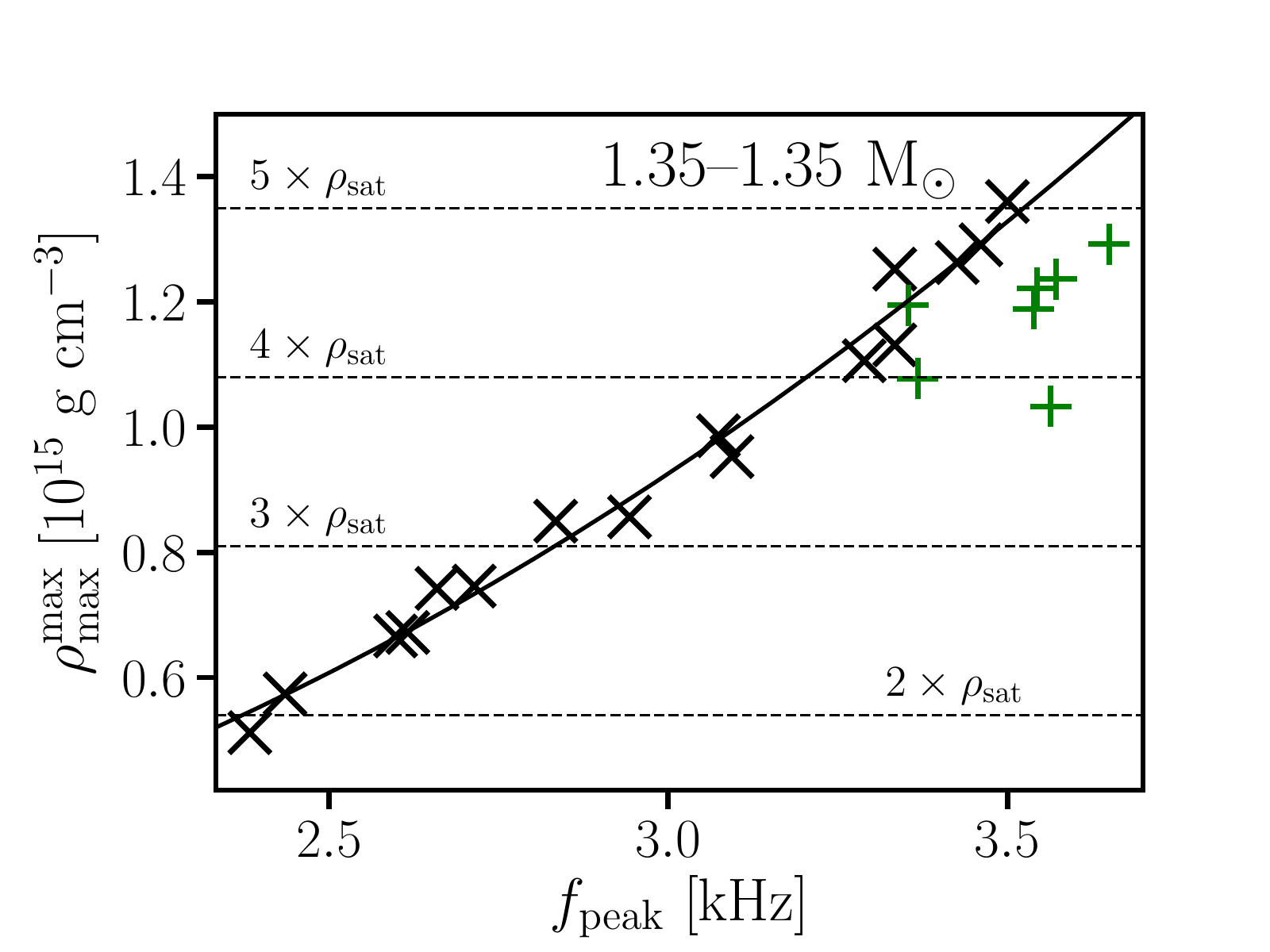}\label{fpeakrhomamaxrelations_b}} 
\\
\subfigure[]{\includegraphics[width=0.48\linewidth]{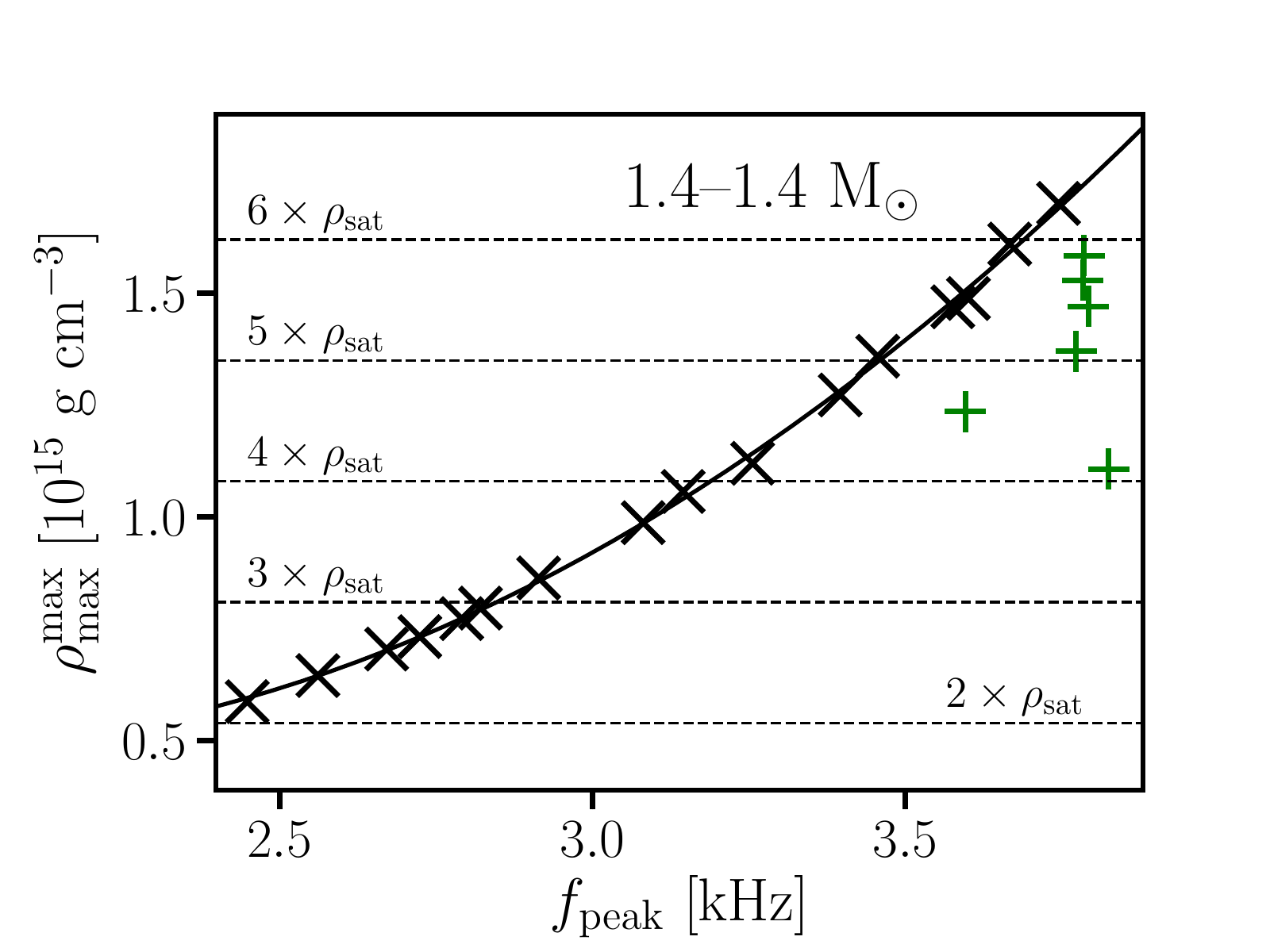}\label{fpeakrhomamaxrelations_c}} \hfill
\subfigure[]{\includegraphics[width=0.48\linewidth]{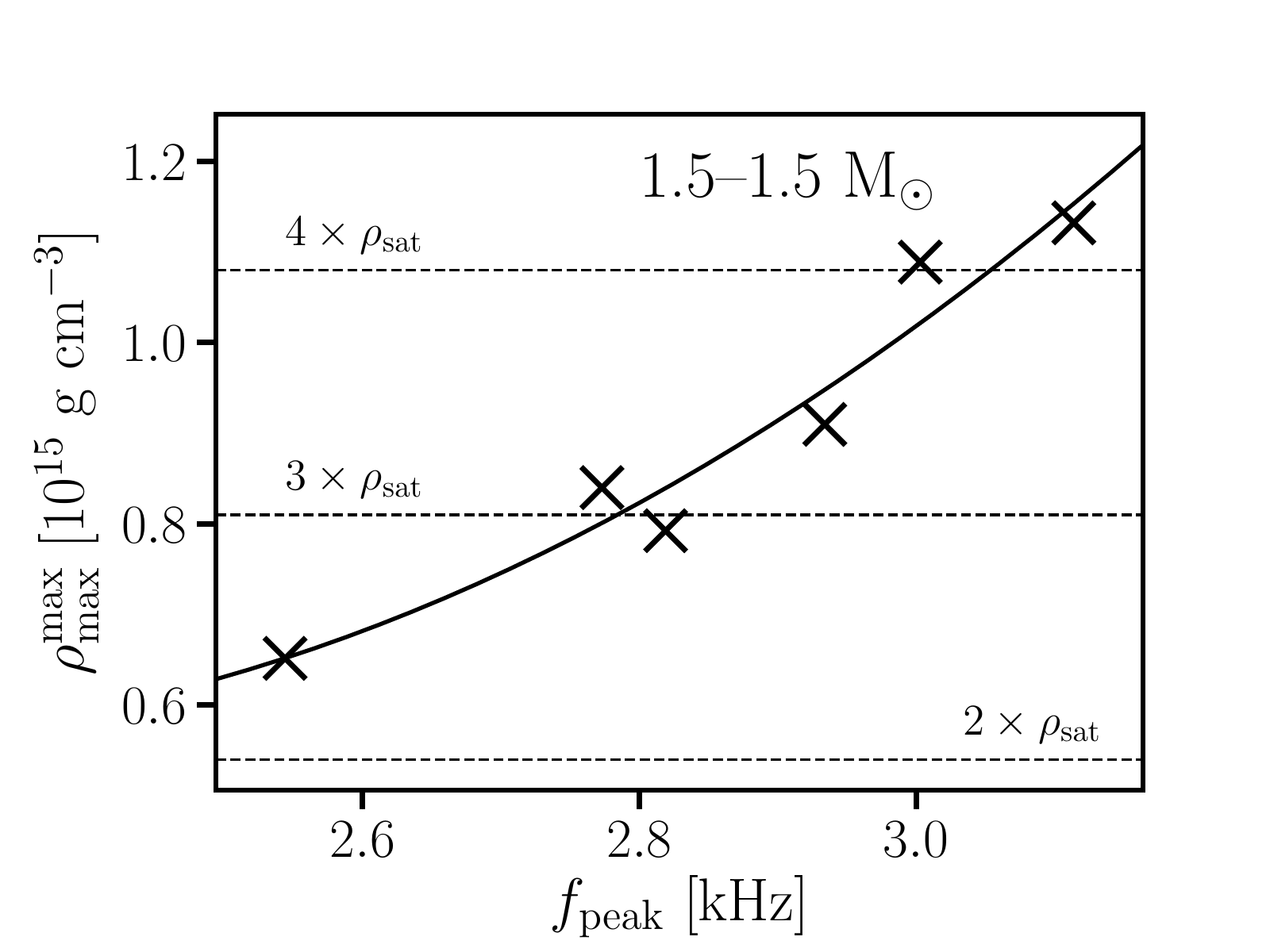}\label{fpeakrhomamaxrelations_d}} 
\caption{Maximum rest-mass density $\rho_{\mathrm{max}}^{\mathrm{max}}$ in the remnant during the first 5 milliseconds after merging as a function of the dominant postmerger GW frequency $f_{\mathrm{peak}}$ for 1.2--1.2~M$_{\odot}$ (graph~(a)), 1.35--1.35~M$_{\odot}$ (graph~(b)), 1.4--1.4~M$_{\odot}$ (graph~(c)) and 1.5--1.5~M$_{\odot}$ (graph~(d)) mergers with different microphysical EoSs. Black crosses show results with purely hadronic EoSs, while green plus signs depict results with hybrid DD2F-SF models. Solid curves display least squares fits to results for purely hadronic EoSs.}
\label{fpeakrhomamaxrelations}
\end{figure*}

The correlation we observed in \cite{Bauswein2019} between $\rho_{\mathrm{max}}^{\mathrm{max}}$ and $f^\mathrm{had}_{\mathrm{peak}}$ for 1.35--1.35~M$_{\odot}$ mergers is shown in Fig.~\ref{fpeakrhomamaxrelations_b}.

Black crosses exhibit data from simulations with purely hadronic EoSs. The solid black line displays the least squares fit to those data of purely hadronic models with a second-order polynomial.
\begin{align}
\rho_{\mathrm{max}}^{\mathrm{max}}(f^\mathrm{had}_{\mathrm{peak}})=(a_{M}(f^\mathrm{had}_{\mathrm{peak}})^2+b_{M}f^\mathrm{had}_{\mathrm{peak}} +c_{M})~\mathrm{g}~ \mathrm{cm}^{-3}\label{QuadraticrhoFit}
\end{align}
with $f^\mathrm{had}_{\mathrm{peak}}$ in kHz. The corresponding other panels in Fig.~\ref{fpeakrhomamaxrelations} show the $\rho_{\mathrm{max}}^{\mathrm{max}}$-$f^\mathrm{had}_{\mathrm{peak}}$-relations with the fits for the binary mass configurations 1.2--1.2~M$_{\odot}$ (Fig.~\ref{fpeakrhomamaxrelations_a}), 1.4--1.4~M$_{\odot}$ (Fig.~\ref{fpeakrhomamaxrelations_c}) and 1.5--1.5~M$_{\odot}$ (Fig.~\ref{fpeakrhomamaxrelations_d}). Generally, we find that the range of postmerger densities we observe in Fig.~\ref{fpeakrhomamaxrelations} is similar to postmerger densities reported in other works (see e.g. \cite{Radice2017,Hotokezaka2011,Hanauske2017}).\\

Not unexpectedly, we find that the maximum density is higher for high postmerger frequencies. This is understandable, since high $f_{\mathrm{peak}}$ result from soft EoSs, which lead to more compact remnants and hence to larger postmerger densities.

The fit parameters $a_{M},~b_{M},~c_{M}$ as well as the mean and the maximum deviations of hadronic models from the fits are provided in Tab.~\ref{fpeakrhofits}.

\begin{table*}
\begin{tabular}{c c c c c c}
\hline\hline
$M_{\mathrm{tot}}$ & $a_{M}$ & $b_{M}$ & $c_{M}$ & mean dev.  & max dev.\\
$[\mathrm{M}_\odot]$ & $[10^{14}~\mathrm{kHz}^{-2}]$ & $[10^{14}~\mathrm{kHz}^{-1}]$ & $[10^{14}]$ & [$10^{15}~\mathrm{g}~\mathrm{cm}^{-3}$]& [$10^{15}~\mathrm{g}~\mathrm{cm}^{-3}$]\\
\hline
2.4 & $2.331$ & $-7.717$ & $10.82$ & 0.017 & 0.034 \\
2.7 & $1.689$ & $-2.927$ & $2.837$ & 0.029 & 0.067 \\
2.8 & $3.418$ & $-12.73$ & $16.65$ & 0.011  & 0.023\\
3.0 & $6.705$ & $-29.14$ & $37.25$ & 0.053  & 0.067 \\
\hline
\hline
\end{tabular}
\caption{Fit parameters $a_{M},~b_{M},~c_{M}$ as defined by Eq.~\eqref{QuadraticrhoFit} for the empirical relations shown in Figs.~\ref{fpeakrhomamaxrelations_a}--\ref{fpeakrhomamaxrelations_d} together with the mean and the maximum deviation of the data from the fit. These fits and the resulting residuals include only data from purely hadronic EoSs.}
\label{fpeakrhofits}
\end{table*}

These relations imply that for a given binary mass the maximum density occurring during the early remnant evolution can be estimated by $f_{\mathrm{peak}}$. The data points for the hybrid DD2F-SF models are mostly shifted towards higher frequencies for binary masses, where they clearly deviate from the $f^\mathrm{had}_{\mathrm{peak}}$-$\Lambda$-relation shown in Figs.~\ref{fpeaklampdarelations_b} and ~\ref{fpeaklampdarelations_c} (1.35--1.35~M$_{\odot}$ and 1.4--1.4~M$_{\odot}$).

Note that for 1.2--1.2~M$_{\odot}$ binaries the situation is different and some hybrid models show larger values of $\rho_{\mathrm{max}}^{\mathrm{max}}$ than expected from their $f_{\mathrm{peak}}$ value. The transition density decreases with temperature and a small fraction of matter undergoes a transition to quark matter even in these low-mass mergers. This increases $\rho_{\mathrm{max}}^{\mathrm{max}}$, but the amount of matter in the quark phase is still too small to strongly affect $f_{\mathrm{peak}}$.

\subsection{Mass-independent relations} \label{fpeakrhomassindep}

As for the $f^\mathrm{had}_{\mathrm{peak}}(\Lambda)$ relations we also construct a mass-independent relation $\rho_{\mathrm{max}}^{\mathrm{max}}(f^\mathrm{had}_{\mathrm{peak}})$.
\begin{figure}
\centering
\includegraphics[width=1.0\linewidth]{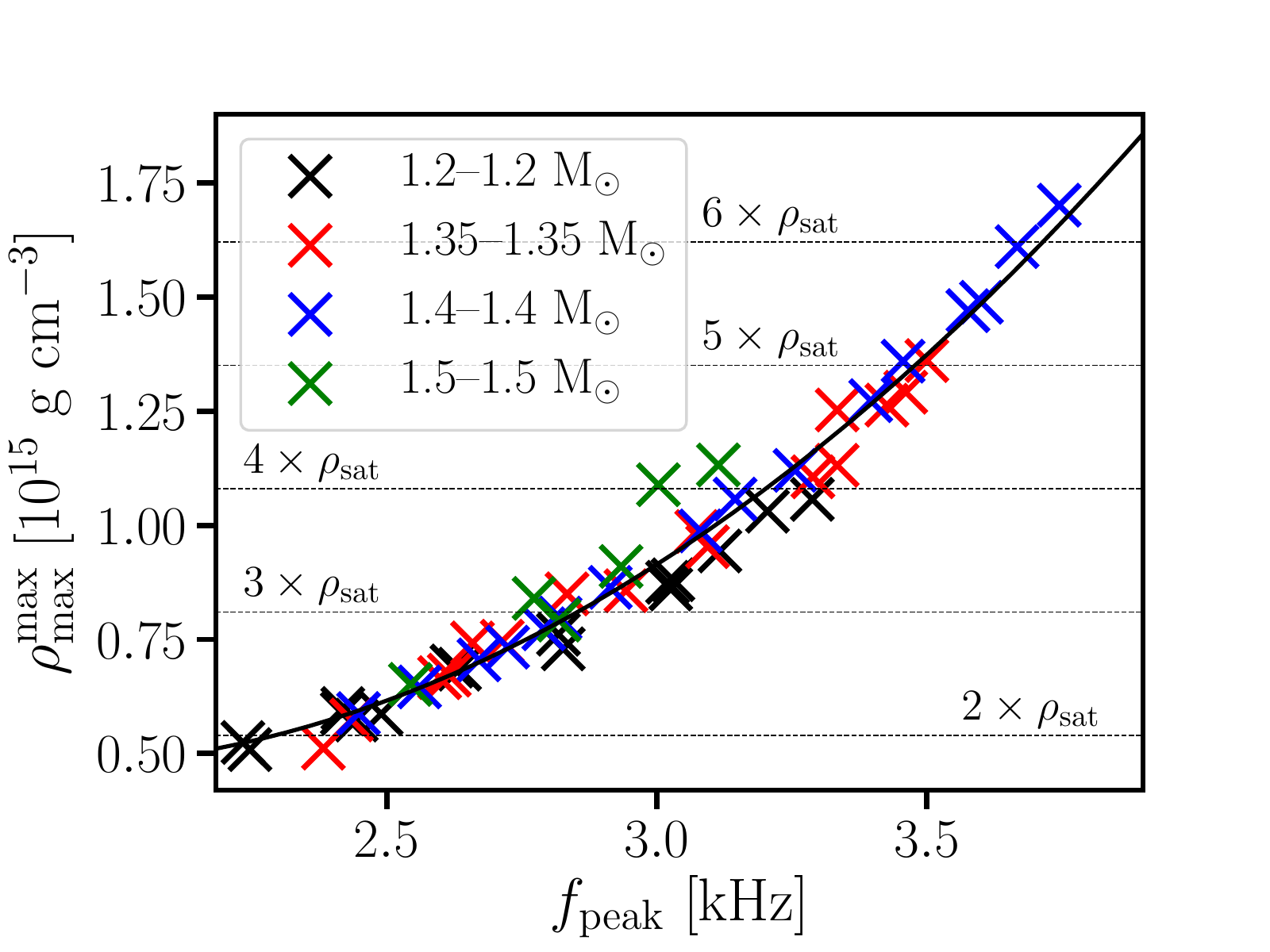}
\caption{Maximum rest-mass density $\rho_{\mathrm{max}}^{\mathrm{max}}$ in the remnant during the first 5 milliseconds after the merger as a function of the dominant postmerger GW frequency $f_{\mathrm{peak}}$ for 1.2--1.2~M$_{\odot}$ (black crosses), 1.35--1.35~M$_{\odot}$ (red crosses), 1.4--1.4~M$_{\odot}$ (blue crosses) and 1.5--1.5~M$_{\odot}$ (green crosses) mergers with different microphysical EoSs. This plot contains the entire data of Figs.~\ref{fpeakrhomamaxrelations_a}--\ref{fpeakrhomamaxrelations_d} excluding the results from the DD2F-SF EoSs. Solid black line shows a least squares fit to all shown datapoints (Eq.~\eqref{QuadraticrhoFit}).}
\label{rhomaxfpeakuniversal}
\end{figure}

Fig.~\ref{rhomaxfpeakuniversal} shows the results from Fig.~\ref{fpeakrhomamaxrelations} combined in a single plot without any further rescaling. The different colors mark data from different binary mass configurations. Data from 1.2--1.2~M$_{\odot}$,~1.35--1.35~M$_{\odot}$,~1.4--1.4~M$_{\odot}$ and 1.5--1.5~M$_{\odot}$ binaries are displayed by black, red, blue and green crosses, respectively. For clarity, the results from the DD2F-SF models are dismissed in Fig.~\ref{rhomaxfpeakuniversal}. Interestingly, we find that the $\rho_{\mathrm{max}}^{\mathrm{max}}(f^\mathrm{had}_{\mathrm{peak}})$ data from different binary mergers follows a nearly universal relation and can be well described by a single quadratic function (solid black line in Fig.~\ref{rhomaxfpeakuniversal}). We obtain the parameters of Eq.~\eqref{QuadraticrhoFit} through a least squares fit with $a_{M}=3.226\times 10^{14}~\mathrm{kHz}^{-2}$, $b_{M}=\mathrm{-}1.178\times 10^{15}~\mathrm{kHz}^{-1}$ and $c_{M}=1.545\times 10^{15}$. The mean and the maximum deviation of the underlying data from this fit is $0.033 \times 10^{15}~\mathrm{g}~\mathrm{cm}^{-3}$ and $0.172 \times 10^{15}~\mathrm{g}~\mathrm{cm}^{-3}$, respectively, i.e. {\raise.17ex\hbox{$\scriptstyle\mathtt{\sim}$}}3\% and {\raise.17ex\hbox{$\scriptstyle\mathtt{\sim}$}}15\% of a typical $\rho_{\mathrm{max}}^{\mathrm{max}}$ value.

The largest values of $\rho_{\mathrm{max}}^{\mathrm{max}}$ are reached in simulations with 1.4--1.4~M$_{\odot}$ binaries and not as one might expect in 1.5--1.5~M$_{\odot}$ mergers. This is due to the fact that for larger binary masses most remnants undergo a prompt collapse to a black hole \cite{Bauswein2013}. In this case no strong GW emission from the postmerger phase occurs. Only simulations with stiff EoSs lead to temporarily stable remnants and hence yield values of $f_{\mathrm{peak}}$. We also point out that for the considered binary masses the highest values of $\rho_{\mathrm{max}}^{\mathrm{max}}$ are of the order of six times nuclear saturation density $\rho_{\mathrm{sat}}$, which is smaller than the central density in isolated, static NSs with masses close to the maximum mass. For example, for the DD2F EoS the largest density in a non-rotating NS is $6.62\times\rho_{\rm sat}$.

As for the $f^\mathrm{had}_{\mathrm{peak}}$($\Lambda$) relation discussed above we observe larger deviations of the data from the universal mass-independent $\rho_{\mathrm{max}}^{\mathrm{max}}(f^\mathrm{had}_{\mathrm{peak}})$ relation than for the mass-dependent relations. Therefore, we again introduce universal relations valid for different mass ranges. 

For this, we follow the same procedure as before. We consider three subsets of data consisting of results from
1.2--1.2~M$_{\odot}$ and 1.35--1.35~M$_{\odot}$, 1.35--1.35~M$_{\odot}$ and 1.4--1.4~M$_{\odot}$ as well as from 1.4--1.4~M$_{\odot}$ and 1.5--1.5~M$_{\odot}$ merger simulations and fit the data in each subset using Eq.~\eqref{QuadraticrhoFit}. The fit parameters as well as the mean and maximum deviation of the data from the fit can be found in Tab.~\ref{fpeakrhoCombifits} for every binary mass range.
 
\begin{table*}
\begin{tabular}{c c c c c c}
\hline\hline
$M_{\mathrm{tot}}$ & $a_{M}$ & $b_{M}$ & $c_{M}$ & mean dev.  & max dev.\\
$[\mathrm{M}_\odot]$ & [$10^{14}~\mathrm{kHz}^{-2}$] & [$10^{14}~\mathrm{kHz}^{-1}$] & [$10^{14}$] & [$10^{15}~\mathrm{g}~\mathrm{cm}^{-3}$]& [$10^{15}~\mathrm{g}~\mathrm{cm}^{-3}$]\\
\hline
2.4--2.7 & $3.079$ & $-11.31$ & $15.12$ & 0.028 & 0.088 \\
2.7--2.8 & $3.275$ & $-12.01$ & $15.71$ & 0.024 & 0.075 \\
2.8--3.0 & $2.255$ & $-5.601$ & $6.065$ & 0.030 & 0.130\\
\hline\hline
\end{tabular}
\caption{Fit parameters $a_{M},~b_{M},~c_{M}$ as defined by Eq.~\eqref{QuadraticrhoFit} for the empirical relations shown in Figs.~\ref{rhomaxfpeakcombirelations_a}--\ref{rhomaxfpeakcombirelations_c} together with the mean and the maximum deviation of the data from the fit. These fits and the resulting residuals include only data from purely hadronic EoSs.}
\label{fpeakrhoCombifits}
\end{table*}
The plots of the three subsets of data together with the respective fit are provided in Appendix~\ref{moreplotsrhofpeak}.

\subsection{$\rho_{\mathrm{max}}^{\mathrm{max}}-\Lambda$ relation}\label{rholambda}
We find that $\rho_{\mathrm{max}}^{\mathrm{max}}$ correlates with $f_{\mathrm{peak}}$ and $f_{\mathrm{peak}}$ scales with $\Lambda$ (for hadronic EoSs). Hence, we also expect $\rho_{\mathrm{max}}^{\mathrm{max}}$ to correlate with $\Lambda$. Since we use second order polynomials to describe the first two relations, we anticipate a combined $\rho_{\mathrm{max}}^{\mathrm{max}}(\Lambda)$ relation to follow a higher order polynomial where the values of $\rho_{\mathrm{max}}^{\mathrm{max}}$ are scaled with the total binary mass $M_{\mathrm{tot}}$.

We find that a third order polynomial of the form
\begin{align}
    \rho_{\mathrm{max}}^{\mathrm{max}}\times M_{\mathrm{tot}}=(a\Lambda^{3}+b\Lambda^{2}+c\Lambda+d)~\mathrm{g}~ \mathrm{cm}^{-3}~ \mathrm{M}_\odot
    \label{rhoOfLambda}
\end{align}
provides a good description of the data. 
\begin{figure}
\centering
\includegraphics[width=1.0\linewidth]{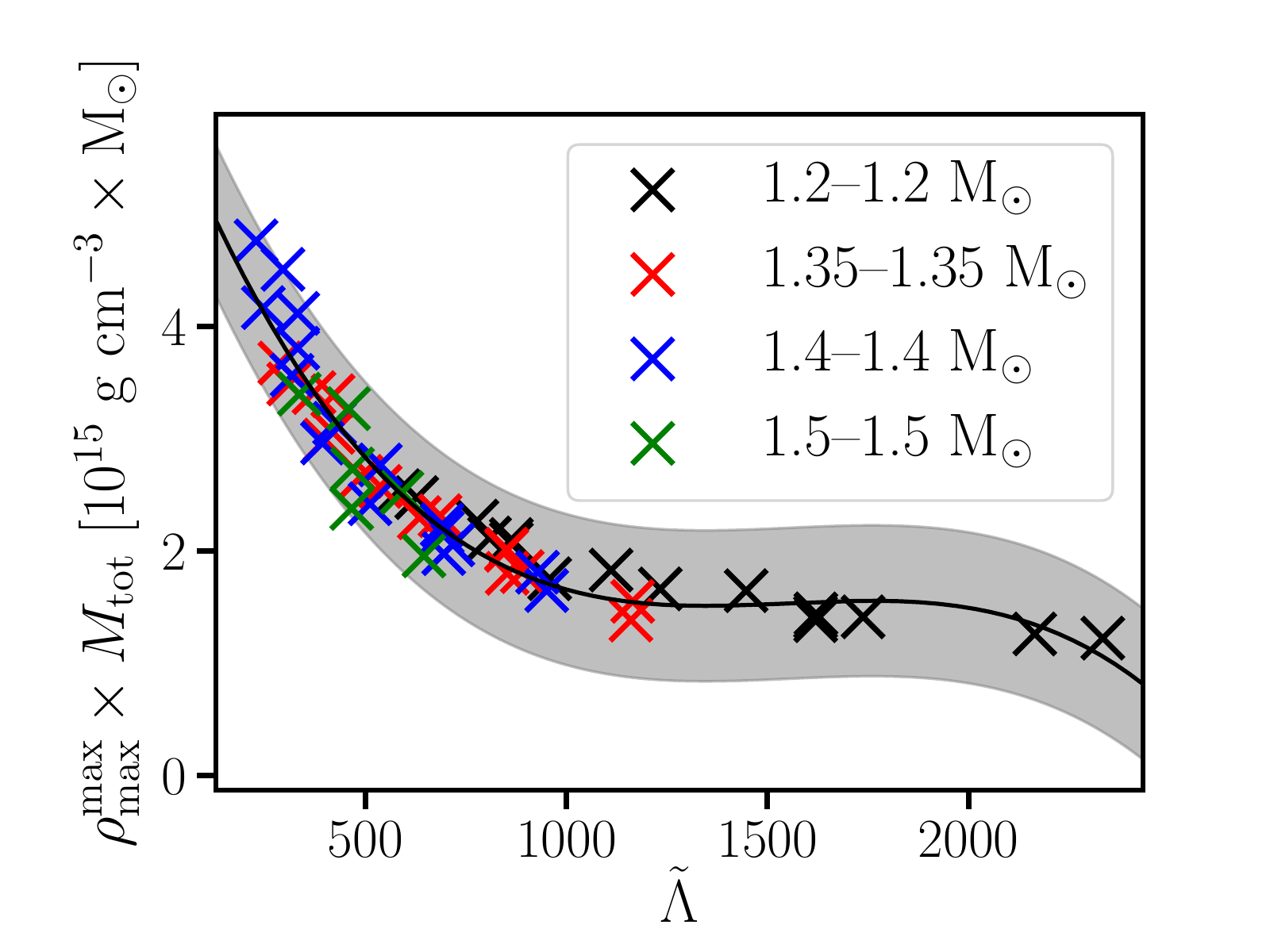}
\caption{Maximum rest-mass density $\rho_{\mathrm{max}}^{\mathrm{max}}$ in the remnant during the first 5 milliseconds after the merger scaled by the total binary mass $M_{\mathrm{tot}}$ as a function of the combined tidal deformability $\tilde{\Lambda}$ for 1.2--1.2~M$_{\odot}$ (black), 1.35--1.35~M$_{\odot}$ (red), 1.4--1.4~M$_{\odot}$ (blue) and 1.5--1.5~M$_{\odot}$ (green) mergers with different purely hadronic microphysical EoSs. The solid black line shows a least squares fit to all shown datapoints (Eq.~\eqref{rhoOfLambda}). The gray shaded area illustrates the maximum deviation of the data from the fit.}
\label{rhomaxlamuniversal}
\end{figure}

This mass-independent relation between $\rho_{\mathrm{max}}^{\mathrm{max}}\times M_{\mathrm{tot}}$ and $\Lambda$ is shown in Fig.~\ref{rhomaxlamuniversal}. Different colored crosses refer to data from hadronic EoSs from different binary masses. Data from 1.2--1.2~M$_{\odot}$,~1.35--1.35~M$_{\odot}$,~1.4--1.4~M$_{\odot}$ and 1.5--1.5~M$_{\odot}$ binaries are displayed by black, red, blue and green signs, respectively. The solid black line shows a least squares fit to the data using Eq.~\eqref{rhoOfLambda}. The gray shaded area depicts the maximum deviation of data from the fit. The fit parameters are given by $a=\mathrm{-}1.260\times 10^{6}$, $b=5.871\times 10^{9}$, $c=\mathrm{-}8.953\times 10^{12}$ and $d=5.996\times 10^{15}$. The mean and the maximum deviation of our data from the fit is $0.178\times 10^{15}~\mathrm{g}~\mathrm{cm}^{-3}~ \mathrm{M}_\odot$ and $0.673\times 10^{15}~\mathrm{g}~\mathrm{cm}^{-3}~ \mathrm{M}_\odot$, respectively.

As before, to increase the accuracy of the relation within individual binary mass ranges, we obtain different parameters of Eq.~\eqref{rhoOfLambda} for different mass ranges by fitting the results from 1.2--1.2~M$_{\odot}$ and 1.35--1.35~M$_{\odot}$, 1.35--1.35~M$_{\odot}$ and 1.4--1.4~M$_{\odot}$ as well as from 1.4--1.4~M$_{\odot}$ and 1.5--1.5~M$_{\odot}$ merger simulations separately. The least squares fit parameters for every mass range are given in Tab.~\ref{rhoOfLambdaCombiFits} together with the mean and the maximum deviation of the data from the fit function. The plots together with the fits for each mass range can be found in the Appendix~\ref{moreplotsrholambda}. A discussion on the effects of asymmetric binaries is provided in Appendix~\ref{asymbinar}.
\begin{table*}
\begin{tabular}{c c c c c c c}
\hline\hline
$M_{\mathrm{tot}}$ & $a$ & $b$ & $c$ & $d$ & mean dev.  & max dev.\\
$[\mathrm{M}_\odot]$ & [$10^{5}$] & [$10^{9}$] & [$10^{12}$] & [$10^{15}$] & [$10^{15}~\mathrm{g}~\mathrm{cm}^{-3} \mathrm{M}_\odot$]& [$10^{15}~\mathrm{g}~\mathrm{cm}^{-3} \mathrm{M}_\odot$]\\
\hline
2.4--2.7 & $-7.085$ & $3.619$ & $-6.286$ & $5.142$ & 0.096 & 0.292 \\
2.7--2.8 & $-64.26$ & $17.40$ & $-16.80$ & $7.559$ & 0.161 & 0.554 \\
2.8--3.0 & $-90.18$ & $22.56$ & $-20.16$ & $8.223$ & 0.207 & 0.492 \\
\hline\hline
\end{tabular}
\caption{Fit parameters $a,~b,~c,~d$ as defined by Eq.~\eqref{rhoOfLambda} for the empirical relations $\rho_{\mathrm{max}}^{\mathrm{max}}(\Lambda)$ shown in Figs.~\ref{rhomaxlambdacombirelations_a}-\ref{rhomaxlambdacombirelations_c} together with the mean and the maximum deviation of the data from the fit. These fits and the resulting residuals include only data from purely hadronic EoSs.}
\label{rhoOfLambdaCombiFits}
\end{table*}

\section{Constraining the onset density}\label{constraining}
In this section we describe, how the empirical relations $f^\mathrm{had}_{\mathrm{peak}}(\Lambda)$ and $\rho_{\mathrm{max}}^{\mathrm{max}}(\Lambda)$ can be employed to constrain the onset density of a strong PT from hadronic to deconfined quark matter. We assume sufficiently accurate measurements of $\Lambda$, $f_{\mathrm{peak}}$ and $M_{\mathrm{tot}}$. For a brief discussion of possible measurement uncertainties see \cite{Bauswein2019} and references therein. The procedure consists of two steps. First, a comparison of the measured values to $f^\mathrm{had}_{\mathrm{peak}}(\Lambda)$ reveals whether or not a strong PT occurred during the merger. Then, the measured $\Lambda$ in combination with the relation $\rho_{\mathrm{max}}^{\mathrm{max}}(\Lambda)$ provides a limit on the onset density of the PT. If there is evidence for a strong PT, $\rho_{\mathrm{max}}^{\mathrm{max}}(\Lambda)$ yields an upper limit on the onset density. In case the $f^\mathrm{had}_{\mathrm{peak}}$-$\Lambda$ comparison does not reveal evidence for a PT, we can exclude a strong PT up to some lower limit. We address possible caveats of this method in Sect.~\ref{conservativelimit} and Sect.~\ref{summary}.
\subsection{Basic procedure}
We demonstrate the basic idea by considering an example of a hypothetical detection of GWs from a NS merger.
For this discussion we adopt a measured total binary mass of $M_{\mathrm{tot}} = 2.65~\mathrm{M}_\odot$. We assume that the measurement provides values $\Lambda$ and $f_{\mathrm{peak}}$.

First, we check whether or not $\Lambda$ and $f_{\mathrm{peak}}$ follow the empirical $f^\mathrm{had}_{\mathrm{peak}}(\Lambda)$ relation. The given total binary mass of 2.65~M$_\odot$ falls in the range of the first $M_\mathrm{tot}$ interval listed in Tab.~\ref{fpeaklambdafits}. We thus compare $f_{\mathrm{peak}}$ with
\begin{align}
f^\mathrm{had}_{\mathrm{peak}}=\frac{1}{2.65}(a\Lambda^{2}+b\Lambda+c)~\rm kHz 
\label{QuadraticLambdaFit2} 
\end{align}
with the parameters $a=1.201\times 10^{-6},~b=\mathrm{-}4.974\times 10^{-3},~c=10.65$ taken from Tab.~\ref{fpeaklambdafits} (first row). Two outcomes are possible.

(1) If the measured $f_{\mathrm{peak}}$ is consistent with Eq.~\eqref{QuadraticLambdaFit2} within the maximum residual of this relation, no PT occurred in the merger remnant. The maximum residual for purely hadronic EoS models for this mass range is 357~Hz$~ $M$_\odot$ (see Tab.~\ref{fpeaklambdafits}). In this case the consistency with Eq.~\eqref{QuadraticLambdaFit2} within at least 357/2.65~Hz implies that the PT did not occur up to the maximum density in the remnant \footnote{Additionally, one has to consider uncertainties of the measured $\Lambda$. For example, we find that using the $\Lambda$ of the DD2F for a NS with 1.325~M$_\odot$ a $\Lambda$ uncertainty of 5\% translates to an additional uncertainty of about 40~Hz for $f^\mathrm{had}_\mathrm{peak}$}. This maximum density is given by the relation $\rho_{\mathrm{max}}^{\mathrm{max}}(\Lambda)$ (Eq.~\eqref{rhoOfLambda}). We thus conclude that the onset density is larger than 
\begin{align}
\rho_{\mathrm{onset}}>\frac{1}{2.65}(a\Lambda^{3}+b\Lambda^{2}+c\Lambda+d)~\rm g~\rm cm^{-3}
\label{rhoOfLambda2}
\end{align}
with the parameters $a=\mathrm{-}7.085\times 10^{5},~b=3.619\times 10^{9},~c=\mathrm{-}6.286\times 10^{12},~d=5.142\times 10^{15}$ taken from Tab.~\ref{rhoOfLambdaCombiFits} (first row) for the corresponding binary mass range. Note that this limit and the limits below can be readily converted to a baryon density $n$ via $\rho=m_\mathrm{u} n$ with $m_\mathrm{u}=931.49432~$MeV.

Alternatively, one can employ $f_{\mathrm{peak}}$ and Eq.~\eqref{QuadraticrhoFit} to constrain $\rho_{\mathrm{onset}}$. Deriving a density limit from the postmerger frequency would actually be a more natural choice to constrain the properties of a PT in the postmerger remnant. However, because of the relatively tight scaling between $\Lambda$ and $f^\mathrm{had}_{\mathrm{peak}}$ the two approaches are equivalent, and in practise one would employ the one resulting in the smallest uncertainties.

(2) If the measured $f_{\mathrm{peak}}$ exceeds Eq.~\eqref{QuadraticLambdaFit2} by more than the maximum residual of 357/2.65~Hz, we would interpret this as evidence of a PT. In this case the $\rho_{\mathrm{max}}^{\mathrm{max}}(\Lambda)$ (Eq.~\eqref{rhoOfLambda}) relation will inform us about the density at which the transition already took place.

For our example the density at which the PT already occurred, has to be smaller than
\begin{align}
\rho_{\mathrm{onset}}<\frac{1}{2.65}(a\Lambda^{3}+b\Lambda^{2}+c\Lambda+d)~\mathrm{g}~ \mathrm{cm}^{-3}
\label{rhoOfLambda3}
\end{align}
with the same parameters as used for Eq.~\eqref{rhoOfLambda2} (adopted for the measured binary mass) but opposite inequality sign. The limit given by Eq.~\eqref{rhoOfLambda3} corresponds to the value of $\rho_{\mathrm{max}}^{\mathrm{max}}$ which we would expect in the remnant if it did not undergo a PT and had remained purely hadronic. The actual densities in the remnant will be larger because of the PT which effectively softens the EoS and thus leads to higher densities.

For both scenarios an error of about 1 to $2\times 10^{14}\mathrm{g}~ \mathrm{cm}^{-3}$ should be adopted which corresponds to the maximum scatter in the employed relations. An additional error from the measurement of $\Lambda$ has to be considered here. 

Note that here $\mathbf{\rho_{\mathrm{onset}}}$ refers to the onset density at zero temperature in beta-equilibrium. Below we discuss the impact of the temperature dependence of the phase boundaries.

As a second example we consider results from GW170817. We adopt a total binary mass of 2.74~M$_\odot$ and a combined tidal deformability of 500. For these values we obtain $\rho_\mathrm{max}^\mathrm{max}=0.988\times 10^{15}~\mathrm{g}~\mathrm{cm}^{-3}$ from Eq.~\eqref{rhoOfLambda}. The scatter of our relation for this total binary mass is $0.202\times 10^{15}~\mathrm{g}~\mathrm{cm}^{-3}$. Assuming a 5\% uncertainty of $\Lambda$ adds an additional scatter of about $0.04\times 10^{15}~\mathrm{g}~\mathrm{cm}^{-3}$. Hence, if $f_\mathrm{peak}$ had been measured precisely a consistency with Eq.~\eqref{QuadraticLambdaFit2} would lead to a lower limit of $\rho_\mathrm{onset}>0.746\times 10^{15}~\mathrm{g}~\mathrm{cm}^{-3}$ and a significant deviation from Eq.~\eqref{QuadraticLambdaFit2} would lead to an upper limit of $\rho_\mathrm{onset}<1.230\times 10^{15}~\mathrm{g}~\mathrm{cm}^{-3}$.

Obviously, the same procedures can be applied to any other measured total binary mass between 2.4~M$_\odot$ and 3.0~M$_\odot$. Depending on the actual value of $M_\mathrm{tot}$ the parameters in Eqs.~\eqref{QuadraticLambdaFit2},~\eqref{rhoOfLambda2} and \eqref{rhoOfLambda3} have to be replaced and different residuals should be considered, all of which are listed in Tabs.~\ref{fpeaklambdafits} and \ref{rhoOfLambdaCombiFits}.

Two more remarks are important. We here derive a procedure that can be directly applied as soon sufficiently accurate measurements are available. We describe the method for any total binary mass assuming a binary system, which is not too asymmetric. In future, it will be advantageous to perform simulations for the measured binary system, i.e. with the same total binary mass and in particular, the same mass ratio. The resulting empirical relations for fixed binary masses will have smaller maximum residuals (comparable to those for Eq.~\eqref{QuadraticLambdaFit} see Tab.~\ref{fpeaklambdafits1} and Eq.~\eqref{QuadraticrhoFit}, see Tab.~\ref{fpeakrhofits}).
This will improve the sensitivity and the accuracy of the procedure. Also, using only a subset of candidate EoSs which are compatible with the observations, will allow to construct more precise relations with smaller residuals.

\subsection{Conservative limits} \label{conservativelimit}
The procedure above can be directly applied. Here we describe a more conservative constraint on the onset density, which has the following background. In our simulations with the seven DD2F-SF models we have encountered two different scenarios that somewhat complicate the procedure to place constraints on the transition density if one intends a particularly conservative estimate. The two effects are competing, and both can be simultaneously present, which is why the procedure described in the previous subsection yields an accurate limit for most models unless one considers rather extreme cases. It was verified that the procedure described above would yield a correct constraint on $\rho_\mathrm{onset}$ for all hybrid DD2F-SF models for the different total binary masses which were actually simulated in this work. However, since not all possible mass configurations were simulated, it is possible that a narrow mass range for some hybrid models exists, where the simple procedure would yield a slightly incorrect limit on $\rho_\mathrm{onset}$. This would not be the case for the more conservative procedure described here.

The first complication arises from the fact that only a sufficiently large core of quark matter leads to a significant shift of the postmerger frequency relative to the tidal deformability. Within our sample of simulations with the DD2F-SF EoSs we observe systems where $\rho_{\mathrm{max}}^{\mathrm{max}}$ exceeds the onset density of the PT, but the postmerger frequency is only slightly or marginally affected. The quark core in these systems is too small to significantly alter the stellar structure of the remnant and thus its oscillation frequency.

Hence, a small amount of quark matter may not necessarily leave a significant and thus observable imprint on the GW signal, i.e. a relative shift of $f_\mathrm{peak}$ which is indicative of a PT. We explicitly refer to Fig.~\ref{avQuarksAll} and the discussion of Appendix~\ref{detDeltaM} for a more detailed analysis of this point. In this scenario, however, a slightly more massive binary system would lead to a sizeable quark matter core and, consequently, an observable signature of quark matter as discussed above. Hence, we can accommodate this situation by an effective prescription, which introduces a shift to rule out quark matter in a fiducial system of somewhat lower mass.

The argument works as follows. Supposed we observed a binary merger $X$ with total mass $M_\mathrm{tot}^\mathrm{X}$ without finding evidence of a sufficiently large quark matter core. Then, we cannot exclude small amounts of quark matter in this system $X$. However, we can rule out the existence of quark matter in a system $Y$ with $M_\mathrm{tot}^\mathrm{Y}=M_\mathrm{tot}^\mathrm{X}-\Delta M$ because if a small fraction of quark matter was present in $Y$, the quark core in $X$ would be much larger and had consequently lead to an observable shift of the postmerger frequency.

This hypothetical system $Y$ would have a somewhat larger tidal deformability, which can be estimated by $\Lambda^\mathrm{y}=\Lambda(M_\mathrm{tot}^\mathrm{y})= \Lambda(M_\mathrm{tot}^\mathrm{X})-\frac{d\,\Lambda}{d\,M_\mathrm{tot}}\,\Delta M$ (note that $\frac{d\,\Lambda}{d\,M_\mathrm{tot}}$ is negative).

In \cite{De2018} the authors describe that generally $\Lambda$ varies as $M^{-6}$. We find that for our sample of hadronic EoSs the slope $\frac{d\,\Lambda}{d\,M_\mathrm{tot}}$ can be well described by
\begin{align}
    \frac{d\,\Lambda}{d\,M_\mathrm{tot}}=z \frac{\Lambda}{M_\mathrm{tot}}
    \label{slopeLambdaoverM}
\end{align}
with $z=\mathrm{-}5.709$. Details can be found in Appendix~\ref{detslope}.\\

Hence, we infer a safe lower bound on $\rho_\mathrm{onset}$ by inserting $\Lambda^{Y}=\Lambda(M_\mathrm{tot}^\mathrm{X})-\frac{d\,\Lambda}{d\,M_\mathrm{tot}}\,\Delta M$ and $M_\mathrm{tot}-\Delta M$ instead of just $M_\mathrm{tot}$ in Eq.~\eqref{rhoOfLambda2}. We estimate an appropriate $\Delta M$ below.

An opposite effect can lead to a second complication. In this paper we intend to constrain $\rho_\mathrm{onset}$ at beta-equilibrium and zero temperature from an estimate of the maximum density in the merger remnant at finite temperature. As mentioned above, temperature effects can lead to a reduction of the onset density for the models considered in this study. Therefore, quark matter may be present in some systems even with $\rho_\mathrm{max}^\mathrm{max}<\rho_\mathrm{onset}$ and might already lead to a strong shift of $f_\mathrm{peak}$. Also, composition effects may in principle lead to the appearance of quark matter at somewhat lower densities.

Again, these effects can be captured by introducing a fiducial binary system with slightly different total binary mass. We thus devise the following procedure.

Suppose we observed a binary merger $X$ with total mass $M_\mathrm{tot}^\mathrm{X}$ revealing clear evidence of a sufficiently large quark matter core. In principle, the temperature dependence of the phase boundary could trigger the occurrence of quark matter and a corresponding GW signal, although the density in the merger remnant did not reach the onset density of the PT at zero temperature. Therefore, if temperature effects in the merger remnant strongly lowered the transition density, our inferred upper bound on $\rho_\mathrm{onset}$ at T=0 might be too small. However, a fiducial, slightly more massive system Z with $M_\mathrm{tot}^\mathrm{Z}=M_\mathrm{tot}^\mathrm{X}+\Delta M$ would yield a correct upper bound for $\rho_\mathrm{onset}$ at zero temperature.

If the appearance of quark matter in the more massive system Z was purely caused by the lowering of the onset density due to thermal effects, no clear signs of a PT would have occurred in system X. But, since X showed evidence for quark matter, the more massive system Z must have reached sufficiently high densities to provide a safe upper limit.

The hypothetical system Z would then have a somewhat smaller tidal deformability $\Lambda^{Z}= \Lambda(M^{X}_\mathrm{tot})+\frac{d\,\Lambda}{d\,M_\mathrm{tot}}\,\Delta M$.
Hence, we infer a safe upper bound on $\rho_\mathrm{onset}$ by inserting $\Lambda^{Z}= \Lambda(M^{X}_\mathrm{tot})+\frac{d\,\Lambda}{d\,M_\mathrm{tot}}\,\Delta M$ and $M_\mathrm{tot}+\Delta M$ instead of simply $M_\mathrm{tot}$ in Eq.~\eqref{rhoOfLambda3}.

We find that a value of $\Delta M$=0.2~M$_{\odot}$ is sufficient to safely capture both effects for all hybrid models tested in this study. The exact determination of $\Delta M$ is described in Appendix~\ref{detDeltaM}. Considering the strong variations among the different quark matter models in this work, we expect that this value suffices for extreme hybrid models. Future work should solidify these findings.

\subsection{Ready-to-use procedure for constraints}\label{procedure}

To summarize the results of the previous discussion we here provide ready-to-use formulas for conservative constraints on the transition density to deconfined quark matter.
We adopt sufficiently accurate measurements of the tidal deformability $\Lambda$, the dominant postmerger GW frequency $f_{\mathrm{peak}}$ and the total mass of the binary $M_\mathrm{tot}$ assuming that the binary is sufficiently symmetric such that mass ratio effects do not play a significant role. We here employ the universal relations which are valid for certain ranges in $M_\mathrm{tot}$. 

The first step is to calculate the value of $f_{\mathrm{peak}}^{\mathrm{had}}$ which is expected for a purely hadronic NS merger based on the measured tidal deformability $\Lambda$ and $M_\mathrm{tot}$. It is given by
\begin{align}
f_{\mathrm{peak}}^{\mathrm{had}}(\Lambda,M_{\mathrm{tot}})=\frac{1}{M_{\mathrm{tot}}}~ (a\Lambda^{2}+b\Lambda+c)~\mathrm{kHz}
\label{fhypo}
\end{align}
with $M_{\mathrm{tot}}$ in M$_{\odot}$. The parameters $a,~b$ and $c$ depend on $M_\mathrm{tot}$ and are given by
\begin{align}
a,b,c=
\begin{cases}
1.201\times 10^{-6},-4.974\times 10^{-3},10.650 \\
\mathrm{for}~2.4~\mathrm{M}_{\odot} \leq M_{\mathrm{tot}}<2.7~\mathrm{M}_{\odot}\\
3.405\times 10^{-6},-8.620\times 10^{-3},12.014 \\
\mathrm{for}~2.7~\mathrm{M}_{\odot} \leq M_{\mathrm{tot}}<2.8~\mathrm{M}_{\odot}\\
4.608\times 10^{-6},-1.012\times 10^{-2},12.472 \\
\mathrm{for}~2.8~\mathrm{M}_{\odot} \leq M_{\mathrm{tot}} \leq 3~\mathrm{M}_{\odot}\\
\end{cases}  \label{fhypopar}
\end{align}
The maximum density during the early postmerger evolution can be well estimated by
\begin{align}
    \rho_{\mathrm{max}}^{\mathrm{max}}(\Lambda,M_{\mathrm{tot}})=\frac{1}{M_{\mathrm{tot}}}(a\Lambda^{3}+\Lambda^{2}+c\Lambda+d)~\mathrm{g}~ \mathrm{cm}^{-3}
    \label{rhomaxhypo}
\end{align}
with $M_{\mathrm{tot}}$ in M$_{\odot}$. The parameters $a,~b,~c$ and $d$ are
\begin{align}
\begin{split}
    &a, b,\\
    &c, d
\end{split}
=
\begin{cases}
-7.085\times 10^{5},3.619\times 10^{9},-6.286\times 10^{12},\\
5.142\times 10^{15}~~ \mathrm{for}~2.4~\mathrm{M}_{\odot} \leq M_{\mathrm{tot}}<2.7~\mathrm{M}_{\odot}\\
-6.426\times 10^{6},1.740\times 10^{10},-1.680\times 10^{13},\\
7.559\times 10^{15}~~\mathrm{for}~2.7~\mathrm{M}_{\odot} \leq M_{\mathrm{tot}}<2.8~\mathrm{M}_{\odot}\\
-9.018\times 10^{6},2.256\times 10^{10},-2.016 \times 10^{13},\\
8.223\times 10^{15}~~ \mathrm{for}~2.8~\mathrm{M}_{\odot} \leq M_{\mathrm{tot}} \leq 3~\mathrm{M}_{\odot}\\
\end{cases}  \label{rhomaxhypopar}
\end{align}
If $f_{\mathrm{peak}}-f_{\mathrm{peak}}^{\mathrm{had}}(\Lambda,M_{\mathrm{tot}})<$ 0.2~kHz there is no clear evidence of a PT. In this case a conservative lower limit on $\rho_\mathrm{onset}$ is given by
\begin{align}
    \rho_{\mathrm{onset}}> \rho_{\mathrm{max}}^{\mathrm{max}}(\Lambda^{X},M_{\mathrm{tot}}-0.2~\mathrm{M}_{\odot})-\Delta
    \label{rhoonsetlowconservative}
\end{align}
with $\Lambda^{X}=\Lambda+5.709\frac{\Lambda}{M_{\mathrm{tot}}}\times 0.2~$M$_{\odot}$. The additional term $\Delta$ corresponds to the largest deviation we observe in our empirical $\rho_{\mathrm{max}}^{\mathrm{max}}(\Lambda)$ relation. It depends on the considered mass and is given by
\begin{align}
\Delta=
   \begin{cases} 
   0.292/(M_{\mathrm{tot}}-0.2~\mathrm{M}_{\odot})~\mathrm{g}~ \mathrm{cm}^{-3}\\
   \mathrm{for}~2.4~\mathrm{M}_{\odot} \leq (M_{\mathrm{tot}}-0.2~\mathrm{M}_{\odot})<2.7~\mathrm{M}_{\odot}\\
   0.554/(M_{\mathrm{tot}}-0.2~\mathrm{M}_{\odot})~\mathrm{g}~ \mathrm{cm}^{-3}\\
   \mathrm{for}~2.7~\mathrm{M}_{\odot} \leq (M_{\mathrm{tot}}-0.2~\mathrm{M}_{\odot})<2.8~\mathrm{M}_{\odot}\\
   0.492/(M_{\mathrm{tot}}-0.2~\mathrm{M}_{\odot})~\mathrm{g}~ \mathrm{cm}^{-3}\\
   \mathrm{for}~2.8~\mathrm{M}_{\odot} \leq (M_{\mathrm{tot}}-0.2~\mathrm{M}_{\odot})<3~\mathrm{M}_{\odot}\\
   \end{cases}
\end{align}

If $f_{\mathrm{peak}}-f_{\mathrm{peak}}^{\mathrm{had}}>$ 0.2~kHz there is strong evidence for the occurrence of a PT. In this case a conservative upper limit on $\rho_\mathrm{onset}$ can be obtained by
\begin{align}
    \rho_{\mathrm{onset}}< \rho_{\mathrm{max}}^{\mathrm{max}}(\Lambda^{X},M_{\mathrm{tot}}+0.2~\mathrm{M}_{\odot})+\Delta
    \label{rhoonsetupconservative}
\end{align}
with $\Lambda^{X}=\Lambda-5.709\frac{\Lambda}{M_{\mathrm{tot}}}\times 0.2$~M$_{\odot}$.

\begin{figure}
\centering
\includegraphics[width=1.0\linewidth]{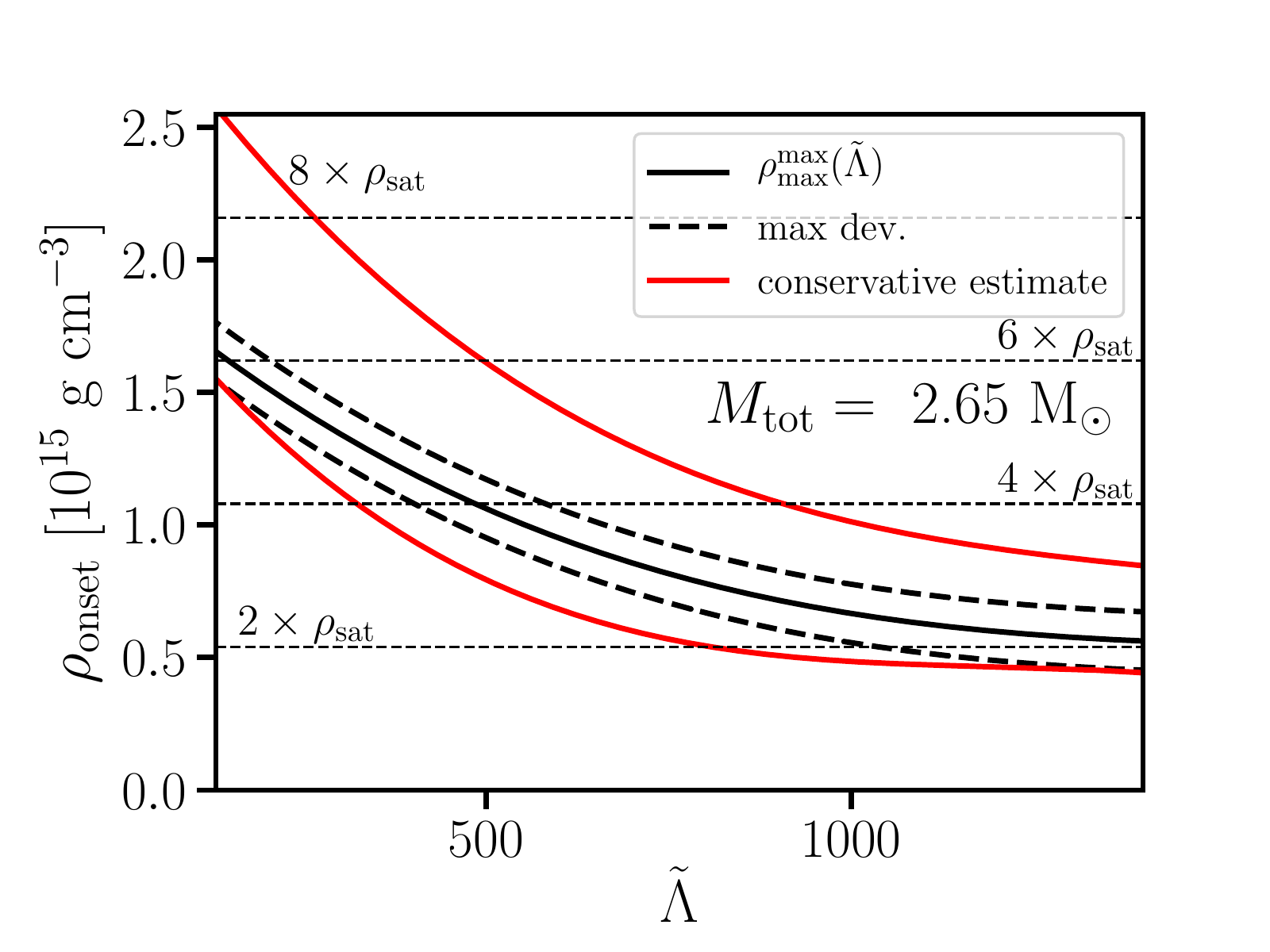}
\caption{Constraints on $\rho_{\mathrm{onset}}$ as a function of $\Lambda$ for a hypothetical 2.65~M$_{\odot}$ binary. The black solid line shows the empirical $\rho_{\mathrm{max}}^{\mathrm{max}}(\Lambda,M_{\mathrm{tot}})$ relation (Eq.~\eqref{rhomaxhypo}) with the parameters from Eq.~\eqref{rhomaxhypopar}. The dashed lines display the uncertainty of the $\rho_{\mathrm{max}}^{\mathrm{max}}(\Lambda,M_{\mathrm{tot}})$ relation. Depending on the consistency of $f_{\mathrm{peak}}$ with Eq.~\eqref{fhypo} these curves illustrate upper or lower limits on $\rho_{\mathrm{onset}}$. The red lines show more conservative constraints introduced in Sect.~\ref{conservativelimit}. See text for more details.}
\label{ProcedurePlot}
\end{figure}

Fig.~\ref{ProcedurePlot} illustrates possible outcomes of this procedure for a total binary mass of $M_{\mathrm{tot}}=2.65~$M$_{\odot}$. It shows a possible limit on $\rho_{\mathrm{onset}}$ as a function of $\Lambda$. The solid black line simply depicts the empirical $\rho_{\mathrm{max}}^{\mathrm{max}}(\Lambda,M_{\mathrm{tot}})$ relation (Eq.~\eqref{rhomaxhypo}) with the parameters from Eq.~\eqref{rhomaxhypopar} (compare with the plots in Appendix~\ref{moreplotsrholambda}). This would be the maximum density we would expect in a purely hadronic remnant to occur as a function of $\Lambda$ without considering any uncertainties.

The dashed lines illustrate the uncertainty of the $\rho_{\mathrm{max}}^{\mathrm{max}}(\Lambda,M_{\mathrm{tot}})$ relation quantified by the maximum scatter of the simulation data. If the value of $f_{\mathrm{peak}}$ is consistent with our $f^\mathrm{had}_{\mathrm{peak}}(\Lambda)$ relation (Eq.~\eqref{fhypo}) then the lower dashed line illustrates the lower limit on $\rho_{\mathrm{onset}}$. The upper dashed line visualizes an upper limit on $\rho_{\mathrm{onset}}$ if the value of  $f_{\mathrm{peak}}$ is not consistent with Eq.~\eqref{fhypo} within about 200~Hz, which indicates that a strong PT has occurred in the remnant.

The red lines show the more conservative constraints that involve extrapolating to a binary of slightly different mass as introduced in Sect.~\ref{conservativelimit}. Again, the upper line displays the upper limit on $\rho_{\mathrm{onset}}$ (Eq.~\eqref{rhoonsetupconservative}) if $f_{\mathrm{peak}}$ deviates strongly from Eq.~\eqref{fhypo}. The lower line depicts the lower limit on $\rho_{\mathrm{onset}}$ (Eq.~\eqref{rhoonsetlowconservative}) assuming $f_{\mathrm{peak}}$ is consistent with Eq.~\eqref{fhypo} \footnote{A small $\Lambda$ uncertainty has little influence on the procedures (both regular and conservative). For example an uncertainty of 5\% leads to about 2\%-4\% additional scatter of $\rho_{\mathrm{onset}}$}. Plots of our procedure at different total binary masses are provided in Appendix~\ref{moreplotsconst}.

\section{Conclusions}\label{summary}
\subsection{Summary}

In this paper we elaborate on a method to detect an unambiguous and measurable signature of the QCD PT in NSs, which we brought forward in \cite{Bauswein2019}. Moreover, we devise a method to constrain the onset density of quark deconfinement. To this end we have performed hydrodynamical simulations of NS mergers with microphysical, temperature-dependent EoSs, which undergo a PT to deconfined quark matter. In this study we consider NS mergers within a wide range of total binary masses. We also corroborate that our findings are not strongly depending on the binary mass ratio by performing simulations for asymmetric binaries. 

The identification of a PT requires the simultaneous measurement of the total binary mass $M_{\mathrm{tot}}$, the tidal deformability $\Lambda$ and the dominant postmerger oscillation frequency $f_{\mathrm{peak}}$, which have all been shown to be measurable with good accuracy in future GW detections. A characteristic increase of the dominant postmerger GW frequency  $f_{\mathrm{peak}}$ relative to a fiducial value derived from the tidal deformability (measured during the inspiral phase) is indicative of a strong PT. The absence of such a frequency shift, i.e. the consistency with an empirical relation  $f^\mathrm{had}_{\mathrm{peak}}(\Lambda)$ which holds for purely hadronic EoS models, may imply that the densities in the merger remnant are not high enough to reach the regime where quark deconfinement occurs.

These findings are explained as follows. Before merging the densities in the progenitor stars are relatively low and no quark matter is present. Note that in this study we consider mostly binary systems where the mass of the initial stars is below the mass where quark matter appears~\footnote{The only exceptions are the binaries with 1.3--1.4~M$_\odot$ and two 1.4~M$_\odot$ NSs with the DD2F-SF-2 EoS, which contain tiny amounts of deconfined quark matter in the cores of the 1.4~M$_\odot$ stars. The combined tidal deformability of this configuration is hardly affected by the presence of small amounts of quark matter}. Hence, the inspiral GW signal is shaped by the purely hadronic regime of the EoS at lower densities and the measured tidal deformability does not contain information about a possible PT at higher densities. After merging the densities increase and some fraction of matter in the remnant possibly undergoes a PT to quark matter. The occurrence of quark matter effectively leads to a strong softening of the EoS beyond the transition density. This results in a more compact remnant, which oscillates at higher frequencies. Purely hadronic EoSs without PTs cannot produce such a strong and prompt softening to increase the postmerger frequencies in such a drastic manner. 

Postmerger frequencies extracted from purely hadronic models and hybrid models can in principle be comparable, and only the comparison between the tidal deformability and the postmerger frequency reveals a strong softening of the EoS and provides the unambiguous signature of a PT. Generally, this effect represents an instructive example of different information contained in the inspiral and the postmerger phase, which is a consequence of the different regimes of the EoS probed in the different phases of the merger.

We demonstrate that the value of $f_{\mathrm{peak}}$ also yields information on the density regime of the NS EoS probed in NS mergers. Specifically, we find that the maximum rest-mass density during the early evolution of the postmerger remnant scales tightly with $f_{\mathrm{peak}}$ for purely hadronic EoSs~\cite{Bauswein2019}. The dominant postmerger GW frequency $f_{\mathrm{peak}}$ can thus be employed to determine which density regime is probed by the merger remnant. For purely hadronic EoSs $f_{\mathrm{peak}}$ and the tidal deformability are strongly correlated and consequently the tidal deformability informs about the remnant's maximum density as well.

Using these relations we devise a ready-to-use procedure to place constraints on the onset density $\rho_\mathrm{onset}$ of a strong PT, which is generated by quark deconfinement. This method is immediately applicable after a GW detection with sufficiently accurate measurements of the tidal deformability $\Lambda$ and the dominant postmerger GW frequency $f_\mathrm{peak}$. If indications for a PT are found from a comparison between $f_{\mathrm{peak}}$ and $\Lambda$, the measured values of $f_{\mathrm{peak}}$ or $\Lambda$ place an upper limit on the onset density of deconfinement. If no signature of a PT is identified, the same relations result in a lower limit of the onset density and quantify in particular up to which density nuclear physics methods are applicable. The simultaneous measurement of inspiral and postmerger GWs is thus of utmost importance to understand the properties of high-density matter.

We address in detail the impact of two effects which potentially complicate the identification of a PT and the exact bound on the onset density. If the system only marginally exceeds the transition density, the quark matter core in the remnant is too small to significantly alter the postmerger frequency. In such systems, which exist only in a very narrow parameter space, the impact of the PT is not yet observable and could potentially lead to an overestimation of the lower limit of the onset density. A counteracting effect is caused by the temperature dependence of the phase boundary of the transition. At finite temperatures deconfinement can take place at lower densities resulting in an underestimation of the upper limit on the onset density. Both effects are relatively weak and are incorporated by an effective scheme such that the identification of the PT and the resulting constraints on the onset density are safe and conservative.

\subsection{Discussion}

We conclude with a couple of additional remarks.

(1) First we note as already discussed in \cite{Bauswein2019} that a significant postmerger frequency increase solely occurs if the PT is sufficiently strong. Only under this condition the transition can be identified and constraints on its onset density can be obtained. Although we only test hybrid models with a first-order PT, very likely the transition does not necessarily need to be first order. Based on our calculations it is conceivable that any transition strong enough to leave an impact on the stellar structure, does affect the postmerger frequency $f_{\mathrm{peak}}$ in the described manner because of the stronger compactification of the remnant.

This said it is clear that our method to detect the onset of quark deconfinement is insensitive and uninformative about the order and type of the transition. Moreover, the described signature does not reveal the underlying mechanism of the PT, e.g., whether the transition is in fact caused by quark deconfinement or by any other mechanism which can introduce a strong softening of the EoS. Arguably, only the hadron-quark PT can be sufficiently strong. We test three microphysical models with a PT to hyperonic matter and find that for these systems the softening of the EoS is not sufficient to change the postmerger frequency in the same way as hybrid models with a hadron-quark PT.

(2) If the transition to quark matter proceeds in a more continuous manner, e.g. through a cross-over without strongly softening the EoS, the PT may not be detectable by the features which we discussed here. Generally, this issue is known as the masquerade problem \cite{Alford2005} since the properties of quark matter may be such that they mimic the behavior of purely hadronic matter. A detailed investigation of the masquerade problem will be addressed in future work.

In this regard we also mention the finding in \cite{Bauswein2019} that the frequency shift is larger if the jump across the PT is larger (for roughly similar stiffness of the quark phase). This finding indicates that there may be a possibility to extract more detailed properties of the PT in the future.

(3) In the other extreme, the PT to quark matter may be too strong, i.e. the density jump may be too large or the stiffness of quark matter may be relatively low. In this case hybrid stars and merger remnants containing a quark core cannot be stabilized against the gravitational collapse. Then a postmerger frequency can only be observed if the density in the remnant is below the transition density and the system is purely hadronic (with at most a tiny admixture of quark matter as in \cite{Most2018a}, which does not alter the postmerger GW signal in a significant and characteristic way). If for more massive binary systems larger amounts of quark matter occur, the remnant collapses and does not emit postmerger GWs that could indicate the occurrence of quark matter. In this scenario it will only be possible to directly obtain lower limits on $\rho_\mathrm{onset}$.

(4) Recently, \cite{Weih2019} reported on a simulation using piecewise polytropes (i.e. a simplified model of a barotropic EoS) and treating temperature effects in a approximate way. This calculation revealed a transition phase from a hadronic postmerger remnant to a remnant with quark core and an associated transition of the postmerger frequency. However, it still has to be clarified whether this scenario does occur in a more microphysical setup including temperature effects consistently. In fact, comparing simulations with approximate temperature treatment and with a consistent thermal description, we find significant differences with regard to the transition to the quark phase and the resulting GW signal for the same underlying EoS model. In particular, the temperature dependence of the phase boundaries plays an important role and, in our microphysical models, triggers the direct formation of quark matter instead of a delayed transition in the case of the simplified thermal treatment.

Also, it remains to be seen if such a signature is easily detectable since any initial hadronic postmerger phase will diminish the power of the later GW emission which is indicative of the presence of quark matter. If either of the postmerger phases (initial hadronic or later quark phase) will be short or if the transition between both phases will be longer, the GW spectrum will not feature pronounced frequency peaks that can be associated with the different stages. Even in an optimal case it will be very challenging to identify and interpret different peaks since the GW spectra feature a lot of subdominant structures even for purely hadronic systems. While resolving frequency evolutions may generally provide further insights in merger dynamics and the EoS, their detection clearly poses additional challenges for GW data analysis, since it would require either an unmodeled search or the inclusion of additional model parameters. For both such search strategies even higher signal-to-noise ratios would be needed independent and in addition to the weaker signal in the case of a delayed occurrence of the PT.

Moreover, it is also unclear whether this scenario would occur in a considerable parameter range of total binary masses. It is likely that a significant initial hadronic transition phase before a quark matter core develops, occurs only for very fine-tuned setups in the binary mass configurations. Then, the scenario effectively resembles one of the above cases, i.e. prompt collapse due to the onset of the PT or a single pronounced shifted postmerger frequency as already discussed in \cite{Bauswein2019} and in this study. We thus do not agree with the claim in \cite{Weih2019} that a preceding hadronic transition period before a quark matter core forms, would lead to a cleaner and stronger signature. On the contrary, a potential preceding hadronic phase with a sudden transition, if it can be at all realized in a more realistic set up with temperature effects, will decrease the power of the characteristic GW peak which is the crucial indicator for the presence of quark matter. (The presence of a peak produced by the early hadronic postmerger phase does not add any information about the presence of quark matter.) It is thus natural to expect that a stronger and much cleaner signature of quark matter will arise in a scenario as brought forward in \cite{Bauswein2019} and further discussed here, where quark matter shapes the GW emission from the beginning of the postmerger evolution.

(5) We also remark that there is the possibility that a PT occurs in static NSs but not in temporarily stable merger remnants. The densities might not be high enough to trigger a transition in the remnant. In particular for stiff EoSs the remnant densities do not increase strongly. The maximum densities found in metastable merger remnants are typically smaller than the maximum central density of static stars (see Fig.~\ref{rhomaxfpeakuniversal} for the highest densities occurring in NS remnants). If the transition density is relatively high, the remnant would rather undergo a direct or quick collapse than reaching the PT regime. In that case one can probe the PT only in very massive NSs \cite{Chen2019,Chatziioannou2019,Pang2020}. As has been argued before it can be very challenging to measure finite size effects during the inspiral of very massive stars. Apart from the difficulties to measure the relatively weak finite-size effects such massive systems may not be very frequent. Hence, it will not be straightforward to detect a clear signature of a PT in this scenario. Note, however, that in this case our procedure to determine a lower limit on the onset density is fully applicable.

(6) Furthermore, we note that within this work we mostly consider systems in which the PT occurs after merging and not yet during the inspiral. If the transition density is relatively low, the progenitor stars would in fact be hybrid stars with quark cores. In this case the tidal deformability is affected by the presence of quark matter and it remains to be seen whether a comparison of a measurement with the $f^\mathrm{had}_\mathrm{peak}(\Lambda)$ relation reveals the presence of a PT. In any case, since quark matter appears only beyond some threshold density, for very low binary masses one would encounter the same scenario as described in this paper, i.e. an inspiral of purely hadronic stars and the appearance of quark matter during merging. This would thus lead to the same strong and unambiguous signature we discussed. Only for the extreme case that even the lightest possible NS already contains quark matter~\cite{Blaschke2020} a characteristic shift of the dominant postmerger GW peak might not occur. However, we note that at least for a few simulations with inspiraling hybrid stars conducted so far we find evidence for a characteristic frequency shift. Moreover, it is conceivable that an extreme model with a very early onset of deconfinement would lead to other very characteristic features in the GW spectrum indicating the presence of quark matter. We will further investigate this scenario in future work, see e.g. \cite{Bauswein2020a}. An early deconfinement transition may also lead to other very prominent signatures, e.g. in heavy-ion collisions or core-collapse supernovae.

Similarly, for asymmetric mergers one can encounter the situation that the more massive binary component is a hybrid star, whereas the lighter star is purely hadronic. In this case the measured combined tidal deformability carries information about the PT and the $\Lambda-f_\mathrm{peak}$ comparison might not easily reveal the presence of a PT. This issue should be addressed in future work focusing on very asymmetric binaries where such mixed configurations occur already for relatively low total binary masses. Stronger mass asymmetries may lead to a small quantitative shift of the empirical $f^\mathrm{had}_\mathrm{peak}(\Lambda)$ relation for hadronic stars. This is why we do not discuss this scenario in greater detail here and restrict the applicability of our method to symmetric and moderately asymmetric systems. We remark that mixed systems, composed of a hybrid star and a purely hadronic star, with a small mass asymmetry generally have to have a relatively high total mass, i.e. $M_\mathrm{tot}/2$ close to the mass where quarks start to appear in static stars. This implies that these mixed binaries undergo a prompt collapse instead of forming a NS postmerger remnant, which emits GWs with a characteristic frequency $f_\mathrm{peak}$. Therefore, mixed systems which have a binary mass ratio close to unity, do not occur in our $\Lambda-f_\mathrm{peak}$ diagrams and thus do not spoil our method of identifying a PT~\footnote{An exception are EoS models like DD2F-SF-2 with a low onset density and a high maximum mass $M_\mathrm{max}$, i.e. EoSs with an extended hybrid branch in the mass-radius relation. For these models mixed systems with small mass asymmetry do not directly form a black hole. For the specific case of DD2F-SF-2 we confirm that the shift towards smaller combined tidal deformabilities because of the presence of a hybrid star is very small and is largely exceeded by the frequency increase of the postmerger GW emission. Hence, these systems do lead to the same signature as purely hadronic progenitors.}.

(7) On the more technical aspects we remark that our empirical relations for fixed binary masses generally show smaller scattering than our mass-independent relation. Once pre- and postmerger GW signals of a binary NS merger have been detected and the values of $\Lambda$, $f_{\mathrm{peak}}$ and $M_{\mathrm{tot}}$ have been measured, it will therefore be desirable to obtain the empirical $f^\mathrm{had}_{\mathrm{peak}}(\Lambda)$ and $\rho_{\mathrm{max}}^{\mathrm{max}}(\Lambda)$ relations for this specific binary mass setup to get a tighter constraint on $\rho_\mathrm{onset}$. This would be readily achieved within a relatively short time by simulating the merger with the observed mass configuration for a large sample of EoSs. It will also be advantageous to consider more hybrid EoSs with different models for both the hadronic and the deconfined quark phase. This will help solidifying the findings of this work and perhaps lead to a more advanced description of temperature effects on the phase boundaries. Currently these effects are captured by an effective parameter $\Delta M=0.2~$M$_\odot$. 

(8) Although this work focuses on the impact of PTs on NS mergers, we emphasize that our finding of empirical relations between GW observables and the maximum density encountered in the postmerger phase is in a more general sense useful. The relations $\rho_{\mathrm{max}}^{\mathrm{max}}(\Lambda)$ and $\rho_{\mathrm{max}}^{\mathrm{max}}(f_\mathrm{peak})$ determine which density regime is actually realized in the postmerger remnant and thus which part of the EoS is shaping the characteristic features of GW signal. Notably, these relationships are relatively tight and nearly binary mass-independent (universal). This finding is complementary to our discussion of \cite{Bauswein2014a} where we showed that the GW signal provides information on the highest densities that can be reached in isolated, static NSs. With regard to the technical challenges we remark the following. While GW frequencies are relatively insensitive to the numerical treatment and the inclusion of different physical effects, future work should solidify the precise values of $\rho_{\mathrm{max}}^{\mathrm{max}}$, which may be more affected by numerical details of the simulations and the exact physical model.

As already indicated under (5), we remark that from our set of calculations we realize that the highest densities which are reached during the early evolution of NS merger remnants are significantly below the maximum density of the most massive non-rotating NSs. The maximum densities are about six times nuclear saturation density for some specific binary setups (possibly one could reach somewhat higher densities for some configurations which we did not simulate). Reaching even higher densities in an at least temporarily stable system is prevented by the prompt gravitational collapse of the remnant. 

(9) We would further like to highlight another important result of our work. We obtain a number of relations between the tidal deformability and the dominant postmerger frequency for fixed binary mass configurations for a large set of hadronic EoSs. We found these relations to be very tight with maximum residuals of only about 100~Hz. To our knowledge it is the first time that such fits are provided in the literature for fixed masses. Such relations are important because binary masses (including the mass ratio) will be measurable with even higher precision in the future when postmerger GW emission becomes detectable. These fits and the corresponding residuals may prove useful in a re-analysis of future signals after an initial mass determination. 

Moreover, if a PT during merging can be excluded, the EoS information extracted from the inspiral phase and from the postmerger phase should agree to within the quoted residuals of roughly 100~Hz. This is an important comparison considering that the extraction of EoS effects from both phases relies one different procedures, which potentially suffer from different systematic uncertainties.

We would also like to stress that our work highlights the importance of dedicated GW instruments with good sensitivity in the frequency range of a few kHz. Our findings demonstrate that GWs in this frequency range carry important information, which is complementary to the data from the inspiral. The basic reason is that compared to the premerger- the postmerger stage probes the EoS regime at higher densities. Apart from the experimental efforts \cite{LIGOScientificCollaboration2015,Acernese2015,Collaboration2017b,Punturo2010,Hild2011,Miller2015} to develop such detectors, it will be also crucial to further develop GW data analysis methods which are designed to extract most information from the postmerger GW signal.

Moreover, existing (HADES \cite{Hades2019}) and future (NICA \cite{Blaschke2016}, FAIR \cite{Friman2011}) heavy-ion experiments can provide useful insights to understand the onset of quark deconfinement and complement the interpretation of data from NS mergers recalling that our signature is only sensitive to the bulk thermodynamical features of the transition but does not reveal the underlying microphysical mechanisms of a PT. Also, one cannot exclude that modifications of General Relativity could in principle mimic the occurrence of a strong PT even if matter in a merger remnant is purely hadronic. These issues can be addressed by future theoretical  work, but in any case it highlights the importance of complementary information from heavy-ion collisions.

\begin{appendix}

\section{Matter distribution in the $\rho$--T plane} \label{apprhot}
In this section we provide plots of the mass distribution in the $\rho$--T plane for a merger simulation of two 1.35 NSs described by the DD2F-SF-6 EoS at the same time steps as the snapshots in Fig.~\ref{densityevolution}. The dashed, black lines mark the onset of the hadron-quark phase transition while the solid black lines are the boundaries of the regions containing pure quark matter.
\begin{figure*}[ht]  
\centering
\subfigure[]{\includegraphics[width=0.48\linewidth]{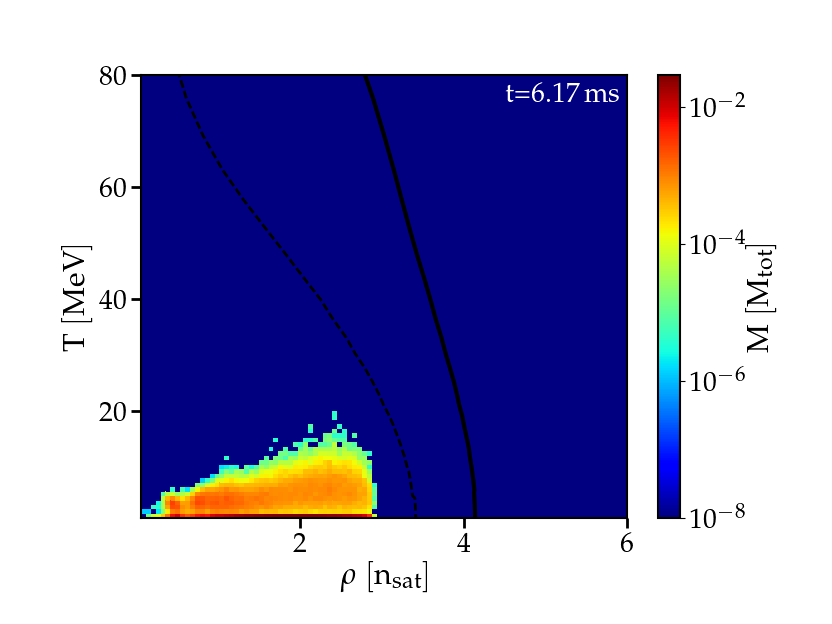}\label{trhoevolution_a}}
\hfill
\subfigure[]{\includegraphics[width=0.48\linewidth]{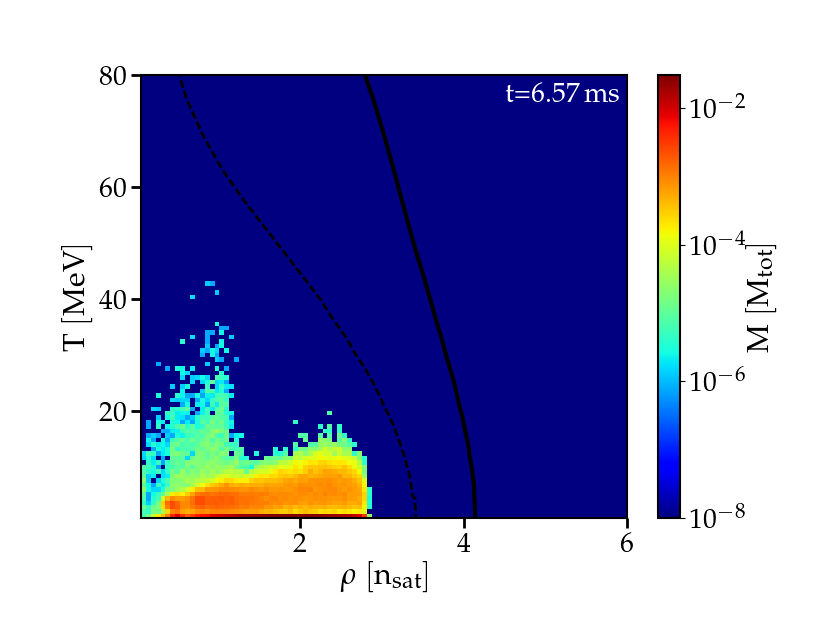}\label{trhoevolution_b}}
\\
\subfigure[]{\includegraphics[width=0.48\linewidth]{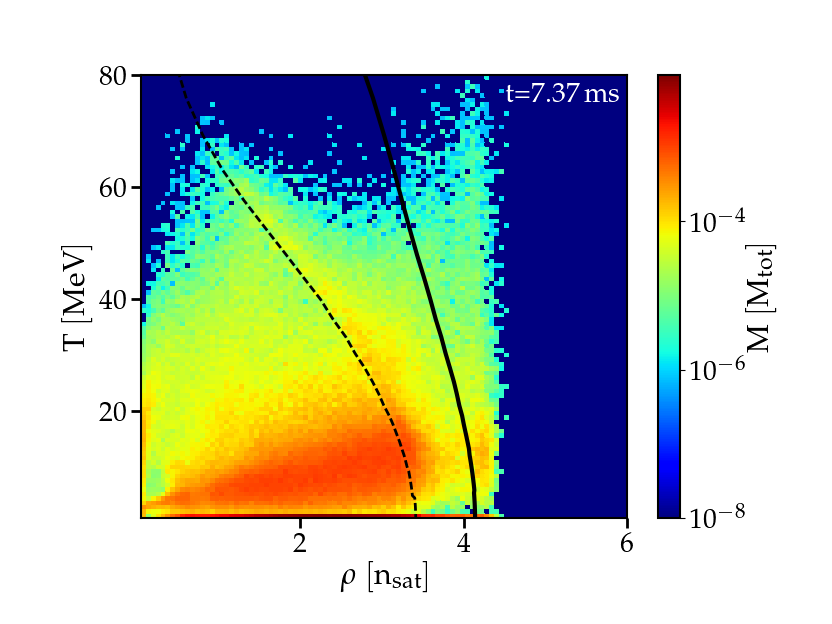}\label{trhoevolution_c}}
\hfill
\subfigure[]{\includegraphics[width=0.48\linewidth]{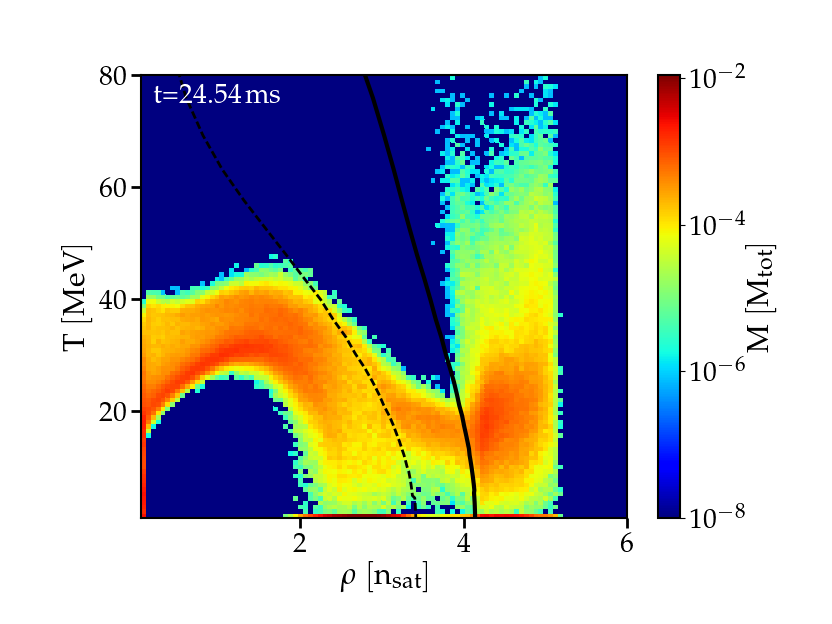}\label{trhoevolution_d}}
\caption{Rest mass distribution (color-coded) of matter in the density-temperature plane for the merger simulation from Fig.~\ref{densityevolution} normalized to the total mass of the system. Note the logarithmic mass scale. The dashed, black line displays the temperature dependent onset density of the hadron-quark phase transition. The solid black line is the boundary of the pure quark matter phase. The total amounts of matter at temperatures lower than 5\,MeV in the graphs (a)-(d) are about 92\%, 92\%, 42\% and 19\%, respectively.}
\label{trhoevolution}
\end{figure*}

Figure \ref{trhoevolution_a} shows the distribution of matter shortly before the merger. One can see that all the material is still in the hadronic phase and that the stellar material is mostly cold. The small amount of matter at somewhat increased temperatures is caused by numerical heating.\\
The matter distribution at the merger is depicted in Fig.~\ref{trhoevolution_b}. Still, there are no deconfined quarks present in the system but during the merging the areas where the stars first come in contact get heated up. This material has rather low densities, as can be seen in Fig.~\ref{trhoevolution_b}.\\
Shortly after the merger a hot, rapidly rotating remnant forms. The matter distribution at this stage is shown in the lower left panel. As already displayed in Fig.~\ref{denistyevolution_c} the remnant now also contains deconfined quark matter in the mixed as well as in the pure quark phase.\\
Figure \ref{trhoevolution_d} shows the matter distribution at a later time after the merger when the remnant has settled down into a more axial-symmetric state. In this phase the system contains a hot core of pure quark matter surrounded by a thin shell of hot matter in the mixed phase. The outer hadronic parts of the merger remnant reach even higher temperatures on average and do not contain any cold matter.

\section{Effects of slightly asymmetric binaries}\label{asymbinar}
In this section we further discuss the impact of asymmetric mergers on our empirical relations. We also include results from another hydrodynamical code to assess the impact of different simulation tools.

\subsection{$f_\mathrm{peak}-\Lambda$ relations}\label{asymfpeakoflambda}
In section \ref{fpeaklambdaasym} we already showed that the effects of slightly asymmetric binaries on the $f^\mathrm{had}_\mathrm{peak}-\tilde{\Lambda}$ relation are small at a fixed chirp mass. Here we additionally illustrate how the mass-independent relation between $f_\mathrm{peak}\times M_{\mathrm{tot}}$ and $\Lambda$ (see section~\ref{fpeaklambdamassindep}) is affected if we also consider binaries with $q<1$. For this we include data from 1.3--1.4~M$_{\odot}$ binaries into the sample shown in Fig.~\ref{fpeaklampdauniversal} and derive a fit using Eq.~\eqref{QuadraticLambdaFitUni}. We use the combined tidal deformability $\tilde{\Lambda}$ here, which is measured in a GW detection. The distinction between $\Lambda$ and $\tilde{\Lambda}$ was not relevant for equal-mass binaries discussed in the main text because both quantities are identical for symmetric systems.
\begin{figure}
\centering
\includegraphics[width=1.0\linewidth]{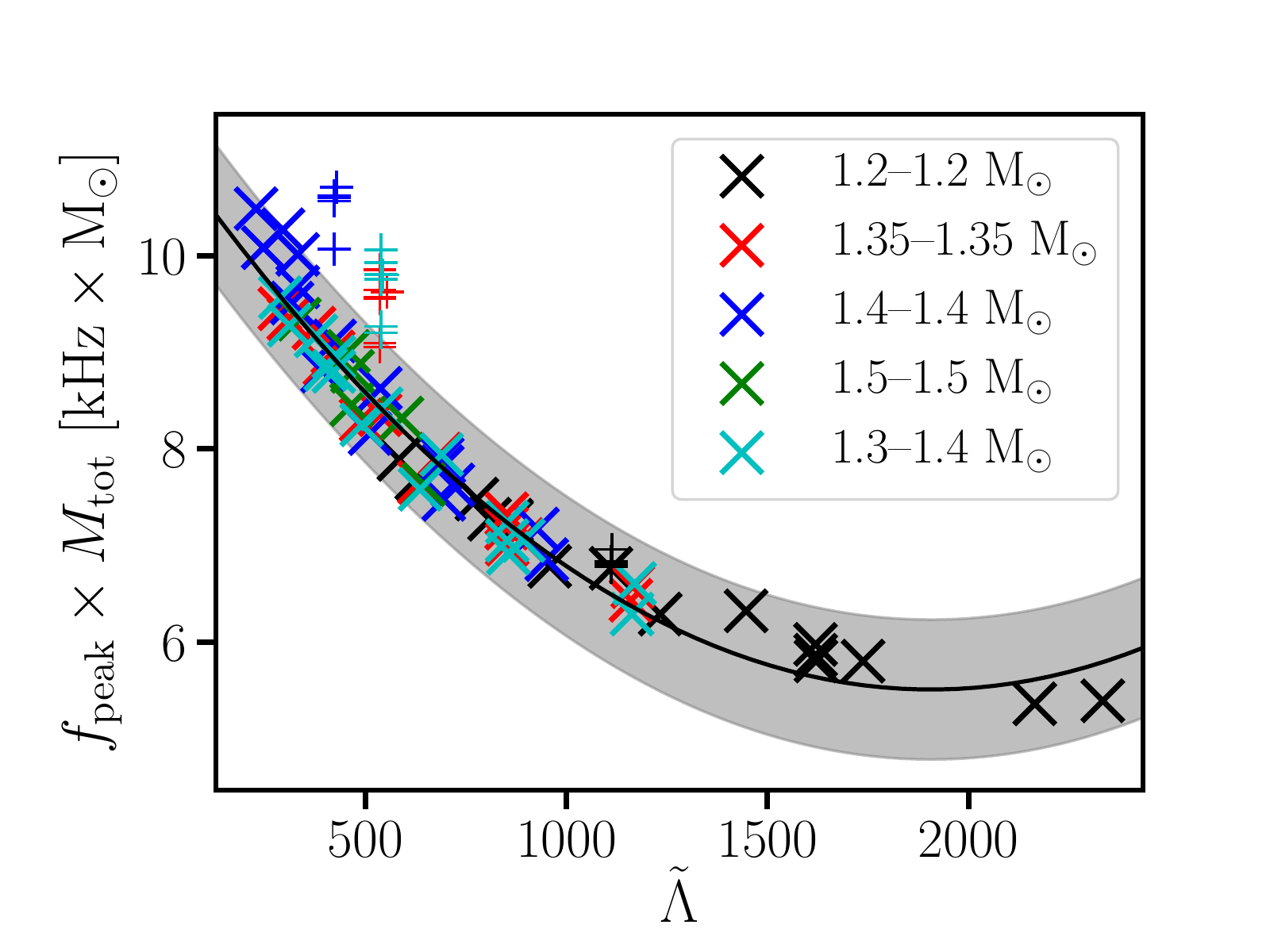}
\caption{Dominant postmerger GW frequency $f_{\mathrm{peak}}$ scaled by the total binary mass $M_{\mathrm{tot}}$ as a function of the combined tidal deformability $\tilde{\Lambda}$. Different colors refer to data from different total binary masses. Crosses refer to data from purely hadronic models while plus signs represent data with hybrid DD2F-SF models. The solid black line is a least squares fit with a second order polynomial to the data (excluding the DD2F-SF models). The gray shaded area illustrates the maximum deviation of the data from hadronic models from the fit. Compared to Fig.~\ref{fpeaklampdauniversal} this plot also contains data from 1.3--1.4~M$_{\odot}$ binaries.}
\label{fpeaklampdauniversalwithasym}
\end{figure}
The result is shown in Fig.~\ref{fpeaklampdauniversalwithasym}. As before, different colored crosses refer to data from hadronic EoSs from different binary mass configurations. Colored plus signs represent results from the hybrid DD2F-SF models. The solid black line shows a least squares fit to the data with a second order polynomial
excluding the data from the hybrid DD2F-SF models. The gray shaded area illustrates the maximum deviation of data from hadronic models from the fit. The fit parameters in Eq.~\eqref{QuadraticLambdaFitUni} are given by $a=1.552\times 10^{-6}$, $b=\mathrm{-}5.920\times 10^{-3} $ and $c=11.15$. The mean deviation of the data from the fit is 194~Hz$~ \mathrm{M}_\odot$ and the maximum residual is 724~Hz$~ \mathrm{M}_\odot$.

As apparent from the figure and the residuals the data including asymmetric mergers also follows the $f^\mathrm{had}_\mathrm{peak}\times M_{\mathrm{tot}}-\tilde{\Lambda}$ relation and the precision is hardly affected. We therefore conclude that this relation also holds for slightly asymmetric binaries.

\subsection{$\rho_{\mathrm{max}}^{\mathrm{max}}-f_\mathrm{peak}$ relations}\label{asymrhofpeak}

\begin{figure}
\centering
\includegraphics[width=1.0\linewidth]{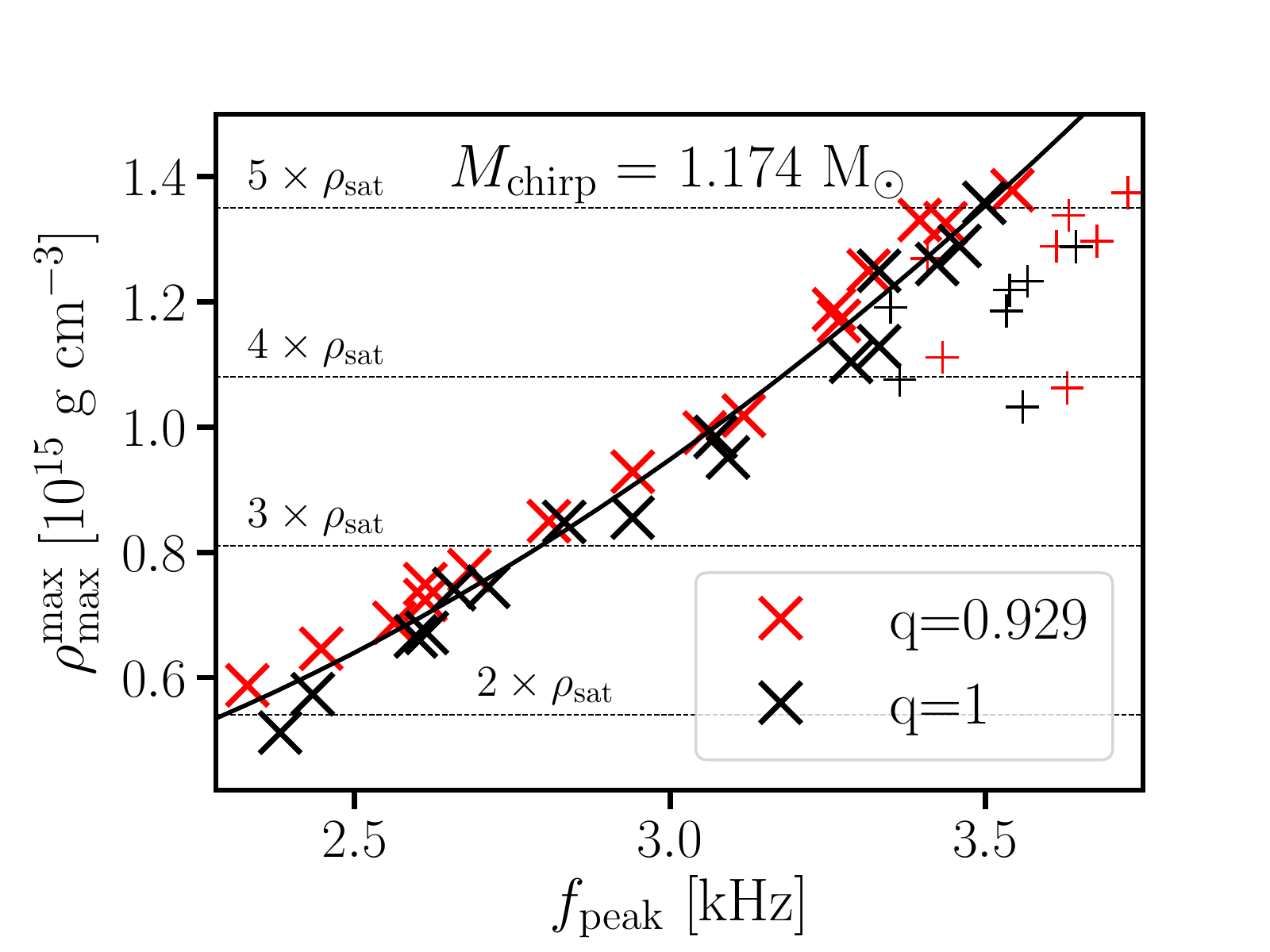}
\caption{Maximum rest-mass density $\rho_{\mathrm{max}}^{\mathrm{max}}$ in the remnant during the first 5 milliseconds after the merger as a function of the dominant postmerger GW frequency $f_{\mathrm{peak}}$. Red symbols refer to data from 1.3--1.4~M$_{\odot}$ binaries, while black symbols refer to data from equal-mass binaries with the chirp mass of a 1.3--1.4~M$_{\odot}$ binary. Crosses represent data from purely hadronic models while plus signs illustrate data with hybrid DD2F-SF EoSs. The solid black line is a least squares fit with a second order polynomial to the data (excluding the hybrid models).}
\label{rhomaxfpeakchirp}
\end{figure}

In Fig.~\ref{rhomaxfpeakchirp} we address the impact of asymmetric mergers on the relation between $\rho_{\mathrm{max}}^{\mathrm{max}}$ and $f_\mathrm{peak}$.
The red symbols display data from 1.3--1.4~M$_{\odot}$ binary simulations, while black symbols refer to interpolated data from symmetric binaries. Crosses depict data obtained with purely hadronic EoSs while plus signs refer to data obtained with hybrid DD2F-SF EoSs. The black solid line shows a least squares fit to all hadronic data with a second order polynomial (see Eq.~\eqref{QuadraticrhoFit}). The fit parameters are given by 
$a_{M}=1.936\times 10^{14},~b_{M}=\mathrm{-}4.477\times 10^{14}$ and $c_{M}=5.490\times 10^{14}$, while the mean and the maximum deviation of hadronic data from the fit are given by $0.033\times 10^{15}~\mathrm{g}~\mathrm{cm}^{-3}$ and $0.077\times 10^{15}~\mathrm{g}~\mathrm{cm}^{-3}$, respectively.

We thus conclude the relation between $f^\mathrm{had}_\mathrm{peak}$ and $\rho_{\mathrm{max}}^{\mathrm{max}}$ still holds for a fixed chirp mass and varying mass ratio. However, the results from asymmetric binaries are generally shifted towards slightly higher densities. Including very asymmetric binaries might therefore result in a less tight relation.

We also investigate the effect of asymmetric binaries on the \textit{mass-independent} $\rho_{\mathrm{max}}^{\mathrm{max}}-f^\mathrm{had}_\mathrm{peak}$ relation. We include data from 1.3--1.4~M$_{\odot}$ binaries and
data from all equal-mass simulations for purely hadronic EoSs.

\begin{figure}
\centering
\includegraphics[width=1.0\linewidth]{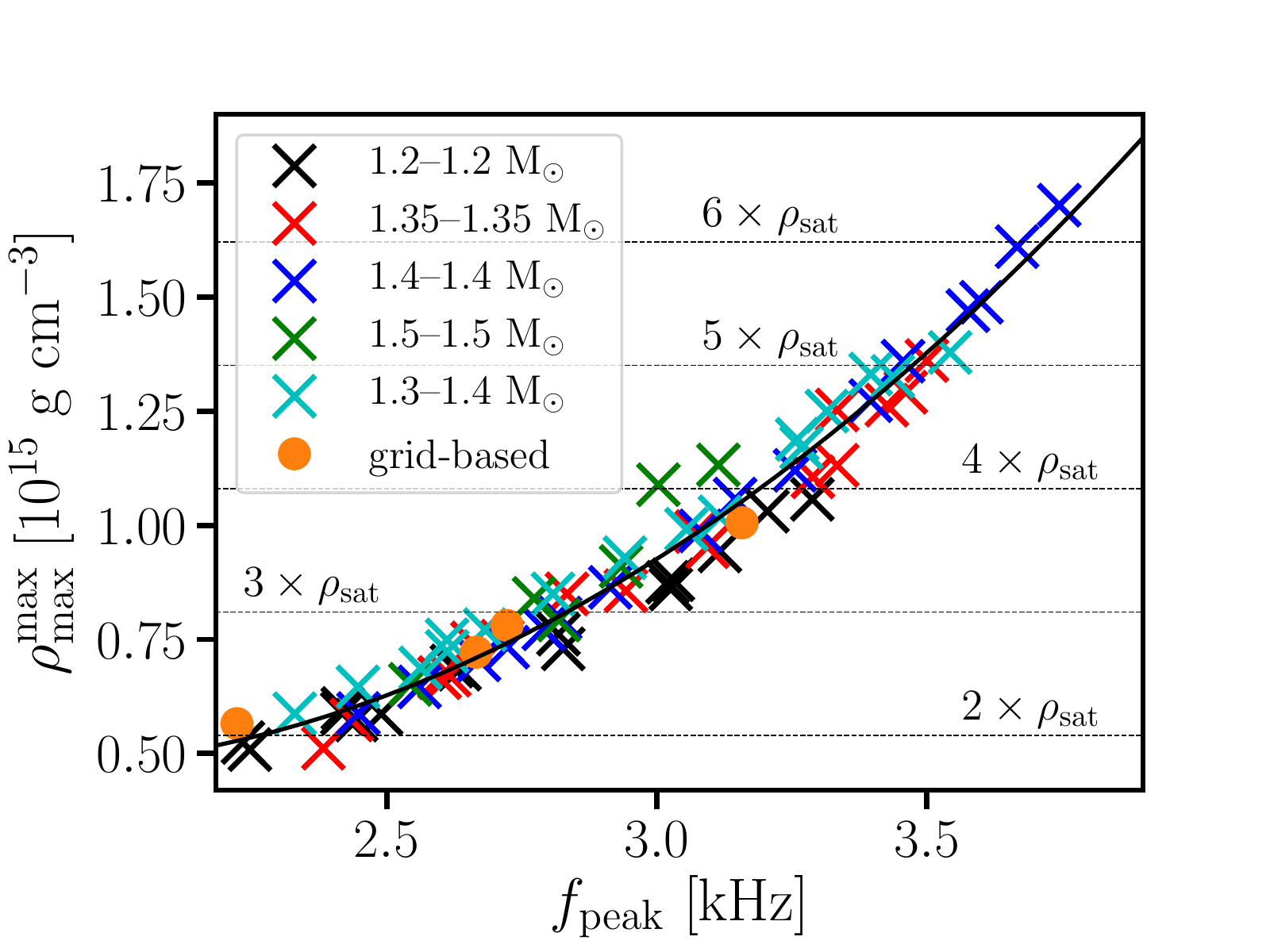}
\caption{Maximum rest-mass density $\rho_{\mathrm{max}}^{\mathrm{max}}$ in the remnant during the first 5 milliseconds after the merger as a function of the dominant postmerger GW frequency $f_{\mathrm{peak}}$ for 1.2--1.2~M$_{\odot}$ (black crosses), 1.35--1.35~M$_{\odot}$ (red crosses), 1.4--1.4~M$_{\odot}$ (blue crosses) 1.5--1.5~M$_{\odot}$ (green crosses) and 1.3--1.4~M$_{\odot}$ (cyan crosses) mergers with different microphysical EoSs. This plot contains the entire data of Fig.~\ref{rhomaxfpeakuniversal} together with data from asymmetric mergers and additional results from grid-based calculations (orange circles). The solid black line is a least squares fit to all shown datapoints (Eq.~\eqref{QuadraticrhoFit}) excluding results from grid-based calculations.}
\label{rhomaxfpeakuniversalwithasym}
\end{figure}

The relation is shown in Fig.~\ref{rhomaxfpeakuniversalwithasym}.
Different colors mark data from different binary mass configurations. Data from 1.2--1.2~M$_{\odot}$, 1.35--1.35~M$_{\odot}$, 1.4--1.4~M$_{\odot}$, 1.5--1.5~M$_{\odot}$ and 1.3--1.4~M$_{\odot}$ binaries are displayed by black, red, blue, green and cyan crosses, respectively. The solid black line shows a least squares fit to the data using Eq.~\eqref{QuadraticrhoFit}. The obtained fit parameters are $a_{M}=3.055\times 10^{14}~\mathrm{kHz}^{-2}$, $b_{M}=\mathrm{-}1.083\times 10^{15}~\mathrm{kHz}^{-1}$ and $c_{M}=1.425\times 10^{15}$. The mean and the maximum deviation of the underlying data from this fit is given by $0.036 \times 10^{15}~\mathrm{g}~\mathrm{cm}^{-3}$ and $0.160 \times 10^{15}~\mathrm{g}~\mathrm{cm}^{-3}$, respectively.

The additional data in Fig.~\ref{rhomaxfpeakuniversalwithasym} also follows the universal $\rho_{\mathrm{max}}^{\mathrm{max}}-f^\mathrm{had}_\mathrm{peak}$ relation. We thus conclude that this correlation also holds for slightly asymmetric binaries.

Furthermore, we validate our findings by including results from grid-based calculations with the Einstein Toolkit~\cite{EinsteinToolkit:2019_10,Cactuscode:web,Goodale:2002a,Baiotti:2004wn}. The orange circles in Fig.~\ref{rhomaxfpeakuniversalwithasym} show the postmerger frequency and the maximum densities from simulations of 1.2--1.2~M$_\odot$ mergers with the APR4, MPA1 and H4 EoSs and a simulation of a 1.3--1.3~M$_\odot$ merger with the MPA1 EoS~\cite{Akmal1998,Muther1987,Lackey2006}. All EoSs are implemented as piecewise polytropes~\cite{Read2009a}. Providing only barotropic relations between density and pressure the EoSs are supplemented with an approximate treatment of thermal effects with $\Gamma_\mathrm{th}=1.75$ (see \cite{Bauswein2010}). In the calculations we use the HLLE Riemann solver~\cite{Harten:1983on,Einfeldt:1988og} with a WENO reconstruction~\cite{WENO:1994,WENO:1996}. The run employs Z4c formulation~\cite{PhysRevD.83.044045,PhysRevD.88.084057} of the Einstein equations. Initial data are generated with the LORENE code \cite{Lorene:web,Gourgoulhon2001}.
 
We run these setups at two different resolutions (277~m and 369~m on the finest refinement level) and observe some dependence on the grid size at this relatively coarse resolution (as good as 0.1\% in $f_\mathrm{peak}$ and 1\% in $\rho_{\mathrm{max}}^{\mathrm{max}}$ for MPA1 but 1\% and 5\%, respectively, for H4). In Fig.~\ref{rhomaxfpeakuniversalwithasym} we include only the data from the high-resolution calculations. Figure~\ref{rhomaxfpeakuniversalwithasym} demonstrates that the relations presented and employed in this paper are not strongly affected by the simulation tool remarking that the two codes used here differ in various aspects like the hydrodynamics and the gravity solver.

\subsection{$\rho_{\mathrm{max}}^{\mathrm{max}}-\Lambda$ relations}\label{asymrholambda}
Now we consider the mass-independent, empirical relation between $\rho_{\mathrm{max}}^{\mathrm{max}}\times M_{\mathrm{tot}}$ and $\Lambda$~(see Eq.~\eqref{rhoOfLambda}). We verify that this relation does not strongly change by adding asymmetric binaries.

\begin{figure}
\centering
\includegraphics[width=1.0\linewidth]{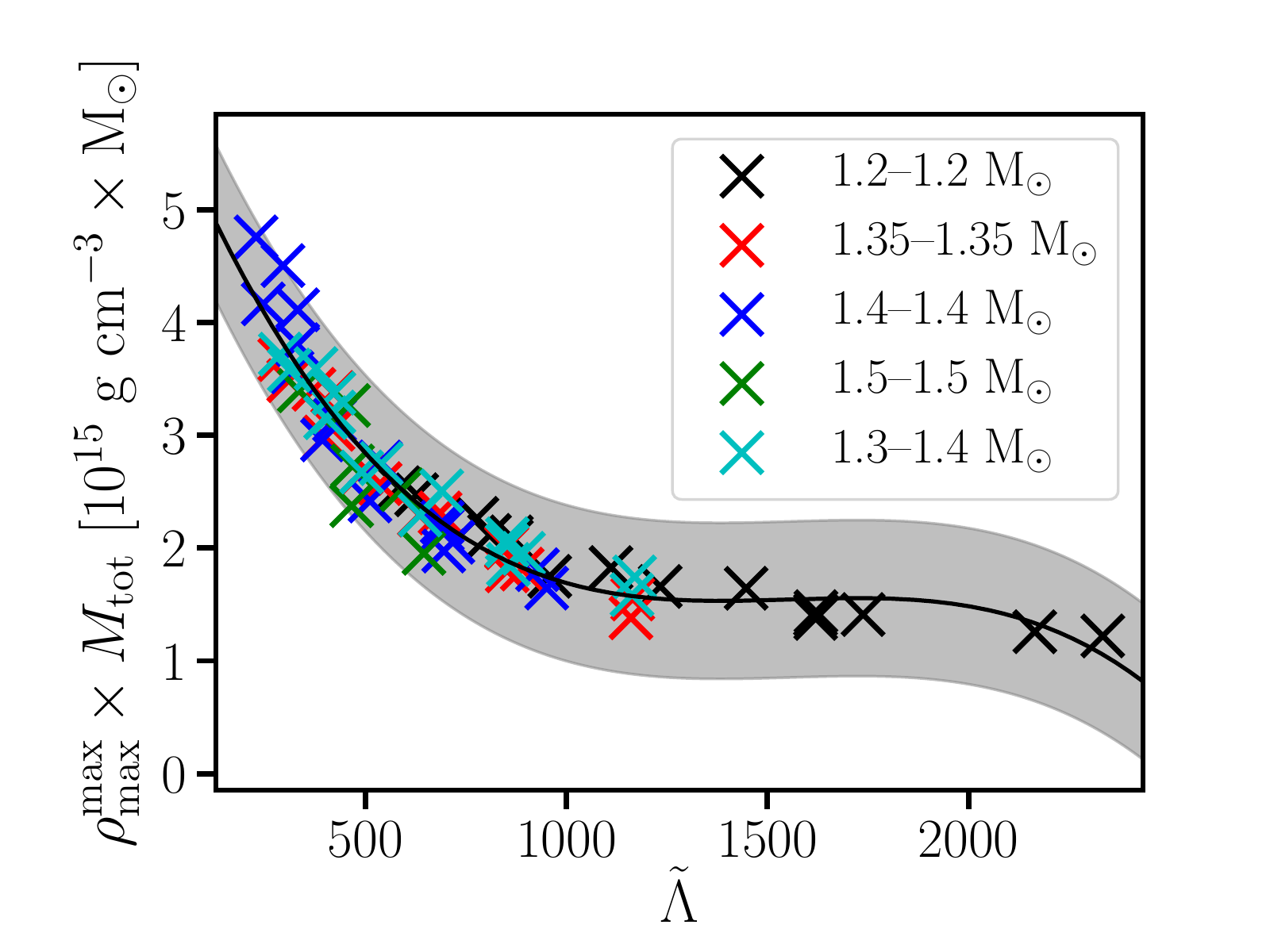}
\caption{Maximum rest-mass density $\rho_{\mathrm{max}}^{\mathrm{max}}$ in the remnant during the first 5 milliseconds after the merger scaled by the total binary mass $M_{\mathrm{tot}}$ as a function of the combined tidal deformability $\tilde{\Lambda}$ for 1.2--1.2~M$_{\odot}$ (black), 1.35--1.35~M$_{\odot}$ (red), 1.4--1.4~M$_{\odot}$ (blue) 1.5--1.5~M$_{\odot}$ (green) and 1.3--1.4~M$_{\odot}$ (cyan) mergers with different microphysical EoSs. This plot contains the entire data of Fig.~\ref{rhomaxlamuniversal} together with results from asymmetric binaries. The solid black line is a least squares fit to all shown data points (Eq.~\eqref{rhoOfLambda}). The gray shaded area illustrates the maximum deviation of the data from the fit.}
\label{rhomaxlamuniversalwithasym}
\end{figure}

We include results from 1.3--1.4~M$_{\odot}$ binaries into the data sample shown in Fig.~\ref{rhomaxlamuniversal} and fit the data using Eq.~\eqref{rhoOfLambda}. The results are summarized in Fig.~\ref{rhomaxlamuniversalwithasym}.
As before, different colored crosses refer to data from hadronic EoSs at different binary masses. Data from 1.2--1.2~M$_{\odot}$,~1.35--1.35~M$_{\odot}$,~1.4--1.4~M$_{\odot}$,~1.5--1.5~M$_{\odot}$ and 1.3--1.4~M$_{\odot}$ binaries are displayed by black, red, blue, green and cyan signs, respectively. The solid black line shows a least squares fit to the data with Eq.~\eqref{rhoOfLambda}. The gray shaded area depicts the maximum deviation of data from the fit. The fit parameters are given by $a=\mathrm{-}1.204\times 10^{6}$, $b=5.614\times 10^{9}$, $c=\mathrm{-}8.618\times 10^{12}$ and $d=5.899\times 10^{15}$. The mean and the maximum deviation of our data from the fit is given by $0.166\times 10^{15}~\mathrm{g}~\mathrm{cm}^{-3}~ \mathrm{M}_\odot$ and $0.693\times 10^{15}~\mathrm{g}~\mathrm{cm}^{-3}~ \mathrm{M}_\odot$, respectively.

Again, we find that the accuracy of this relation is not greatly affected by including slightly asymmetric binaries. Therefore, we conclude that our procedure to constrain the onset density of a strong PT is also applicable to not too asymmetric binary mergers if the mass ratio is not measured precisely.

\section{Mass-independent empirical relations}\label{moreplots}
In this section we provide additional plots of mass-independent relations and our developed procedure to constrain the onset density that were discussed but not employed in the main part of this work. 
\subsection{$\Lambda-f_\mathrm{peak}$ relations}\label{moreplotsfpeaklambda}
\begin{figure*}
\centering
\subfigure[]{\includegraphics[width=0.32\linewidth]{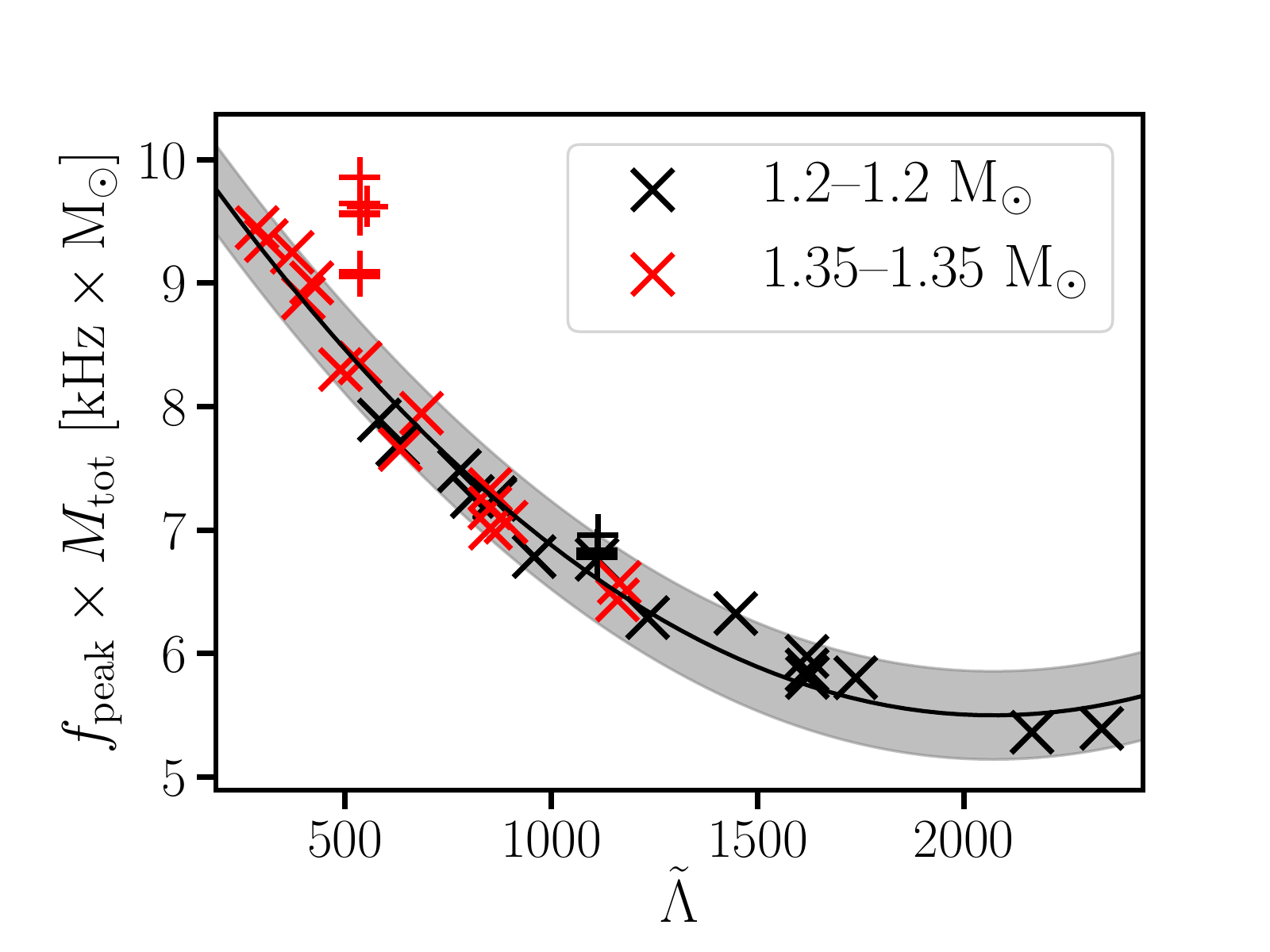}\label{fpeaklampdacombirelations_a}}
\subfigure[]{\includegraphics[width=0.32\linewidth]{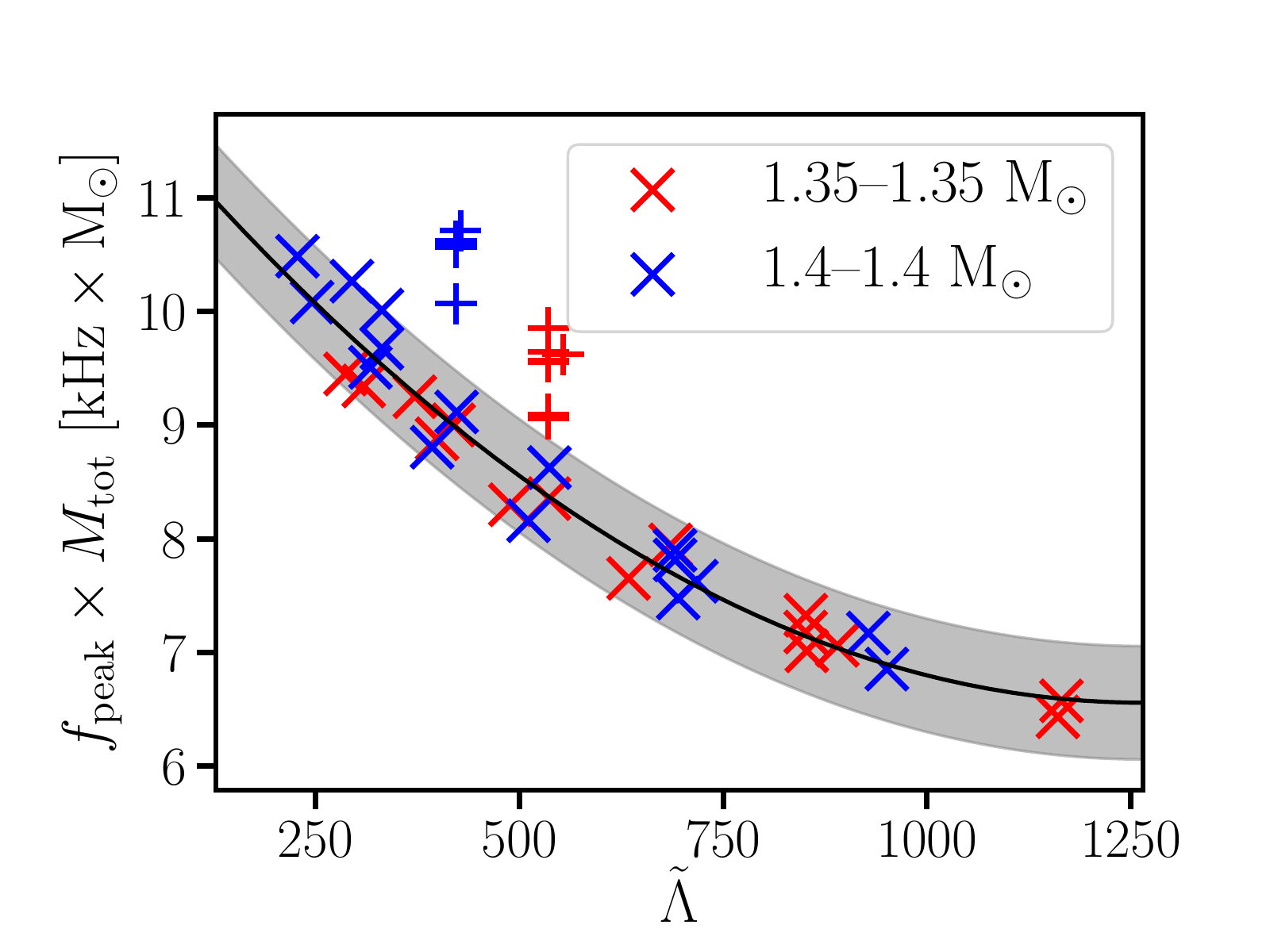}\label{fpeaklampdacombirelations_b}}
\subfigure[]{\includegraphics[width=0.32\linewidth]{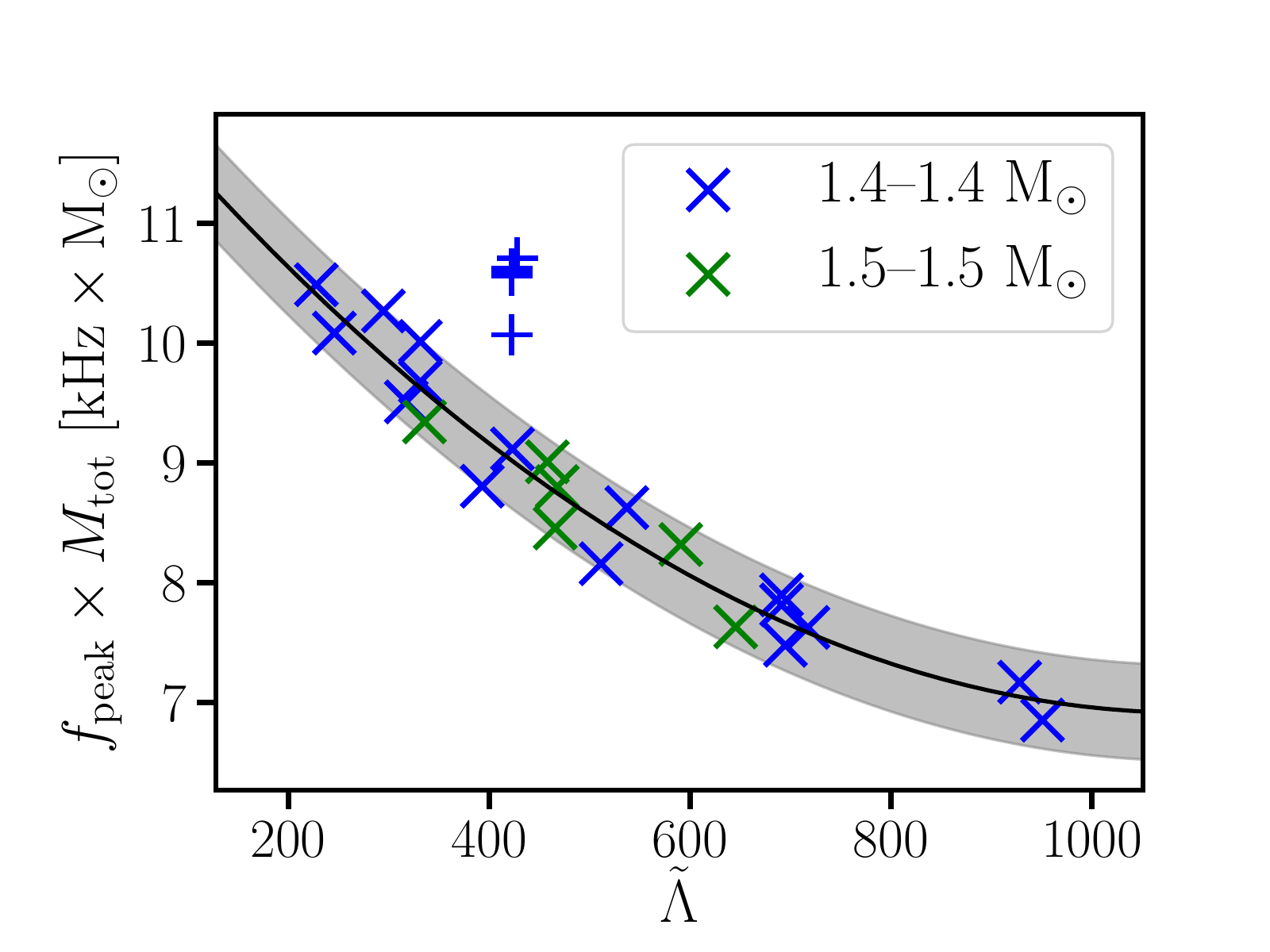}\label{fpeaklampdacombirelations_c}}
\caption{Dominant postmerger GW frequency $f_{\mathrm{peak}}$ scaled by the total binary mass $M_{\mathrm{tot}}$ as a function of the combined tidal deformability $\tilde{\Lambda}$. Plotted are results from 1.2--1.2~M$_{\odot}$ and 1.35--1.35~M$_{\odot}$ binaries (graph~(a)), from  1.35--1.35~M$_{\odot}$ and 1.4--1.4~M$_{\odot}$ binaries (graph~(b)) and from 1.4--1.4~M$_{\odot}$ and 1.5--1.5~M$_{\odot}$ binaries (graph~(c)). Different colors refer to data from different total binary masses. Crosses refer to data from purely hadronic models, while plus signs represent data with hybrid DD2F-SF models. The solid black lines are least squares fits with a second order polynomial to the data (excluding the DD2F-SF models) in the respective plot. The gray shaded areas illustrate the maximum deviation of the data of hadronic models from each fit. At binary masses of 1.35--1.35~M$_{\odot}$ and 1.4--1.4~M$_{\odot}$ DD2F-SF models appear as outliers.}
\label{fpeaklampdacombirelations}
\end{figure*}
Figures \ref{fpeaklampdacombirelations_a}--\ref{fpeaklampdacombirelations_c} display mass-independent relations between the dominant postmerger GW frequency rescaled by $M_\mathrm{tot}$ an the tidal deformability for the individual binaries. As explained in the main text (see Sect.~\ref{fpeaklambdamassindep}), it is advantageous to produce several of these relation each restricted to a smaller range in $M_\mathrm{tot}$. Solid curves are least squares fits listed in Tab.~\ref{fpeaklambdafits}. (excluding the data from the hybrid DD2F-SF models). The gray shaded areas illustrate the maximum deviation of data from hadronic models from the fits.

Comparing Fig.~\ref{fpeaklampdacombirelations} to Fig.~\ref{fpeaklampdauniversal}  one can see that this procedure of defining different relations for different mass ranges reduces the scatter of the data from the fits. In Fig.~\ref{fpeaklampdacombirelations} all data from hybrid DD2F-SF models for binary masses of 1.35--1.35~M$_{\odot}$ and 1.4--1.4~M$_{\odot}$ appear as outliers in all three panels. This was not the case for all hybrid models in the relation shown in Fig.~\ref{fpeaklampdauniversal} in the main part.

\subsection{$\rho_{\mathrm{max}}^{\mathrm{max}}-f_\mathrm{peak}$ relations}\label{moreplotsrhofpeak}
Similarly as for the $f^\mathrm{had}_{\mathrm{peak}}(\Lambda)$ relation we find that the scatter of the data in a mass-independent relation between $\rho_{\mathrm{max}}^{\mathrm{max}}$ and $f_{\mathrm{peak}}$ can be reduced. We introduce different relations for data from 1.2--1.2~M$_{\odot}$ and 1.35--1.35~M$_{\odot}$, 1.35--1.35~M$_{\odot}$ and 1.4--1.4~M$_{\odot}$ as well as from 1.4--1.4~M$_{\odot}$ and 1.5--1.5~M$_{\odot}$ binaries, respectively (see Sect.~\ref{fpeakrhomassindep}). The fit parameters as well as the mean and the maximum deviation from the fit for each relation are shown in Tab.~\ref{fpeakrhoCombifits}.

\begin{figure*}
\centering
\subfigure[]{\includegraphics[width=0.32\linewidth]{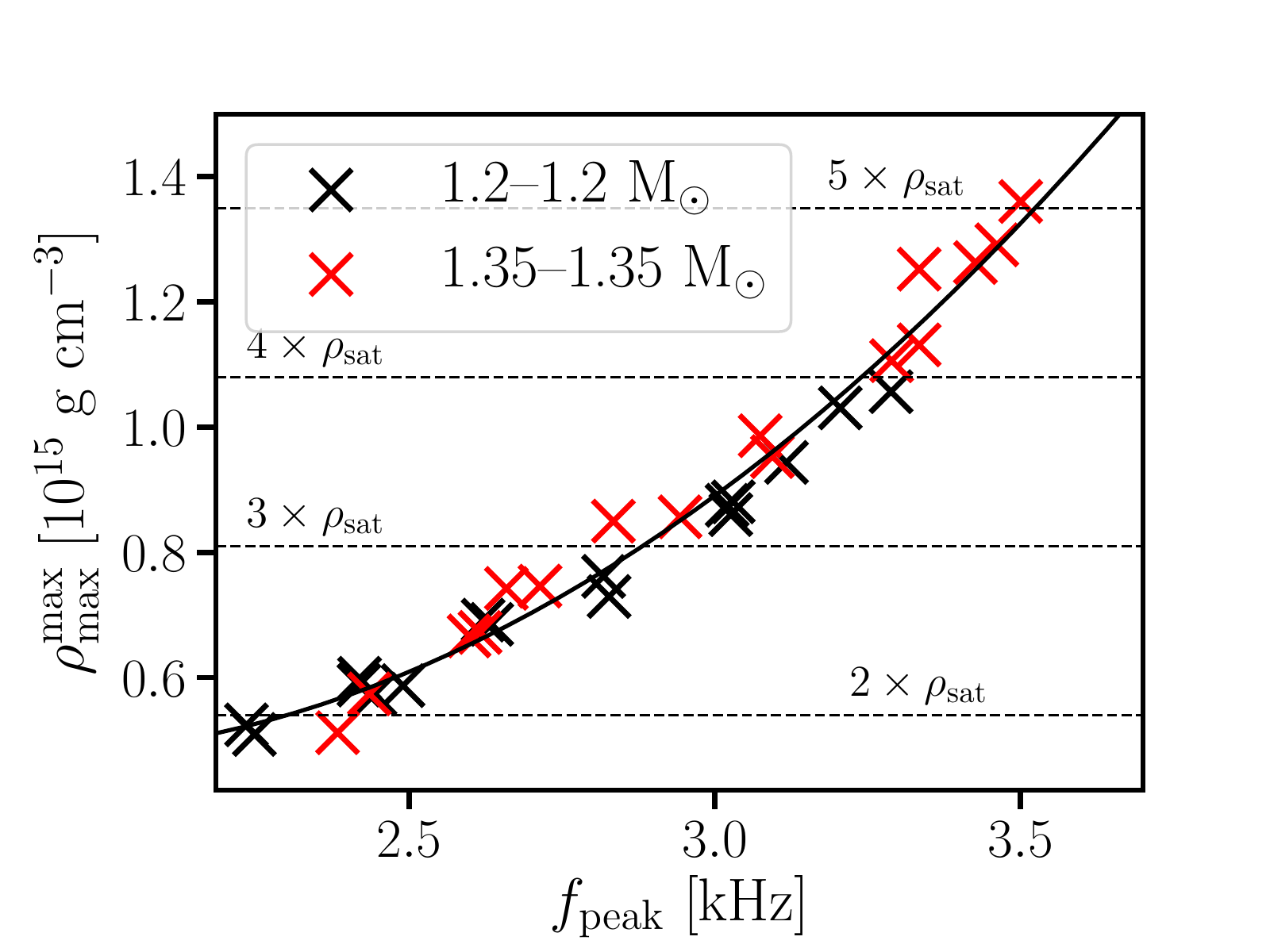}\label{rhomaxfpeakcombirelations_a}}
\subfigure[]{\includegraphics[width=0.32\linewidth]{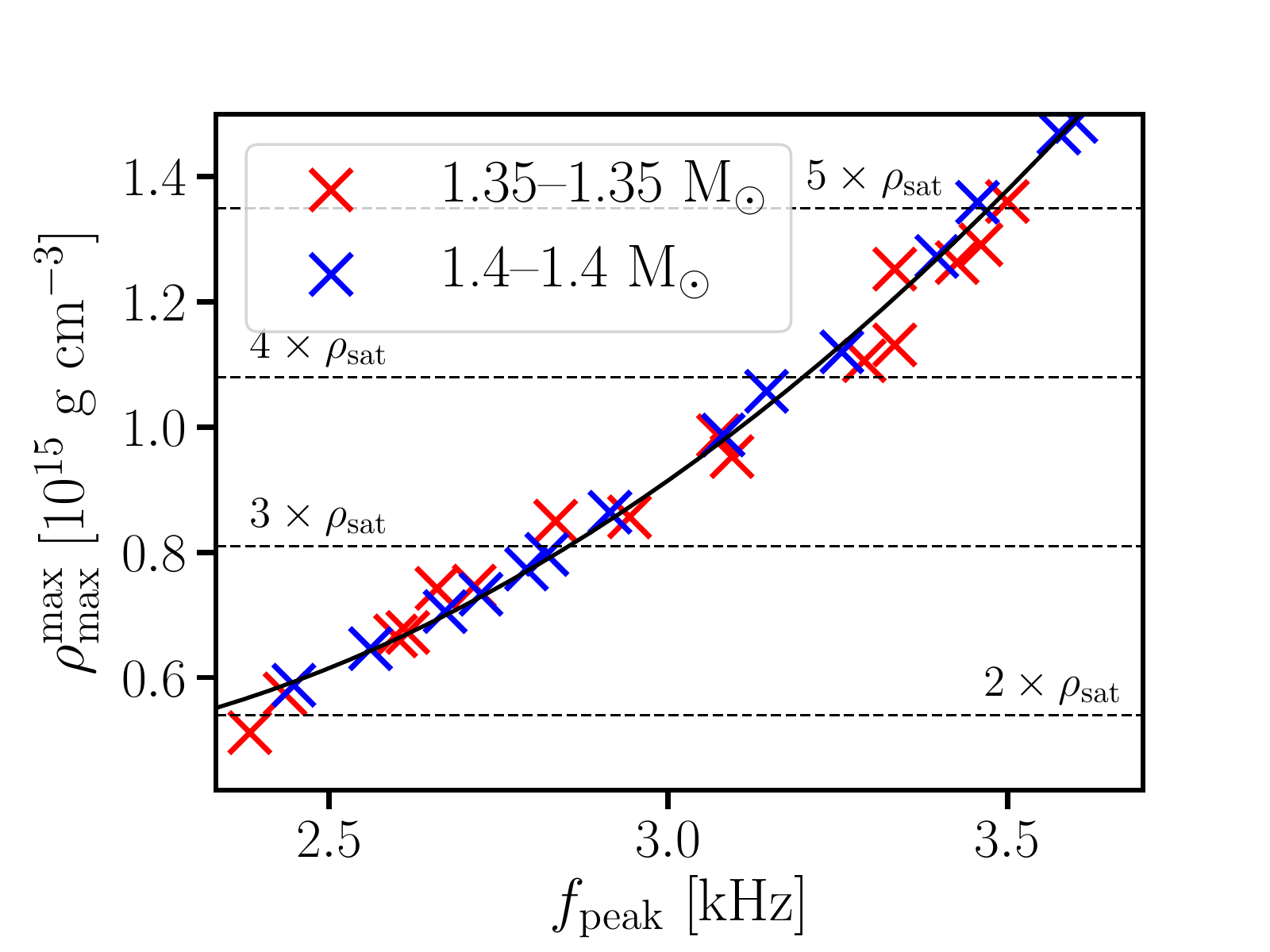}\label{rhomaxfpeakcombirelations_b}}
\subfigure[]{\includegraphics[width=0.32\linewidth]{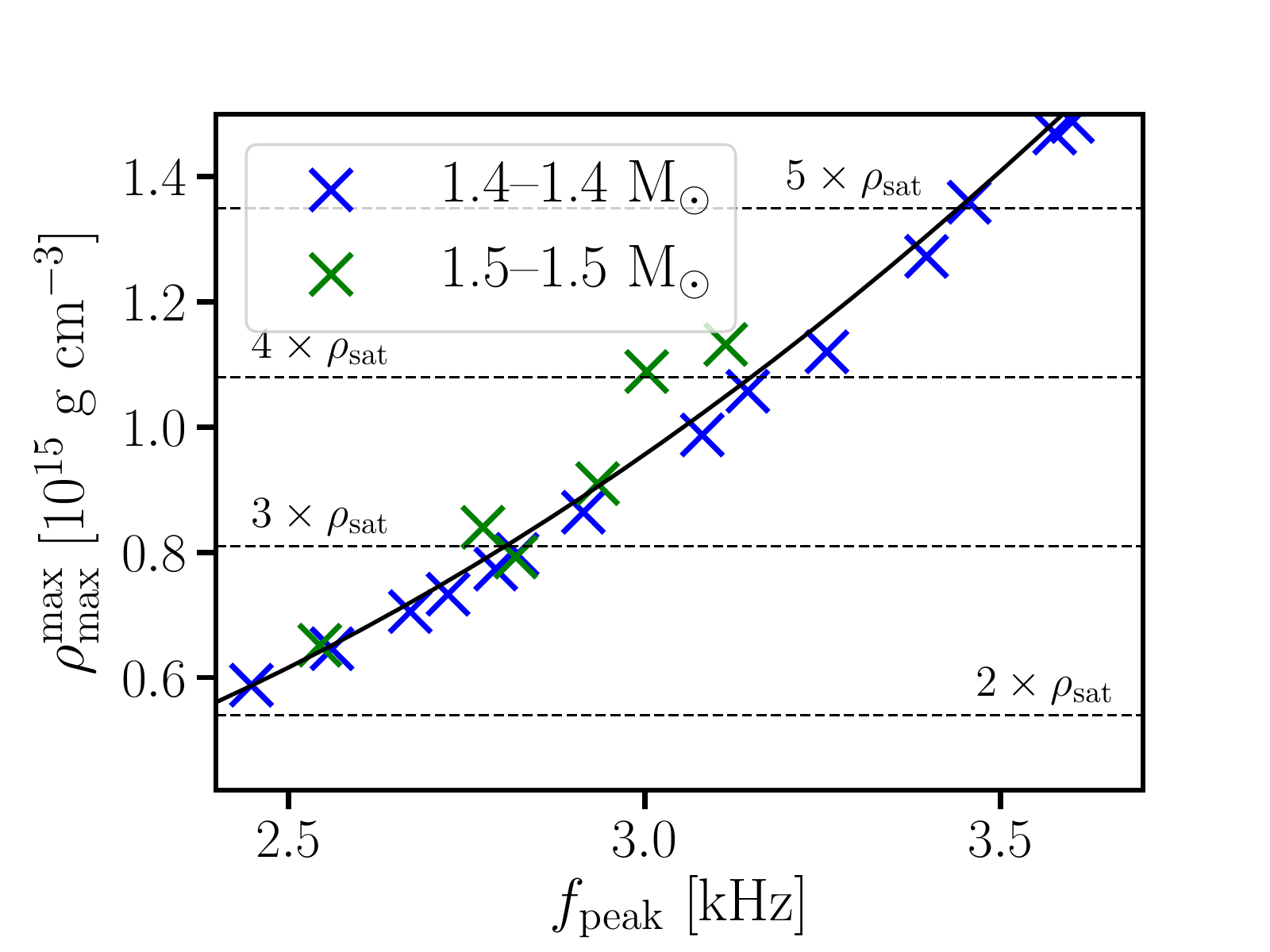}\label{rhomaxfpeakcombirelations_c}}
\caption{Maximum rest-mass density $\rho_{\mathrm{max}}^{\mathrm{max}}$ in the remnant during the first 5 milliseconds after the merger as a function of the dominant postmerger GW frequency $f_\mathrm{peak}$ for 1.2--1.2~M$_{\odot}$ and 1.35--1.35~M$_{\odot}$ binaries (graph~(a)), 1.35--1.35~M$_{\odot}$ and 1.4--1.4~M$_{\odot}$ binaries (graph~(b)) and for 1.4--1.4~M$_{\odot}$ and 1.5--1.5~M$_{\odot}$ binaries (graph~(c)) with purely hadronic EoSs. Different colored crosses refer to data from different total binary masses. The solid black lines are least squares fits with a second order polynomial to the data.}
  \label{rhomaxfpeakcombirelations}
\end{figure*}

These fits are shown in Fig.~\ref{rhomaxfpeakcombirelations} together with data from the total binary masses of 1.2--1.2~M$_{\odot}$ and 1.35--1.35~M$_{\odot}$ (Fig.~\ref{rhomaxfpeakcombirelations_a}), 1.35--1.35~M$_{\odot}$ and 1.4--1.4~M$_{\odot}$ (Fig.~\ref{rhomaxfpeakcombirelations_b}) as well as 1.4--1.4~M$_{\odot}$ and 1.5--1.5~M$_{\odot}$ (Fig.~\ref{rhomaxfpeakcombirelations_c}).
Different colored crosses refer to data from purely hadronic EoSs from different binary masses. The solid black line in each graph displays the fit of Eq.~\eqref{QuadraticrhoFit} to the data in the respective plot.

By comparing Fig.~\ref{rhomaxfpeakcombirelations} to the single universal relation (Fig.~\ref{rhomaxfpeakuniversal}) one can see that this procedure reduces the maximum deviations of the data from the respective fit in every mass range.
\subsection{$\rho_{\mathrm{max}}^{\mathrm{max}}-\Lambda$ relations}\label{moreplotsrholambda}
\begin{figure*}
\centering
\subfigure[]{\includegraphics[width=0.32\linewidth]{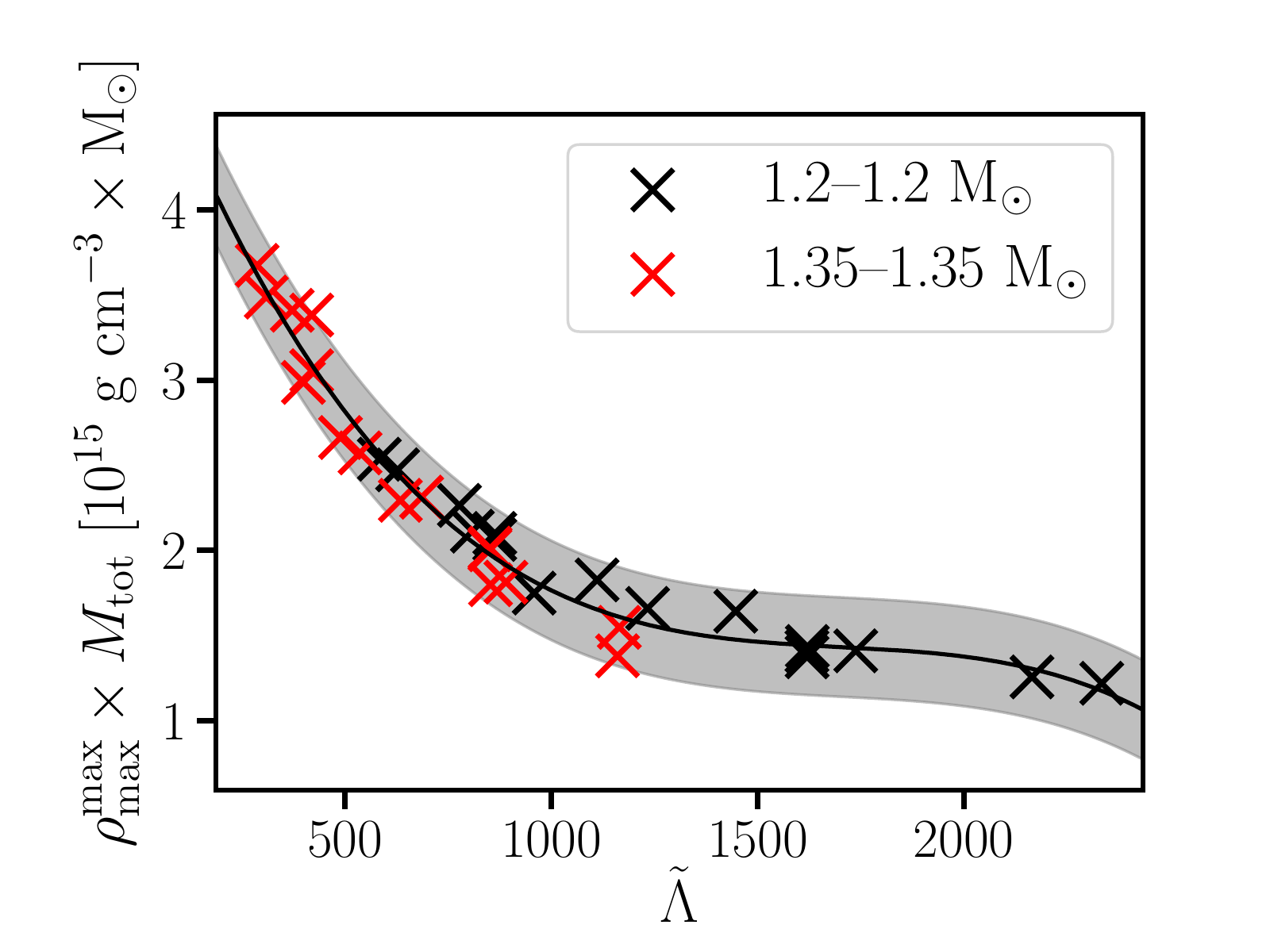}\label{rhomaxlambdacombirelations_a}}
\subfigure[]{\includegraphics[width=0.32\linewidth]{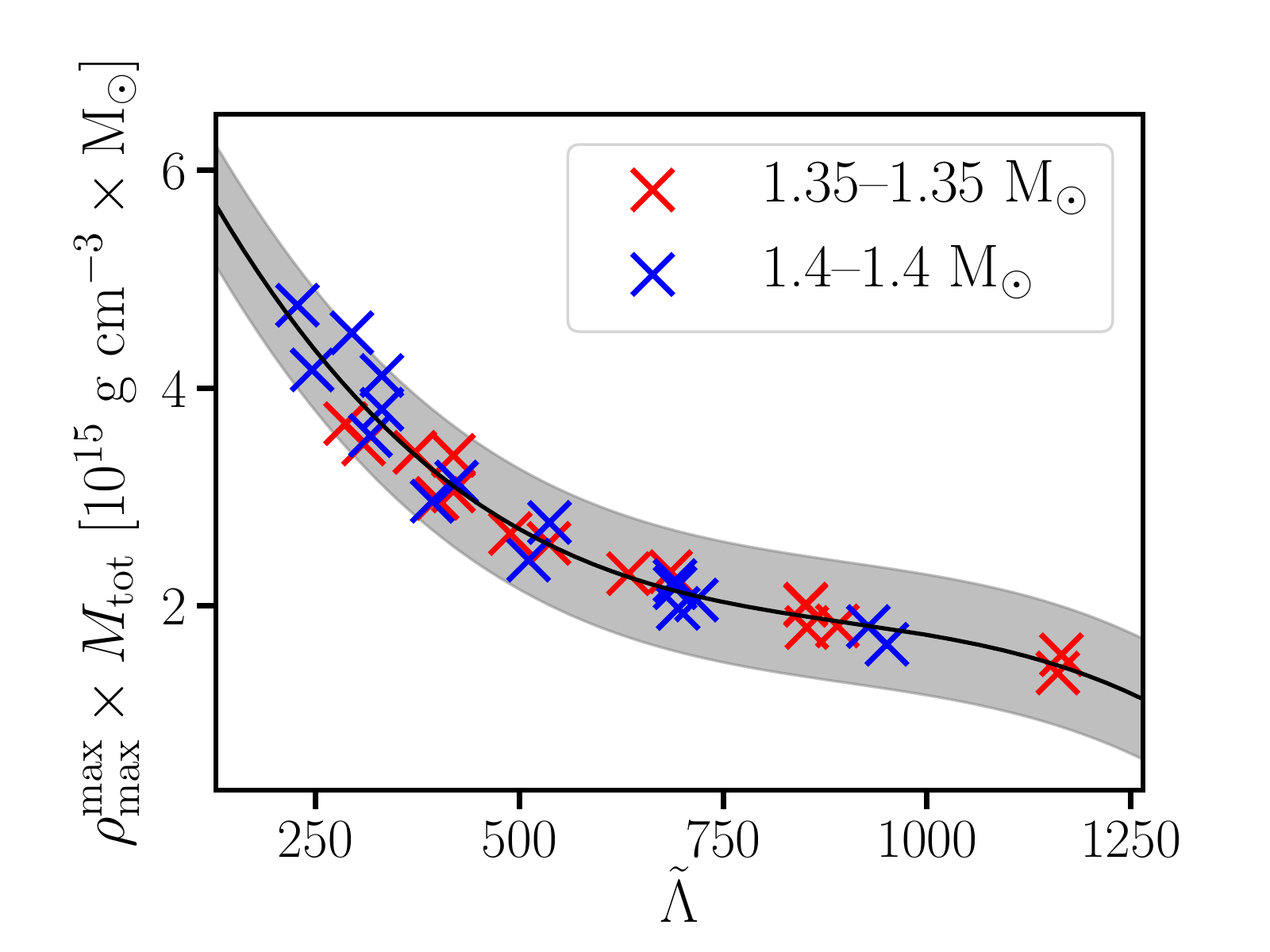}\label{rhomaxlambdacombirelations_b}}
\subfigure[]{\includegraphics[width=0.32\linewidth]{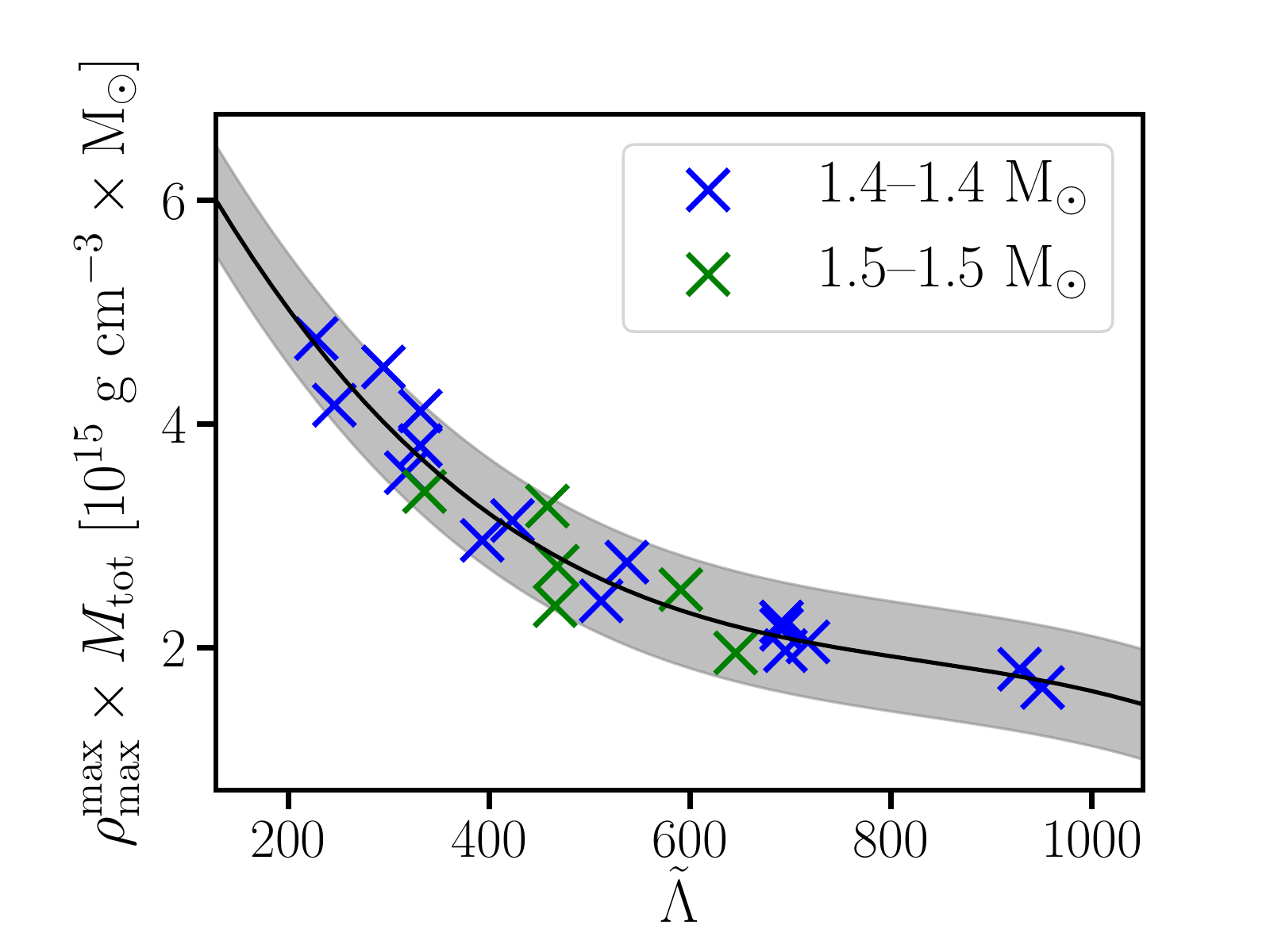}\label{rhomaxlambdacombirelations_c}}
  \caption{Maximum rest-mass density $\rho_{\mathrm{max}}^{\mathrm{max}}$ in the remnant during the first 5 milliseconds after the merger scaled by the total binary mass $M_{\mathrm{tot}}$ as a function of the combined tidal deformability $\tilde{\Lambda}$ for 1.2--1.2~M$_{\odot}$ and 1.35--1.35~M$_{\odot}$ binaries (graph~(a)), 1.35--1.35~M$_{\odot}$ and 1.4--1.4~M$_{\odot}$ binaries (graph~(b)) and for 1.4--1.4~M$_{\odot}$ and 1.5--1.5~M$_{\odot}$ binaries (graph~(c)) with purely hadronic EoSs.  Different colored crosses refer to data from different total binary masses. The solid black lines are least squares fits with Eq.~\eqref{rhoOfLambda} to the data. The gray shaded areas illustrate the maximum deviation of the data from the respective fit.}
  \label{rhomaxlambdacombirelations}
\end{figure*}

Finally, we discuss the same procedure for the relation between $\rho_{\mathrm{max}}^{\mathrm{max}}$ and $\Lambda$ (see Sect.~\ref{rholambda}).
As before, we find that accuracy of this relation can be increased by fitting results from 1.2--1.2~M$_{\odot}$ and 1.35--1.35~M$_{\odot}$, 1.35--1.35~M$_{\odot}$ and 1.4--1.4~M$_{\odot}$ as well as from 1.2--1.2~M$_{\odot}$ and 1.35--1.35~M$_{\odot}$ merger simulations separately. The least squares fit parameters for every mass range are shown in Tab.~\ref{rhoOfLambdaCombiFits}. These fits are shown in Fig.~\ref{rhomaxlambdacombirelations} together with data from the total binary masses of 1.2--1.2~M$_{\odot}$ and 1.35--1.35~M$_{\odot}$ (Fig.~\ref{rhomaxlambdacombirelations_a}), 1.35--1.35~M$_{\odot}$ and 1.4--1.4~M$_{\odot}$ (Fig.~\ref{rhomaxlambdacombirelations_b}) and from 1.4--1.4~M$_{\odot}$ and 1.5--1.5~M$_{\odot}$ (Fig.~\ref{rhomaxlambdacombirelations_c}).
Different colored crosses refer to data from different total binary masses of purely hadronic EoSs. The solid black lines show least squares fits to the data in the respective plot. The gray shaded areas illustrate the maximum deviation of the data from each fit.

A comparison of Fig.~\ref{rhomaxlambdacombirelations} to Fig.~\ref{rhomaxlamuniversal} demonstrates that the procedure of fitting different mass ranges individually decreases the overall maximum deviation of the data from the fits.

\section{Determination of $\frac{d\,\Lambda}{d\,M_\mathrm{tot}}$}\label{detslope}
Here we describe how we obtain the slope $\frac{d\,\Lambda}{d\,M_\mathrm{tot}}$ of the tidal deformability $\Lambda$, which we employ in the main paper to interpolate $\Lambda$ to slightly different total binary masses. Here, $\Lambda$ refers to the tidal deformability of a single star with mass $M$, while $M_\mathrm{tot}$ is the total mass of a NS binary. We only consider equal-mass binaries, i.e. $M_\mathrm{tot}=2M$. In this case $\Lambda$ coincides with the combined tidal deformability $\tilde{\Lambda}$, which is the parameter that describes finite-size effects in waveform models.
 
The tidal deformability is defined by $\Lambda(M)=2/3k_{2}(M)(R(M)/M)^{5}$ with the tidal Love number $k_{2}(M)$ and the stellar radius $R(M)$ both of which are functions of mass. For small changes in $M$ and typical NS masses, the radius and $k_{2}$ do not strongly vary with mass for a hadronic EoS in a range of moderately high masses. Hence, the mass dependence of $\Lambda$ is dominated by the term $(R/M)^{5}$ and we suspect that the slope $\frac{d\,\Lambda}{d\,M}$ can be approximated as
\begin{align}
    \frac{d\,\Lambda}{d\,M}=-\frac{5}{M} \frac{2k_{2}R^{5}}{3M^{5}}=-5\frac{\Lambda}{M}.
\end{align}
With $M_\mathrm{tot}=2M$ we thus expect a relation of the form $\frac{d\,\Lambda}{d\,M_\mathrm{tot}}=z\frac{\Lambda}{M_\mathrm{tot}}$. We obtain the parameter z by calculating $\Lambda$ for every hadronic EoS used in this paper for NS masses of 1.2~M$_{\odot}$, 1.3~M$_{\odot}$, 1.4~M$_{\odot}$ and 1.5~M$_{\odot}$ which is the relevant mass range. We then determine the slope $\frac{d\,\Lambda}{d\,M_\mathrm{tot}}$ through finite differencing.

\begin{figure}[ht]
\centering
\includegraphics[width=1.0\linewidth]{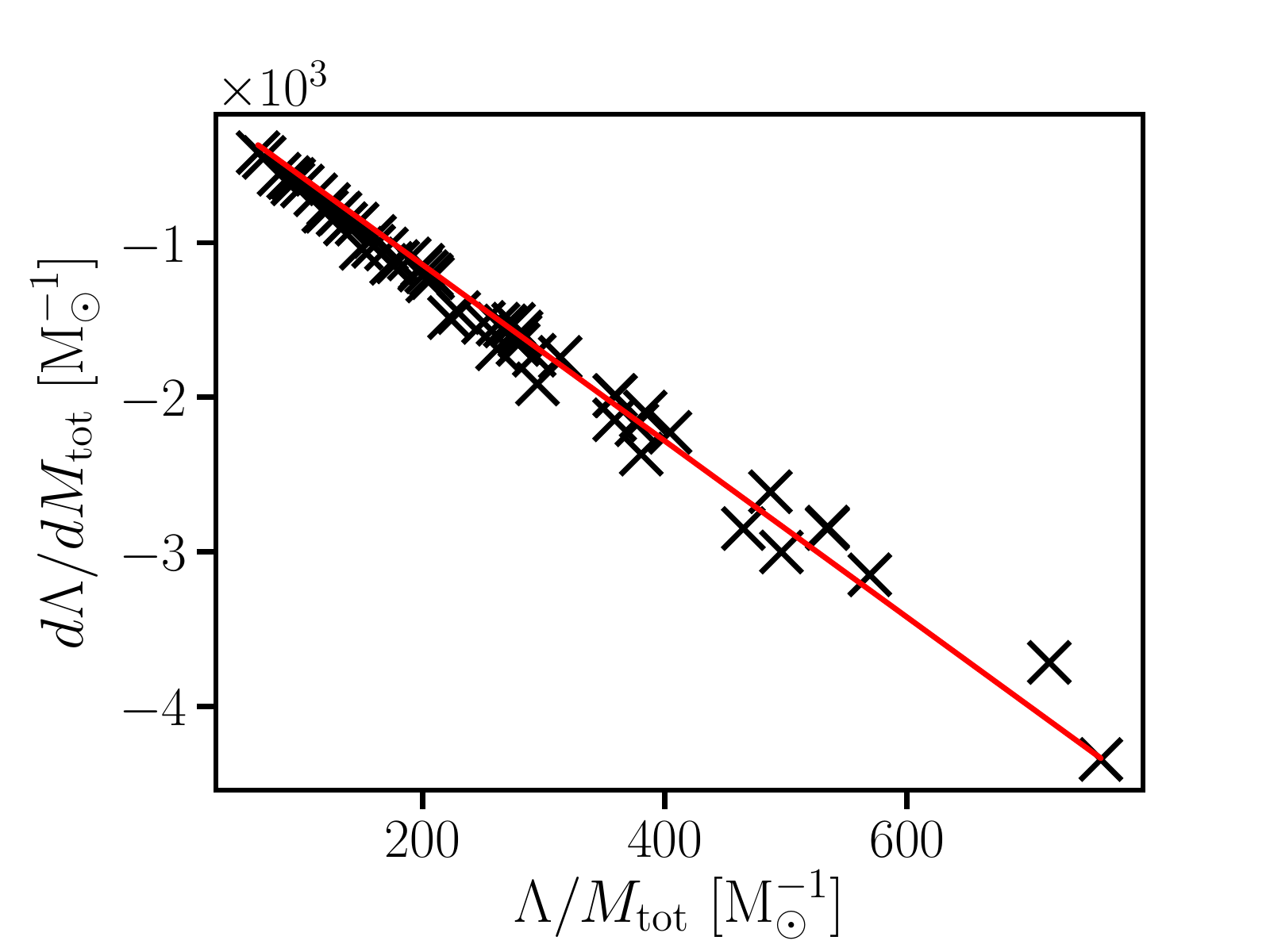}
\caption{The slope $\frac{d\,\Lambda}{d\,M_\mathrm{tot}}$ of the tidal deformability $\Lambda$ with respect to the total binary mass $M_\mathrm{tot}$ as a function of $\frac{\Lambda}{M_\mathrm{tot}}$ for all hadronic EoSs used in this Paper. The red line shows a least squares fit $\frac{d\,\Lambda}{d\,M_\mathrm{tot}}=z\frac{\Lambda}{M_\mathrm{tot}}$.} 
\label{dLambdaoverdM}
\end{figure}

Fig.~\ref{dLambdaoverdM} displays $\frac{d\,\Lambda}{d\,M_\mathrm{tot}}$ as a function of $\frac{\Lambda}{M_\mathrm{tot}}$. We determine the parameter $z$ through a least squares fit and obtain $z=\mathrm{-}5.709$, which is somewhat larger than 5. We also refer to \cite{De2018} which first noted that $\Lambda$ varies roughly with $M^{-6}$. 

\section{Determination of $\Delta M$} \label{detDeltaM}
\begin{figure}
\centering
\includegraphics[width=1.0\linewidth]{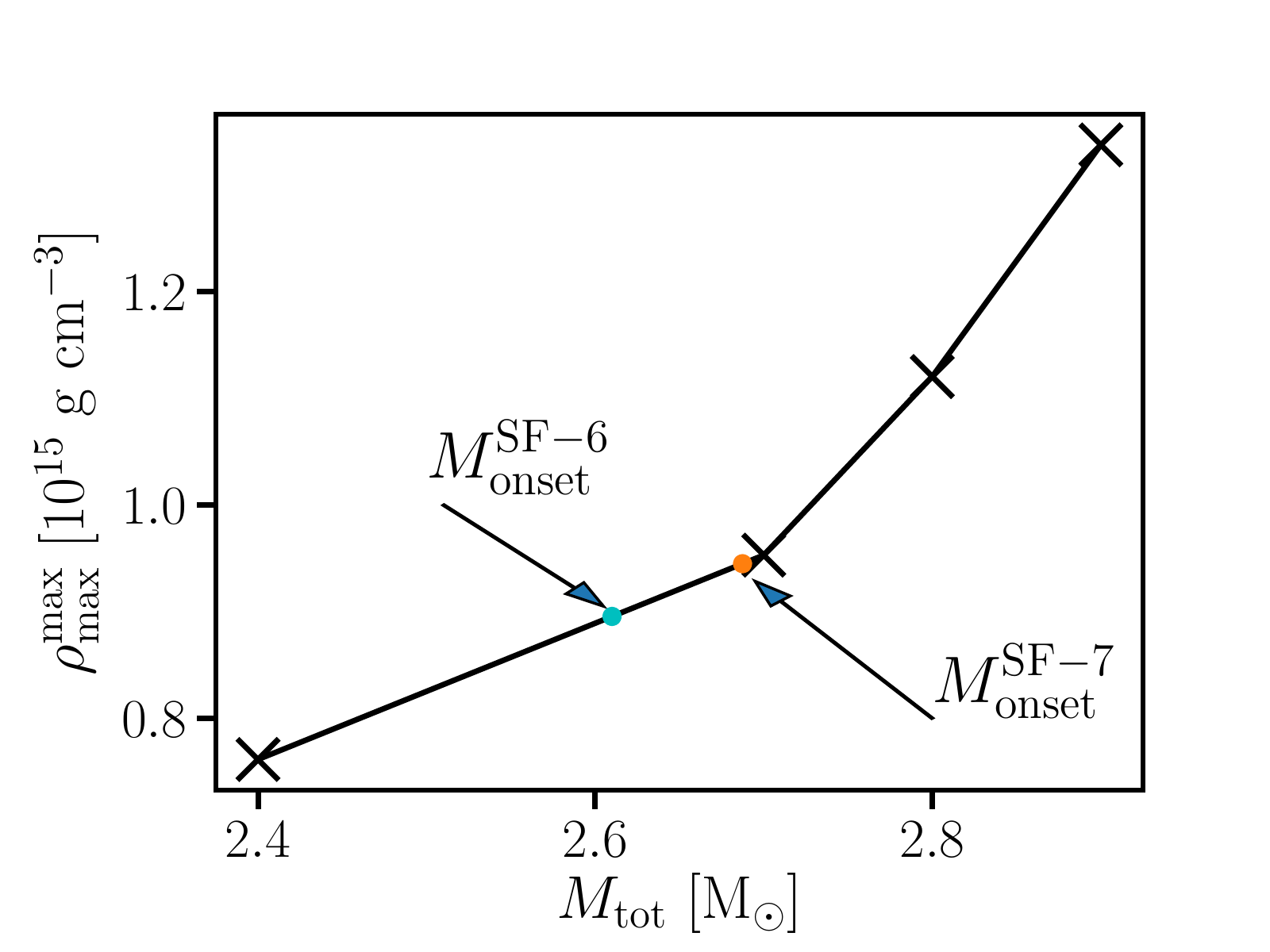}
\caption{Maximum rest-mass density within the first 5~ms after the merger $\rho_{\mathrm{max}}^{\mathrm{max}}$ as a function of the total binary mass $M_{\mathrm{tot}}$ for the purely hadronic DD2F EoS. Crosses show actual simulation data, between these points linear interpolation is used. The cyan (orange) point indicates the $M_{\mathrm{tot}}$ value for which $\rho_{\mathrm{max}}^{\mathrm{max}}$ is equal to the transition density of the hybrid DD2F-SF-6 (DD2F-SF-7) EoS.}
\label{rhomaxmaxmtotdd2f}
\end{figure}

\begin{figure}
\centering
\includegraphics[width=1.0\linewidth]{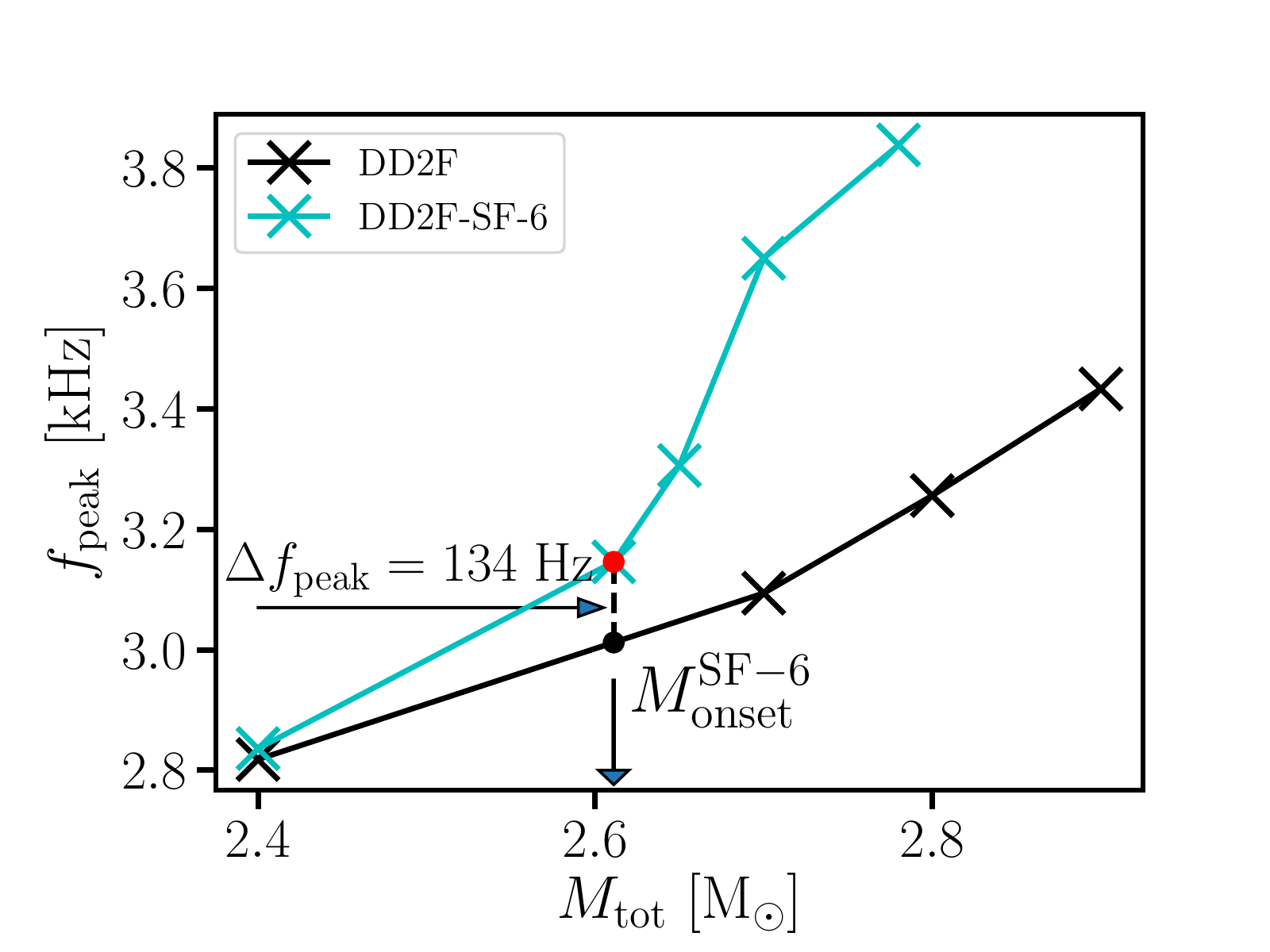}
\caption{Dominant postmerger GW frequency $f_{\mathrm{peak}}$ as a function of the total binary mass $M_{\mathrm{tot}}$ for the purely hadronic DD2F EoS (black) and the hybrid DD2F-SF-6 EoS (cyan). Crosses show simulation data, which we connect with a linear interpolation. The dots highlight data points with a binary mass $M_\mathrm{tot,onset}$ where $\rho_{\mathrm{max}}^{\mathrm{max}}$ in a simulation with the purely hadronic DD2F model is expected to reach the transition density of the DD2F-SF-6 EoS (see Fig.~\ref{rhomaxmaxmtotdd2f}).}
\label{fpeakmtotSF6}
\end{figure}

\begin{figure}
\centering
\includegraphics[width=1.0\linewidth]{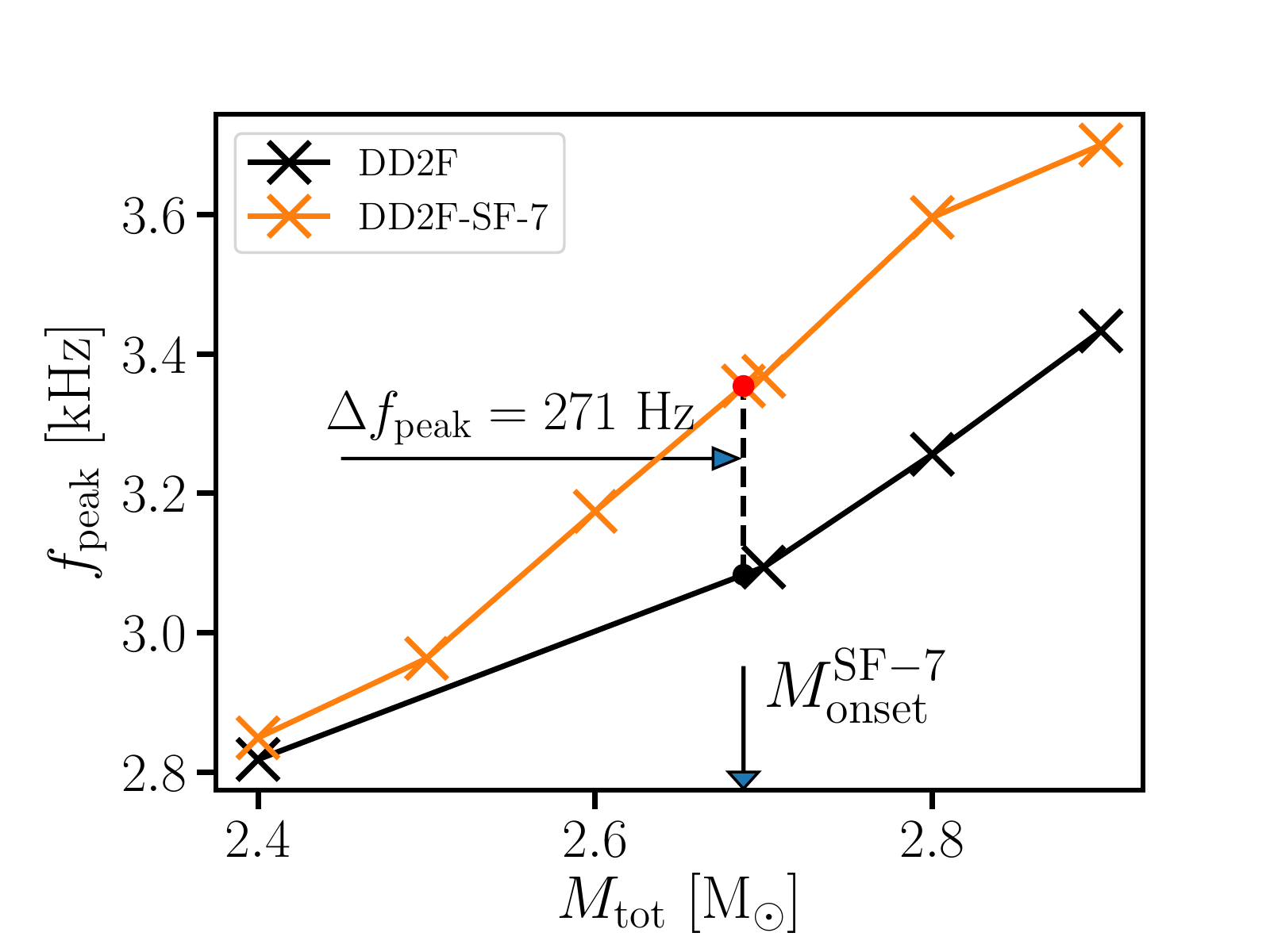}
\caption{Dominant postmerger GW frequency $f_{\mathrm{peak}}$ as a function of the total binary mass $M_{\mathrm{tot}}$ for the purely hadronic DD2F EoS (black) and the hybrid DD2F-SF-7 EoS (orange). Crosses show data points, between these points linear interpolation is used. Dots highlight the data points with a total binary mass $M_\mathrm{tot,onset}$ where $\rho_{\mathrm{max}}^{\mathrm{max}}$ in a simulation with the purely hadronic DD2F model is expected to reach the transition density of the DD2F-SF-7 EoS (see Fig.~\ref{rhomaxmaxmtotdd2f}).}
\label{fpeakmtotSF7}
\end{figure}

\begin{table*}
\begin{tabular}{c c c c c c}
\hline\hline
EoS & $n_{\mathrm{onset}}$ & $M_\mathrm{tot,onset}$ & $M_\mathrm{tot,onset}$ +$\Delta M$ & $\Delta f_{\mathrm{peak}}$  & PT detectable?\\
 & [fm$^{-3}$] & $[\mathrm{M}_\odot]$ & $[\mathrm{M}_\odot]$ & [Hz] & \\
\hline
DD2F-SF-1 & 0.533 & 2.579 & $M_\mathrm{tot,onset}-0.079$ & 24 & no \\ 
DD2F-SF-1 & 0.533 & 2.579 & $M_\mathrm{tot,onset}+0.0$ & 239 & yes \\
DD2F-SF-2 & 0.466 & 2.349 & $M_\mathrm{tot,onset}+0.0$ & 69 & no \\
DD2F-SF-2 & 0.466 & 2.349 & $M_\mathrm{tot,onset}+0.051$ & 82 & no \\
DD2F-SF-2 & 0.466 & 2.349 & $M_\mathrm{tot,onset}+0.151$ & 187 & marginally \\
DD2F-SF-2 & 0.466 & 2.349 & $M_\mathrm{tot,onset}+0.201$ & 295 & yes \\
DD2F-SF-3 & 0.538 & 2.611 & $M_\mathrm{tot,onset}-0.061$ & 84 & no \\
DD2F-SF-3 & 0.538 & 2.611 & $M_\mathrm{tot,onset}+0.0$ & 291 & yes \\
DD2F-SF-4 & 0.580 & 2.706 & $M_\mathrm{tot,onset}-0.106$ & 86 & no \\
DD2F-SF-4 & 0.580 & 2.706 & $M_\mathrm{tot,onset}+0.0$ & 274 & yes \\
DD2F-SF-5 & 0.499 & 2.504 & $M_\mathrm{tot,onset}+0.0$ & 79 & no \\
DD2F-SF-5 & 0.499 & 2.504 & $M_\mathrm{tot,onset}+0.096$ & 355 & yes \\
DD2F-SF-6 & 0.545 & 2.611 & $M_\mathrm{tot,onset}+0.0$ & 134 & no \\
DD2F-SF-6 & 0.545 & 2.611 & $M_\mathrm{tot,onset}+0.039$ & 258 & yes \\
DD2F-SF-7 & 0.562 & 2.688 & $M_\mathrm{tot,onset}-0.188$ & 53 & no \\
DD2F-SF-7 & 0.562 & 2.688 & $M_\mathrm{tot,onset}-0.088$ & 172 & marginally \\
DD2F-SF-7 & 0.562 & 2.688 & $M_\mathrm{tot,onset}+0.0$ & 271 & yes \\
\hline\hline
\end{tabular}
\caption{Results from several merger simulations with different hybrid EoSs and different total binary masses. $n_{\mathrm{onset}}$ is the onset density of the PT for the respective hybrid EoS. $M_\mathrm{tot,onset}$ refers to the total binary mass where simulations with the purely hadronic DD2F EoS reach a maximum density which equals the onset density of the respective PT. The fourth column lists the total binary mass of simulations with the different hybrid EoSs, where we quantify the total binary mass shifted by $\Delta M$ relative to $M_\mathrm{tot,onset}$. $\Delta f_{\mathrm{peak}}$  is the difference in the dominant postmerger GW frequency between the hybrid model and the respective purely hadronic simulation with the same total binary mass. The last column summarizes, whether a PT is detectable according to the criteria that the frequency difference $\Delta f_{\mathrm{peak}}$ should be at least 150~Hz.}
\label{tablehybridruns}
\end{table*}

In this appendix we discuss the determination of the parameter $\Delta M$, which effectively absorbs thermal effects of the phase boundary and introduces a buffer that is necessary because too small quark matter cores do not strongly affect the postmerger GW frequency (see Sect. \ref{conservativelimit}). Figure~\ref{rhomaxmaxmtotdd2f} shows the maximum rest-mass density $\rho_{\mathrm{max}}^{\mathrm{max}}$ during the early postmerger evolution of systems with different total binary masses for our hadronic reference EoS DD2F. Unsurprisingly, the maximum density $\rho_\mathrm{max}^\mathrm{max}$ which is reached in the remnant continuously increases with the total binary mass.

From Fig.~\ref{rhomaxmaxmtotdd2f} we determine for which binary mass $M_\mathrm{tot,onset}$ we expect the purely hadronic model to reach a maximum density that equals the onset density of the PT of a given hybrid EoS. If thermal effects were unimportant, we would expect that increasing the total binary mass, quark matter would appear first in this system. However, because of thermal effects quark matter occurs already in binaries with lower binary mass, i.e. at smaller densities.

At the same time, a small quark core might not result in a strong impact on the postmerger GW frequency, as explained in Sect.~\ref{conservativelimit}

To assess these competing effects, we simulate binary mergers with $M_\mathrm{tot,onset}$ and compare the resulting peak frequency of the respective hybrid model with the one of the purely hadronic model.

An example is shown in Fig.~\ref{fpeakmtotSF6} for the DD2F-SF-6 EoS, which has an onset density of 0.545~fm$^{-3}$ (see Tab.~\ref{tablehybridruns}). 
According to Fig.~\ref{rhomaxmaxmtotdd2f}, a merger with the hadronic DD2F would reach this density for a binary mass of $M_\mathrm{tot,onset}= 2.611~$M$_\odot$.
Figure.~\ref{fpeakmtotSF6} shows the dominant postmerger GW frequency as a function of the total binary mass for the hadronic reference model DD2F (black) and for the DD2F-SF-6 model (cyan). In this figure one can conveniently read off at which total binary mass the appearance of quark matter starts to have a strong impact on the postmerger GW emission. This is the case where the two curves start to deviate by more than {\raise.17ex\hbox{$\scriptstyle\mathtt{\sim}$}}150~Hz
(typical maximum residual in Fig.~\ref{fpeaklampdacombirelations}) because then the hybrid model would occur as an outlier in Fig.~\ref{fpeaklampdacombirelations} (note that in this binary mass range both EoSs yield the same tidal deformability during the inspiral).

In Fig.~\ref{fpeakmtotSF6} the postmerger frequency of the simulation with the hybrid EoS (red dot) is shifted by 134~Hz relative to the hadronic model (black dot) at $M_\mathrm{tot}=M_\mathrm{tot,onset}$, i.e. at the binary mass where the purely hadronic model reaches a maximum density which equates the onset density at zero temperature. The fact that there is an increase of $f_\mathrm{peak}$ is a result of the temperature dependence of the transition density as discussed above.

But, the presence of quark matter is only marginally detectable because the increase of $f_\mathrm{peak}$ relative to DD2F is still relatively small. For this binary mass the quark core is still too small to have a strong impact on the postmerger GW emission. However, if we consider the binary with $M_\mathrm{tot}=M_\mathrm{tot,onset}+0.039~$M$_\odot$, the difference between the hybrid model and the purely hadronic model amounts to 258~Hz, which would be larger than the scatter of the $f^\mathrm{had}_\mathrm{peak}-\Lambda$ relation for any hadronic model. It would thus be indicative of the presence of quark matter. For this specific hybrid model $\Delta M=0.039~\mathrm{M}_\odot$ would thus be sufficient to clearly detect the occurrence of quark matter.

An example of the opposite scenario is given by the DD2F-SF-7 EoS, which has an onset density of 0.562~fm$^{-3}$. The orange dot in Fig.~\ref{rhomaxmaxmtotdd2f} shows that a merger with the hadronic DD2F would reach this density for a total binary mass of $M_\mathrm{tot,onset}= 2.688~$M$_\odot$. Figure~\ref{fpeakmtotSF7} compares the dominant postmerger GW frequency as a function of $M_\mathrm{tot}$ for the DD2F-SF-7 (orange) and the hadronic reference model DD2F (black). For this hybrid EoS the postmerger frequency (red dot) is shifted by 271~Hz with respect to the hadronic model (black dot) at a total binary mass of $M_\mathrm{tot}= M_\mathrm{tot,onset}$. This is a substantial difference and would manifest itself in a deviation from the $f^\mathrm{had}_\mathrm{peak}-\Lambda$ relation. A clear deviation would still be observable for total binary masses slightly below $M_\mathrm{tot,onset}$. For this EoS the onset density decreases sufficiently through thermal effects to produce sufficiently large quark cores at total binary masses slightly smaller than $M_\mathrm{tot,onset}$. 

If we consider a binary with $M_{\mathrm{tot}}=M_\mathrm{tot,onset}-0.188~$M$_\odot$ the difference between hybrid and purely hadronic model drops to 53~Hz, which would be smaller than the scatter of the $f^\mathrm{had}_\mathrm{peak}-\Lambda$ relation and therefore be consistent with the assumption of a purely hadronic EoS. Hence, for this hybrid EoS $\Delta M=0.188~$M$_\odot$ would be sufficient to obtain a lower limit on the transition density.\\

We repeat this setup for all seven hybrid models. The results are summarized in Tab.~\ref{tablehybridruns}. We find that for 4 hybrid EoS models the simulations with $M_\mathrm{tot}=M_\mathrm{tot,onset}$ lead to a sufficiently strong increase of the postmerger GW frequency that the presence of a PT would be identified.

For 3 hybrid EoSs, the postmerger GW frequencies for the models with $M_\mathrm{tot}=M_\mathrm{tot,onset}$ are not too different from the purely hadronic reference model. Hence, the appearance of a PT would not be detected. However, calculations with $M_\mathrm{tot}=M_\mathrm{tot,onset}+ 0.201~$M$_\odot$ do yield a sufficiently strong increase of $f_\mathrm{peak}$ and thus an unambiguous signature for the presence of quark matter.

Overall, for capturing extreme hybrid models, we find than employing $\Delta M \approx 0.2~$M$_{\odot} $ is sufficient to obtain conservative limits on $\rho_\mathrm{onset}$ (see Tab.~\ref{tablehybridruns}).

We remark that our hybrid EoS models are based on only one description of hadronic matter. In principle, the procedure of constraining $\rho_\mathrm{onset}$ (i.e. $\Delta M$ that encodes under which conditions quark matter affects the postmerger GW emission) could also somewhat depend on the chosen hadronic model. We note however, that the hadronic EoS DD2F falls roughly in the center of EoS models which are admitted by the constraints from GW170817. Hence, we do not expect significant deviations and we are confident that our conservative choices above are sufficient.

\begin{figure}[ht]
\centering
\includegraphics[width=1.0\linewidth]{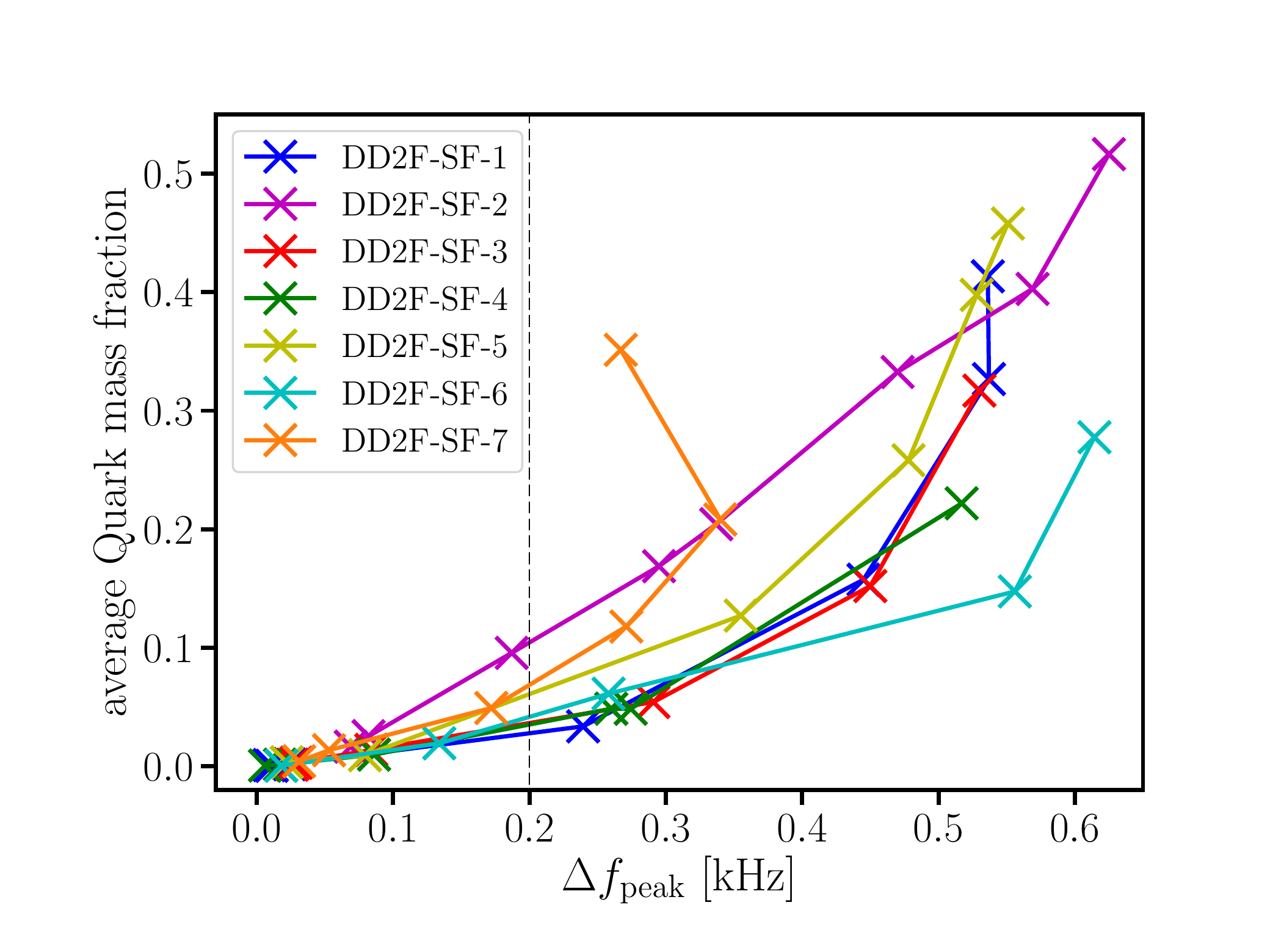}
\caption{Average mass fraction of quark matter in the merger remnant during the first 5\,ms after the merger as a function of the shift in $f_\mathrm{peak}$ with respect to the corresponding purely hadronic merger for binaries with different masses and different hybrid EoSs. Crosses represent actual simulation results, each color refers to a specific hybrid model. The black dashed line marks a shift of $f_\mathrm{peak}$ by 200 Hz caused by the appearance of quark matter, any shift larger than this would be a clear indication of the presence of a PT.}
\label{avQuarksAll}
\end{figure}

Finally, in order to further analyze the behavior of the different models and to summarize the observations (of different $\Delta M$), we show in Fig.~\ref{avQuarksAll} the average mass fraction of quark matter in the first 5~ms after the merger as function of the deviations of $f_\mathrm{peak}$, i.e. the difference between the dominant postmerger GW frequency of the hybrid model and the corresponding purely hadronic system. Crosses represent actual simulation results, which for clarity we connect with straight lines. Generally, for a given EoS the frequency shift increases roughly linearly for not too large quark fractions. For larger quark fractions the frequency shift saturates or even decreases (e.g. for DD2F-SF-7, see also Fig.~\ref{fpeakmtotSF7}). We connect this to the stiffening of the EoS at higher densities in the quark phase.

A difference of $\Delta f_\mathrm{peak}$, which would be indicative of the presence of deconfined quarks, is found for quark mass fraction between roughly 0.05 and 0.1. 
Note that, when considering different quark EoSs, similar shifts in $f_\mathrm{peak}$ can be caused by very different amounts of quark matter. So the increase of $f_\mathrm{peak}$ does not correlate strongly with the amount of quark matter, which indicates that rather the detailed properties of the quark matter EoS are important for shaping the postmerger GW emission and not necessarily the amount and distribution of matter in the quark phase.

\section{Constraints on the onset density}\label{moreplotsconst}

\begin{figure*}[ht]
\centering
\subfigure[]{\includegraphics[width=0.32\linewidth]{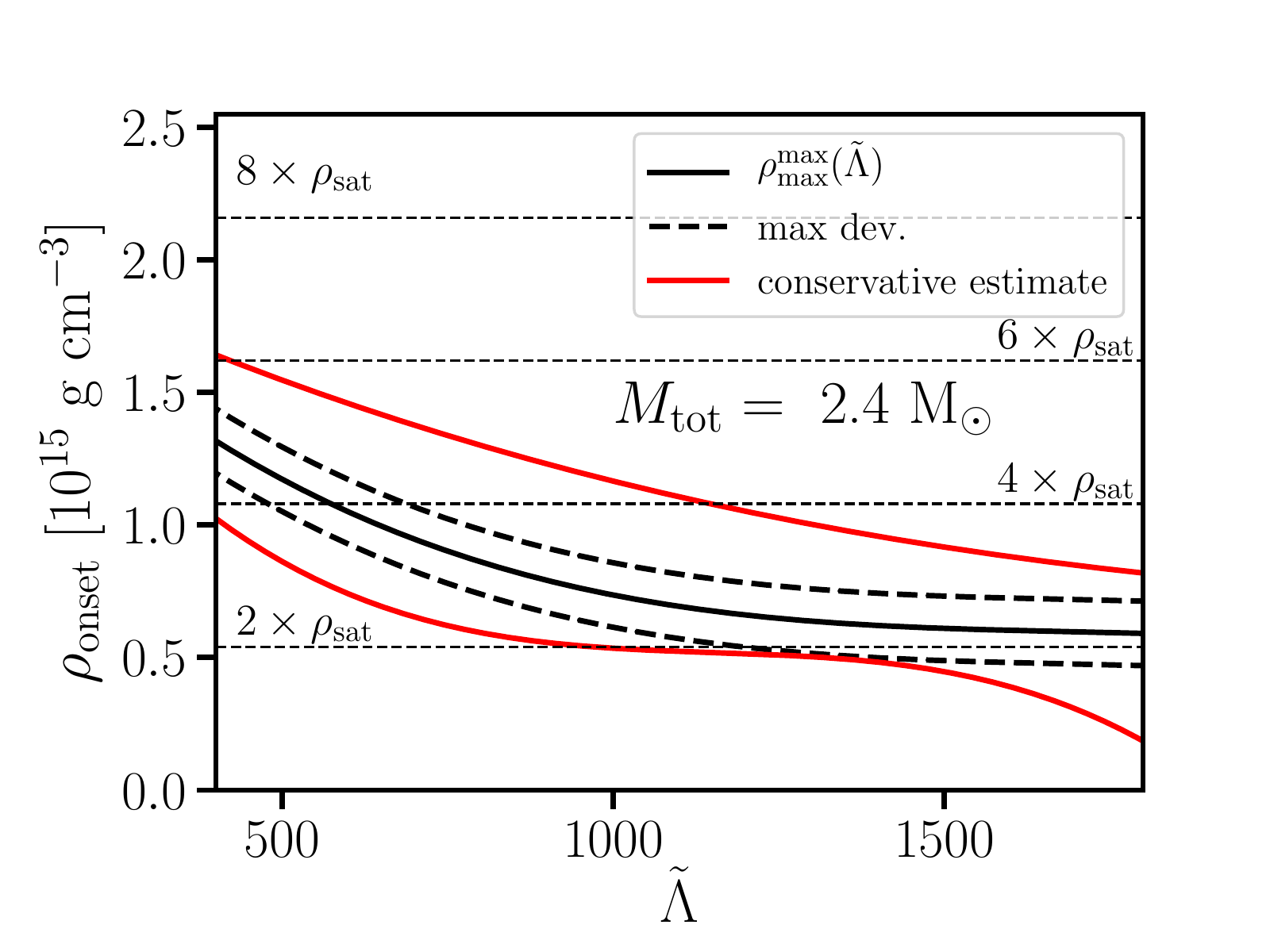}\label{moreconcrete_a}}
\subfigure[]{\includegraphics[width=0.32\linewidth]{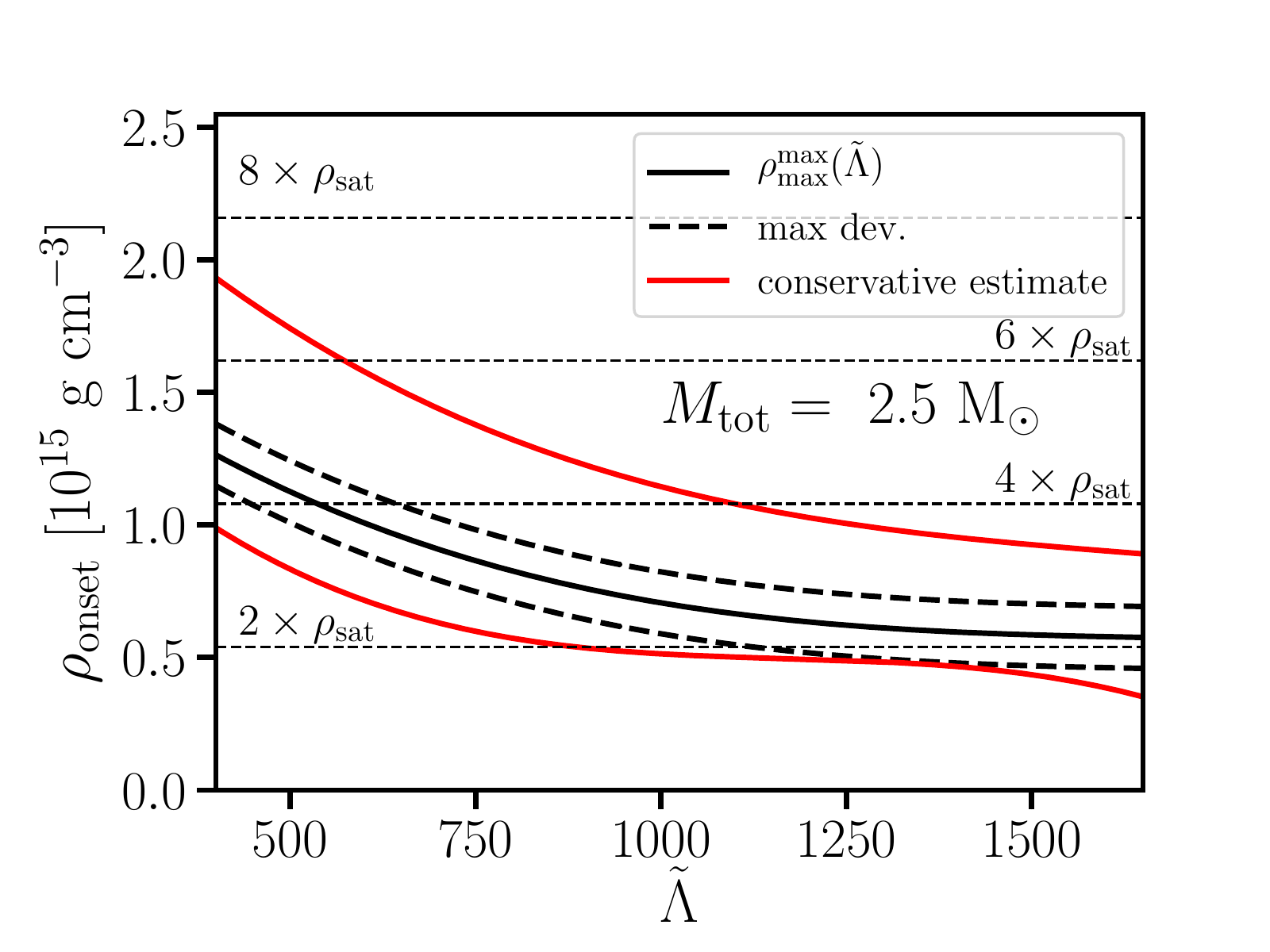}\label{moreconcrete_b}}
\subfigure[]{\includegraphics[width=0.32\linewidth]{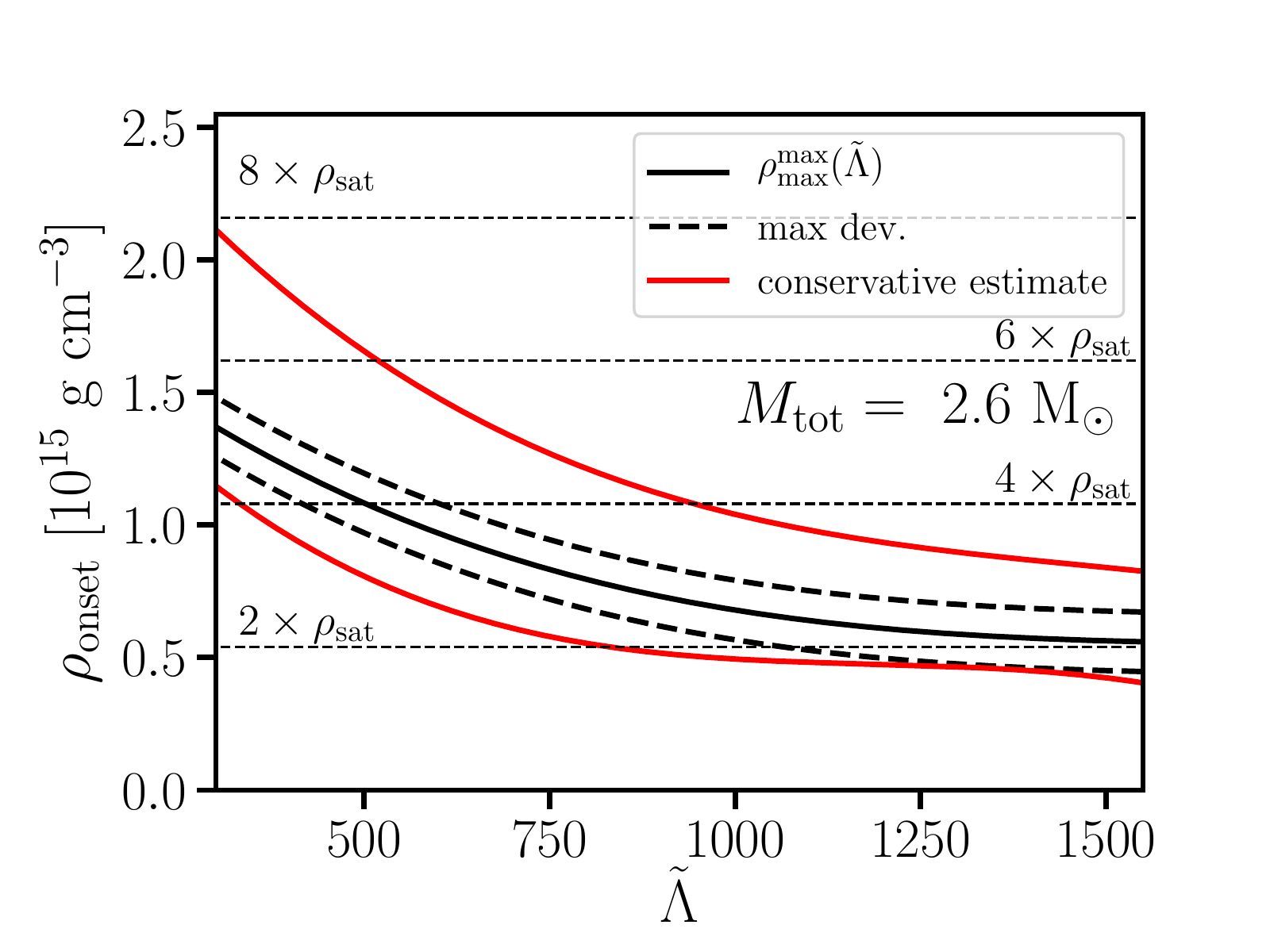}\label{moreconcrete_c}}
\subfigure[]{\includegraphics[width=0.32\linewidth]{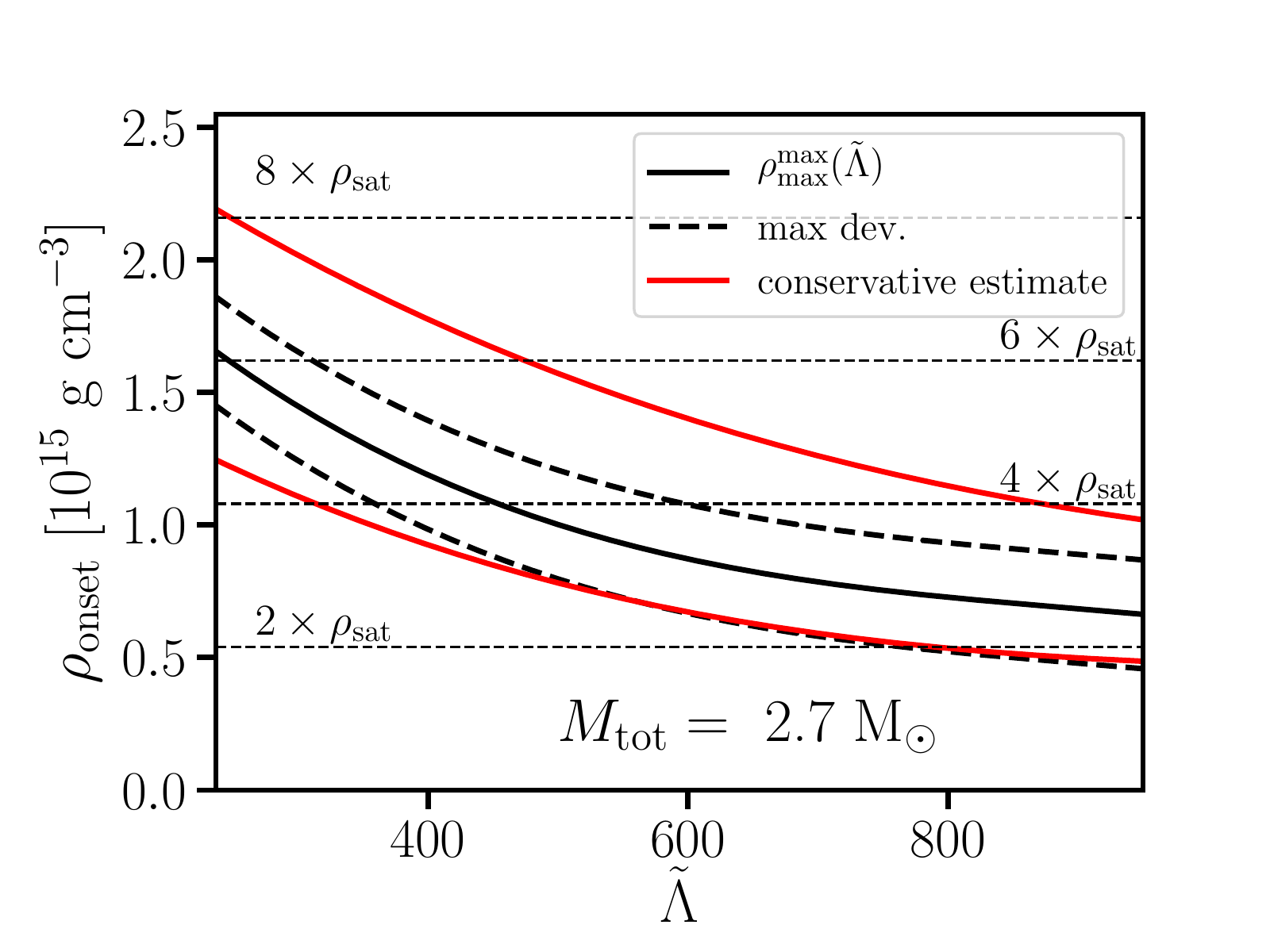}\label{moreconcrete_d}}
\subfigure[]{\includegraphics[width=0.32\linewidth]{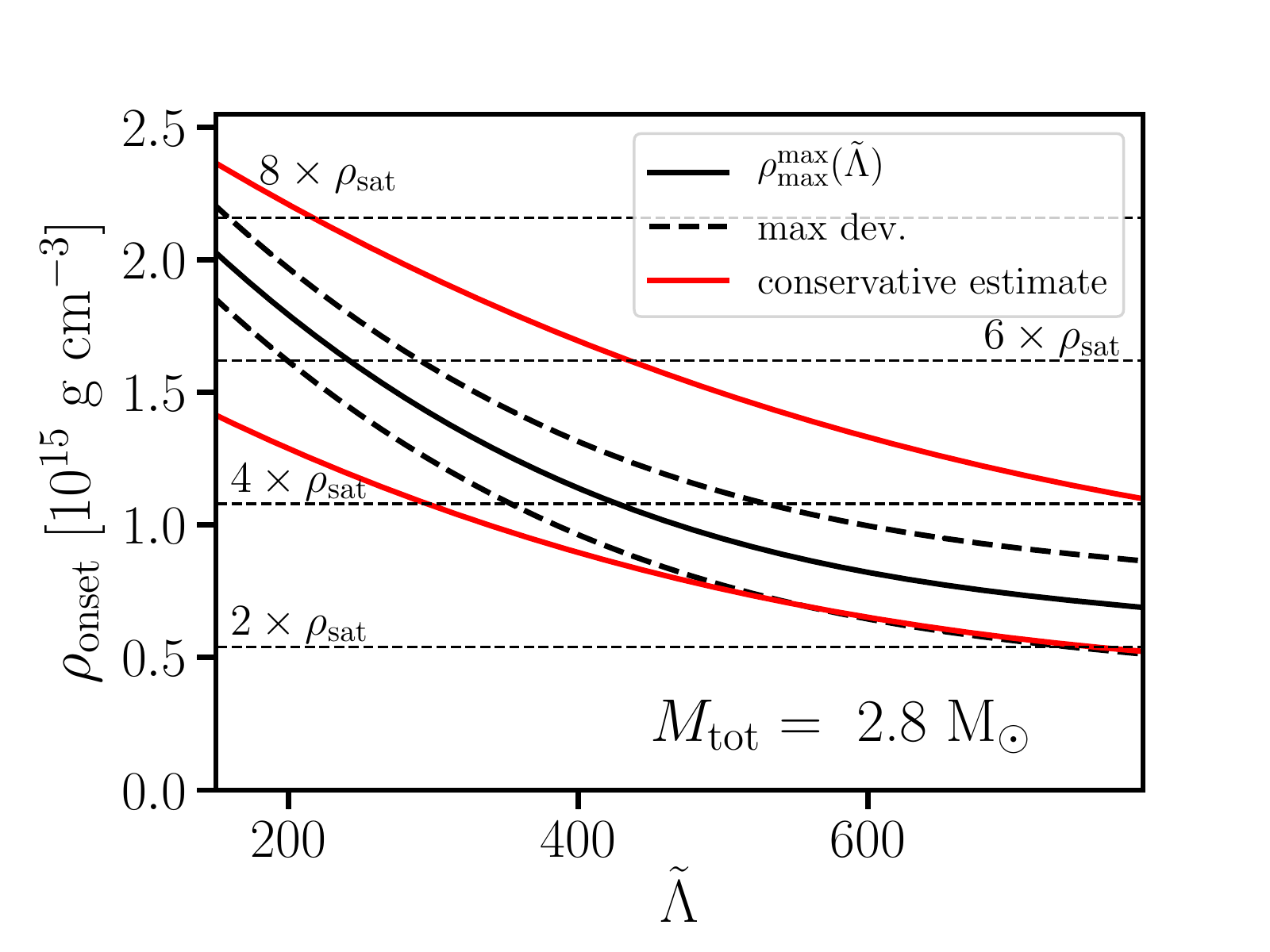}\label{moreconcrete_e}}
\subfigure[]{\includegraphics[width=0.32\linewidth]{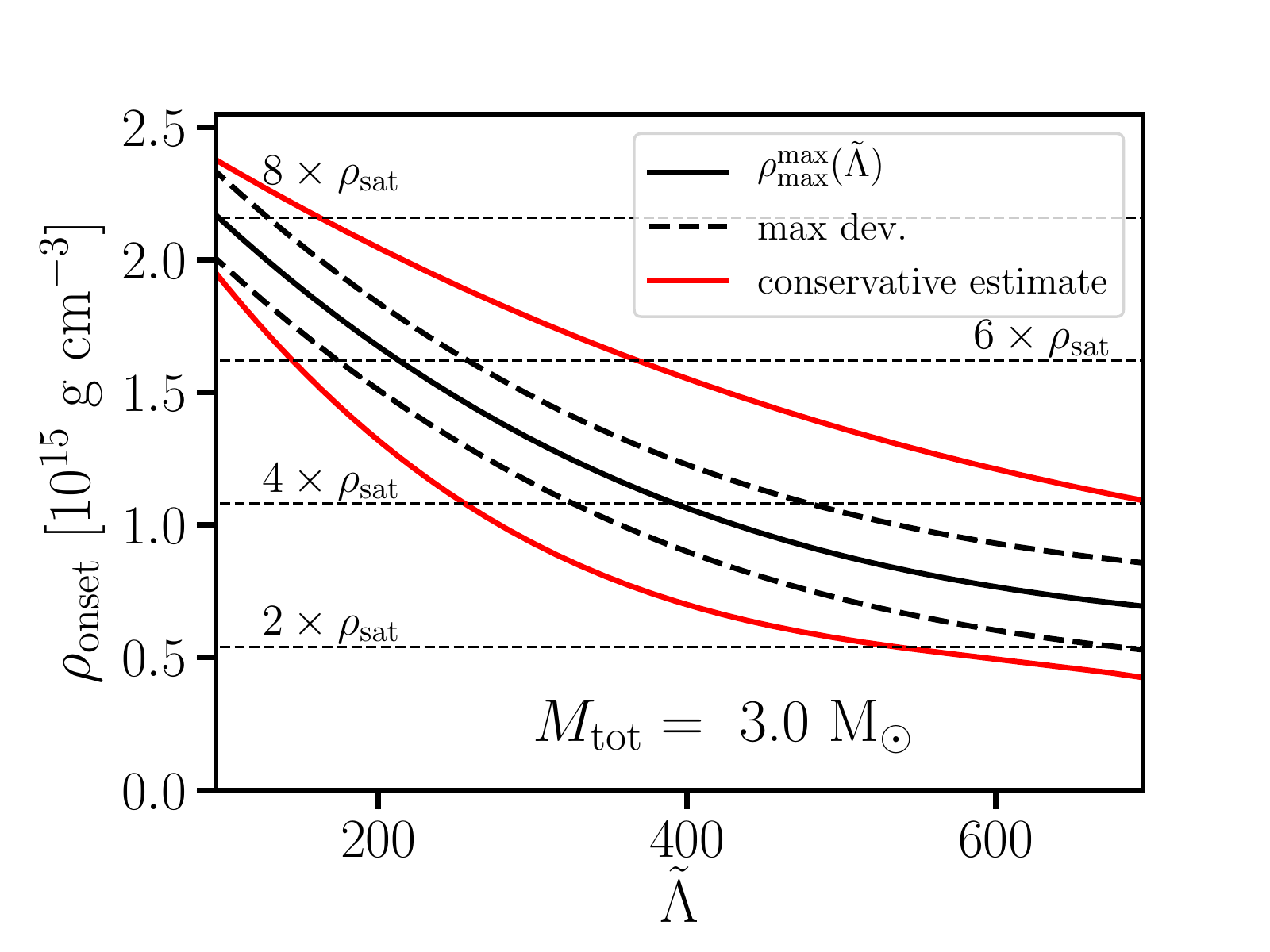}\label{moreconcrete_f}}
\caption{Constraints on the onset density of deconfinement $\rho_{\mathrm{onset}}$ as a function of $\Lambda$ for hypothetical binaries of different total binary masses. The black solid lines show the empirical $\rho_{\mathrm{max}}^{\mathrm{max}}(\Lambda,M_{\mathrm{tot}})$ relation (Eq.~\eqref{rhomaxhypo}) with the parameters from Eq.~\eqref{rhomaxhypopar} for each mass. The dashed lines display the uncertainty of the $\rho_{\mathrm{max}}^{\mathrm{max}}(\Lambda,M_{\mathrm{tot}})$ relation. Depending on the consistency of $f_{\mathrm{peak}}$ with Eq.~\eqref{fhypo} these represent upper or lower limits on $\rho_{\mathrm{onset}}$. The red lines show more conservative constraints introduced in Sect.~\ref{conservativelimit}. See text for more details.}
\label{moreconcrete}
\end{figure*}

In Sect.~\ref{procedure} we show the concrete constraints on the onset density of quark deconfinement which would result from a measurement of $f_\mathrm{peak}$ and $\Lambda$ for a total binary mass of 2.65~M$_\odot$ employing our procedure. In Figs.~\ref{moreconcrete_a}--\ref{moreconcrete_f} we provide the same plot for binary masses of 2.4~M$_\odot$, 2.5~M$_\odot$, 2.6~M$_\odot$, 2.7~M$_\odot$, 2.8~M$_\odot$ and 3.0~M$_\odot$. As before, these plots show possible limits on $\rho_\mathrm{onset}$ as a function of $\Lambda$, where the range of $\Lambda$ is adapted to the respective total binary mass. In each plot the solid black line displays the empirical $\rho_{\mathrm{max}}^{\mathrm{max}}(\Lambda,M_{\mathrm{tot}})$ relation (Eq.~\eqref{rhomaxhypo}) with the corresponding parameters from Eq.~\eqref{rhomaxhypopar} for the given total binary mass. These densities are the maximum densities one would expect in each remnant as a function of $\Lambda$ if no PT occurred. 

The dashed lines illustrate the uncertainty of the $\rho_{\mathrm{max}}^{\mathrm{max}}(\Lambda,M_{\mathrm{tot}})$ relation, which we determine from the maximum scatter within our simulation data. If a measured $f_{\mathrm{peak}}$ is consistent with the $f^\mathrm{had}_{\mathrm{peak}}(\Lambda)$ relation (Eq.~\eqref{fhypo}), the lower dashed lines indicate the lower limit on $\rho_{\mathrm{onset}}$ at this binary mass. The upper dashed lines represent an upper limit on $\rho_{\mathrm{onset}}$ if $f_{\mathrm{peak}}$ is inconsistent with Eq.~\eqref{fhypo} (to within about 200~Hz), which indicates that a strong PT has occurred in the remnant. 

The red lines show the constraints we obtain with our more conservative estimate that involves extrapolating to a binary of slightly different mass introduced in Sect.~\ref{conservativelimit}. Again, the upper lines visualize upper limits on $\rho_{\mathrm{onset}}$ (Eq.~\eqref{rhoonsetupconservative}) if $f_{\mathrm{peak}}$ deviates strongly from Eq.~\eqref{fhypo}. The lower lines depict the lower limits on $\rho_{\mathrm{onset}}$ (Eq.~\eqref{rhoonsetlowconservative}) in cases where $f_{\mathrm{peak}}$ is consistent with Eq.~\eqref{fhypo}.

A refined analysis with improved fit formulae, with fixed binary masses and with a more detailed assessment of the effects which are currently captured by introducing $\Delta M\approx0.2~$M$_\odot$, will likely lead to more stringent constraints for the lower limit in this regime.   

There is also some range of $\Lambda$ for binary masses of 2.7~M$_\odot$ and 2.8~M$_\odot$ where the more conservative procedure leads to larger lower limits than the uncertainty of the $\rho_{\mathrm{max}}^{\mathrm{max}}(\Lambda,M_{\mathrm{tot}})$ relation. This behavior is also an artifact of the different chosen fit formulae, which are employed in the different procedures to derive the limits visualized by the dashed curves and the red curves. In this range of the tidal deformability, which is on the verge of being excluded by GW170817, we include only a few EoS models in our study. Thus, the functional form of the fit is not well constrained. Until a more refined analysis in this parameter range becomes available, one should adopt the dashed curve as the more conservative limit.

Generally, we stress once more that it will be advantageous to simulate a new set of binary mergers once sufficiently accurate measured binary masses are available and to obtain fits for this specific setup. This will lead to tighter constraints on the onset density and generally on the EoSs region probed in the merger remnant.

\end{appendix}

\acknowledgements{We thank B. Friman and T. Galatyuk for helpful discussions. AB acknowledges support by the European Research Council (ERC) under the European Union's Horizon 2020 research and innovation programme under grant agreement No. 759253. This work was funded by Deutsche Forschungsgemeinschaft (DFG, German Research Foundation) - Project-ID 279384907 - SFB 1245 and - Project-ID 138713538 - SFB 881 (``The Milky Way System'', subproject A10). NUFB and TF acknowledge support from the Polish National Science Center (NCN) under grant no. 2019/32/C/ST2/00556 (NUFB) and no. 2016/23/B/ST2/00720 (TF). DB acknowledges support through the Russian Science Foundation under project No. 17-12-01427 and the MEPhI Academic Excellence Project under contract No.~02.a03.21.0005. The work of TS is supported by the Klaus Tschira Foundation. TS is Fellow of the International Max Planck Research School for Astronomy and Cosmic Physics at the University of Heidelberg (IMPRS-HD) and acknowledges financial support from IMPRS-HD. We acknowledge stimulating discussions during the EMMI Rapid Reaction Task Force: The physics of neutron star mergers at GSI/FAIR and the support of networking activities by the COST Actions CA15213 ``THOR'', CA16117 ``ChETEC'' and CA16214 ``PHAROS''.}


\end{document}